\documentclass[fleqn,usenatbib]{mnras}
\usepackage[normalem]{ulem}
\DeclareRobustCommand{\VAN}[3]{#2}
\let\VANthebibliography\thebibliography
\def\thebibliography{\DeclareRobustCommand{\VAN}[3]{##3}\VANthebibliography}

\usepackage{multicol}
\usepackage{afterpage}
\usepackage{lastpage}
\usepackage[super]{nth}
\usepackage{caption}
\usepackage{graphicx}	
\usepackage{amsmath}	
\usepackage{amssymb}	
\usepackage{longtable}
\usepackage{anyfontsize}

\usepackage{newtxtext,newtxmath}
\newcommand{\Rsun}{R$_{\odot}$}

\newcommand{\hvezda}{\ensuremath{\mbox{HD\,34736}}}
\newcommand{\zav}[1]{\left(#1\right)}
\newcommand{\hzav}[1]{\left[#1\right]}

\newcommand{\oc}{\textit{O-C}}

\newcommand{\teff}{$T_{\rm eff}$}
\newcommand{\logg}{$\log g$}
\newcommand{\teffA}{$T_{{\rm eff}A}$}
\newcommand{\loggA}{$\log g_{A}$}
\newcommand{\teffB}{$T_{{\rm eff}B}$}
\newcommand{\loggB}{$\log g_{B}$}

\newcommand{\T}{\mathit{\Theta}}

\newcommand{\bz}{\ensuremath{\langle B_{\rm z}\rangle}}
\newcommand{\ddeg}{$^{\circ}$}
\newcommand{\kms}{km\,s$^{-1}\,$}
\newcommand{\vsini}{$\upsilon_{\rm e}\sin i\,$}
\newcommand{\vr}{$V_{\rm r}$}
\newcommand{\bs}{\ensuremath{\langle B\rangle}}
\def\gtrsim{\mathrel{\hbox{\rlap{\hbox{\lower4pt\hbox{$\sim$}}}\hbox{$>$}}}}
\def\ltsim{\mathrel{\hbox{\rlap{\hbox{\lower4pt\hbox{$\sim$}}}\hbox{$<$}}}}

\defcitealias{2014AstBu..69..191S}{Paper~I}

\title[The magnetic double-lined binary HD\,34736]{HD\,34736: An intensely magnetised double-lined spectroscopic binary with rapidly-rotating chemically peculiar B-type components}
\author[E. Semenko et al.]{E. Semenko,$^{1}$\thanks{E-mail: eugene@narit.or.th}
	O. Kochukhov,$^{2}$
    Z. Mikul\'a\v{s}ek,$^{3}$
	G.~A. Wade,$^{4}$
	E. Alecian,$^{5}$
	D. Bohlender,$^{6}$
	\newauthor
	B. Das, $^{7}$
	D.~L.~Feliz, $^{8}$
	J. Jan\'ik,$^{3}$
    J.~Kol\'{a}\v{r},$^{3}$
    J.~Krti\v{c}ka,$^{3}$
	D.~O. Kudryavtsev,$^{9}$
	J.~M.~Labadie-Bartz,$^{10}$
    \newauthor
	D. Mkrtichian,$^{1}$
	D. Monin,$^{6}$
	V. Petit,$^{11}$
    I.~I. Romanyuk,$^{9}$
	M.~E. Shultz,$^{11}$
    D. Shulyak,$^{12}$
	R.~J.~Siverd,$^{13}$
    \newauthor
	A. Tkachenko,$^{14}$
    I.~A. Yakunin,$^{9,15}$
	M. Zejda,$^{3}$
	 and the BinaMIcS collaboration
	\\
$^{1}$National Astronomical Research Institute of Thailand, 260 Moo 4, T. Donkaew, A. Maerim, 50180, Chiangmai, Thailand \\
$^{2}$Department of Physics and Astronomy, Uppsala University, Box 516, 75120 Uppsala, Sweden \\
$^{3}$Department of Theoretical Physics and Astrophysics, Masaryk University, Kotl\'a\v{r}sk\'a 2, CZ 611 37 Brno, Czech Republic \\
$^{4}$Department of Physics \& Space Science, Royal Military College of Canada, PO Box 17000 Station Forces, Kingston, ON, Canada K7K 0C6 \\
$^{5}$Universit\'e Grenoble Alpes, IPAG, F-38000 Grenoble, France \\
$^{6}$National Research Council of Canada, Herzberg Astronomy and Astrophysics Research Centre, 5071 West Saanich Road, Victoria, \\ BC V9E 2E7, Canada\\
$^{7}$CSIRO, Space and Astronomy, P.O. Box 1130, Bentley WA 6102, Australia\\
$^{8}$American Museum of Natural History, 200 Central Park West, Manhattan, NY 10024, USA \\
$^{9}$Special Astrophysical Observatory, Russian Academy of Sciences, Nizhnii Arkhyz, Russia, 369167 \\
$^{10}$LESIA, Paris Observatory, PSL University, CNRS, Sorbonne University, Universit\'e Paris Cit\'e, 5 place Jules Janssen, 92195 Meudon, France \\
$^{11}$Dept. of Physics and Astronomy \& Bartol Research Institute, University of Delaware, Newark, DE, 19716, USA \\
$^{12}$Instituto de Astrof\'{\i}sica de Andaluc\'{\i}a - CSIC, c/ Glorieta de la Astronom\'{\i}a s/n, 18008 Granada, Spain \\
$^{13}$Institute for Astronomy, University of Hawaii, 2680 Woodlawn, Honolulu, HI 96822, USA \\
$^{14}$Institute of Astronomy, KU Leuven, Celestijnenlaan 200D, 3001 Leuven, Belgium \\
$^{15}$Saint Petersburg State University, Saint Petersburg 199034, Russia \\
}

\date{Accepted XXX. Received YYY; in original form ZZZ}

\pubyear{2024}

\begin{document}
\label{firstpage}
\maketitle

\begin{abstract}
We report the results of a comprehensive study of the spectroscopic binary (SB2) system HD 34736 hosting two chemically peculiar (CP) late B-type stars. Using new and archival observational data, we characterise the system and its components, including their rotation and magnetic fields. Fitting of the radial velocities yields $P_\mathrm{orb}=83\fd219(3)$ and $e=0.8103(3)$. The primary component is a CP He-wk star with \teffA$\;=13000\pm500$\,K and \vsini$\;=75\pm3$\,\kms, while the secondary exhibits variability of Mg and Si lines, and has \teffB$\;=11500\pm1000$\,K and \vsini$\;=110$--180\,\kms. TESS and KELT photometry reveal clear variability of the primary component with a rotational period $P_{\mathrm{rot}A}=1\fd279\,988\,5(11)$, which is lengthening at a rate of $1.26(6)$ s\,yr$^{-1}$. For the secondary, $P_{\mathrm{rot}B}=0\fd522\,693\,8(5)$, reducing at a rate of $-0.14(3)$ s\,yr$^{-1}$. The longitudinal component \bz\ of the primary’s strongly asymmetric global magnetic field varies from $-6$ to +5\,kG. Weak spectropolarimetric evidence of a magnetic field is found for the secondary star. The observed X-ray and radio emission of HD 34736 may equally be linked to a suspected T Tau-like companion or magnetospheric emission from the principal components. Given the presence of a possible third magnetically active body, one can propose that the magnetic characteristics of the protostellar environment may be connected to the formation of such systems.
\end{abstract}



\section{Introduction}

Chemically Peculiar, or CP, stars comprise an important group of upper main sequence objects. The catalogue compiled by \cite{RM2009} lists 8205 known or suspected CP stars in the range of effective temperatures between approximately 7 and 25\,kK. Among them, 3652 stars exhibit abnormal lines of helium, iron-peak elements, and rare-earth elements in their spectra. Such stars are commonly referred to as Ap/Bp or CP2 stars. The latter designation, introduced by \cite{1974ARA&A..12..257P}, is often applied to early-type variable stars with stable magnetic fields, which generally have globally organized (approximately simple dipolar or low-order multipolar) configurations. While these magnetic fields have only been observationally detected in 10--15\% of CP2 stars, all such stars are believed to be magnetic \citep{2002A&A...392..637S}.

The spectral peculiarities of CP2 stars are understood to be a superficial effect resulting from atomic diffusion enabled by their magnetic fields and generally slow rotation.  On a relatively short timescale, the diffusion process produces abnormal vertical and surface distributions of select chemical elements, resulting in a typical (although quite diverse) spectrum of peculiarities \citep{2018CoSka..48...58K}. The presence of regions of chemical constrast in stellar photospheres (sometimes referred to as ``chemical spots'') produce rotationally modulated photometric variability due to flux redistribution. The modern theory of atomic diffusion, developed from the early foundation by \cite{1970ApJ...160..641M}, can explain a wide range of chemical anomalies observed in upper main sequence stars \citep{2015ads..book.....M}.

The CP2 phenomenon first appears in main sequence stars with masses $\sim 1.4~M_\odot$, with an incidence rapidly increasing to 10-15\% for masses of 3.6$M_\odot$ \citep[e.g.,][]{2019MNRAS.483.2300S}. However, there is no direct correlation between the mass and the strength of the measured magnetic field \citep[e.g.,][]{2019MNRAS.490..274S}. Fields up to several tens of kG are not particularly rare, even in quite cool Ap stars (e.g., HD\,154708~-- \citet{2005A&A...440L..37H}, HD\,178892~-- \citet{2006A&A...445L..47R}). A lower field limit of around 100--300\,G \citep[the so-called ``magnetic desert'', ][]{2007A&A...475.1053A,2014IAUS..302..338L,2023MNRAS.521.3480K} probably has a physical meaning. Theoretical studies link the existence of this lower field limit in CP2 stars to specific processes of early stellar formation \citep[e.g., ][]{2020ApJ...900..113J, 2020A&A...641A..13J, 2023MNRAS.521.1415M}. While the properties of fossil magnetic fields are well known, the origin of magnetism in peculiar stars remains unclear.

The evolutionary decay of magnetic field strengths, found, for example, by \citet{2007A&A...470..685L, 2019MNRAS.483.3127S} and \citet{2019MNRAS.490..274S}, suggests that among various hypotheses proposed to explain the phenomenon of magnetic CP2 stars, the most plausible is that of a fossil origin. This hypothesis states that the field observed on the main sequence descends from a seed field acquired during the earlier stages of stellar evolution. The seed can be a local galactic magnetic field in the region of formation of the star \citep{1989MNRAS.236..629M}, amplified through turbulent processes such as a pre-main sequence dynamo or stellar mass transfer or mergers \citep{2019Natur.574..211S}. The latter scenario might explain the observed low incidence of CP2 stars in short-period binary and multiple systems \citep{2015IAUS..307..330A}.

An observational survey aimed at studying the formation and evolution of magnetic fields in CP2 stars of the Orion OB1 stellar association was initiated at the Special Astrophysical Observatory of the Russian Academy of Sciences (SAO) in 2013 \citep{2013AstBu..68..300R}. In the survey, special attention was paid to the completeness of the sample. Individual measurements of the longitudinal magnetic field \bz, i.e. the magnetic field projected to the line of sight and averaged over the visible stellar hemisphere, were obtained using the Main Stellar Spectrograph (MSS) of the 6-m Big Telescope Alt-azimuthal (BTA) installed in the North Caucasus mountains and then published in a series of papers by \citet{2019AstBu..74...55R, 2021AstBu..76...39R, 2021AstBu..76..163R}. In 2022, when the observational component of the survey was completed, \cite{2022MNRAS.515..998S} summarised the results. Altogether, 31 CP2 stars out of 56 were found or confirmed as magnetic. For 14 stars, this status was established for the first time. All programme stars were observed at least four times to avoid potential non-detection due to the rotational variation of the field.

As a member of Orion OB1, \hvezda, was selected for spectropolarimetric observation among the other CP2 stars of subgroup 1c \citep[corresponding to an age $\log t=6.66$,][]{1994A&A...289..101B} of the association. The signatures of a strong, variable magnetic field were detected in the first spectra of HD\,34736 from 2013. The star showed an extraordinarily strong magnetic field \bz\ exceeding 5\,kG. Moreover, the star was recognized as an SB2 system consisting of two early-type stars. The magnetic field was detected only in the dominant spectrum of the narrow-lined component, which we refer to hereafter as the magnetic primary. A short period $P=0\fd3603$ extracted from the \textit{HIPPARCOS} photometry was tentatively considered as the possible period of orbital motion in the system. These results were published by \citet{2014AstBu..69..191S}, or \citetalias{2014AstBu..69..191S} hereinafter. The true period of magnetic field variations, $P=1\fd29$, identified with the rotational period of the primary component, was announced later by \citet{2017ASPC..510..214R}.

A strong magnetic field and the suspected short orbital period made HD\,34736 a fascinating system for detailed study within the framework of the Binarity and Magnetic Interactions in Stars (BinaMIcS) project~\citep{2015IAUS..307..330A}. Compact binaries with magnetic CP components are important laboratories to understanding the origin and evolution of stellar magnetic fields in the upper main sequence. 

Here, we present the results of a comprehensive study of HD\,34736 carried out within the BinaMIcS project. The rest of this paper is organized as follows: In Section \ref{sec:observ}, we describe the observational material obtained for this study and its processing. Data analysis and results are presented in Section \ref{sec:analysis}. Section \ref{sec:discuss} summarises the findings and presents the discussion.

\section{Observations} \label{sec:observ}
For this study, we organised a multi-site spectroscopic and spectropolarimetric monitoring campaign with observational facilities in Europe, Asia, and North America. The observation times are summarised in Table \ref{table:summary}. Photometric variability of the star was studied using archival photometry from ground and space telescopes. The subsequent sections explain the details of data acquisition and processing.

\subsection{Spectroscopy and spectropolarimetry}
\subsubsection{Medium-resolution spectropolarimetry at SAO and DAO}
During the period 2013--2020, \hvezda\ was observed 137 times with the Main Stellar Spectrograph  \citep[MSS,][]{2014AstBu..69..339P} of the 6-m telescope at the Special Astrophysical Observatory (SAO) in the North Caucasian region of Russia. An individual observation consisted of two sub-exposures, normally limited to 10 min. In this case, the mean signal-to-noise ratio of combined spectra measured at 455\,nm varied between 200 and 300 depending on the observational conditions. The data handling and techniques used for the longitudinal magnetic field measurement are described in detail by \citet{2022MNRAS.515..998S}.

Ten medium-resolution spectropolarimetric observations were obtained with dimaPol ($R \approx 10\,000$) installed at the Dominion Astrophysical Observatory (DAO) from Nov. 10 2014 to Mar. 6 2015. The Stokes $V$ observations of the $H_\beta$ line were used to derive longitudinal field measurements\,\citep{2012PASP..124..329M}; the Heliocentric Julian Days and corresponding longitudinal field measurements are listed in Table \ref{table:summary}.

\subsubsection{ESPaDOnS spectropolarimetry}
An Echelle SpectroPolarimetric Device for the Observation of Stars (ESPaDOnS) \citep{2003ASPC..307...41D} is a fibre-fed high-resolution ($R\approx 65\,000$) \'echelle spectrograph, equipped with a polarimeter placed at the Cassegrain focus of the Canada-France-Hawaii Telescope (CFHT).  ESPaDOnS observations were obtained in 2014--2016 over two runs separated by one year within the context of the BinaMIcS Large Program \citep{2015IAUS..307..330A}. The 2014--2015 run was aimed at detecting magnetic fields and following them over the rotational and orbital periods of the system. The second run, scheduled in January 2016, was aimed at dense observations around periastron, where the radial velocity separation of the components is greatest, but also where both components are physically the closest, hence, when maximum interactions (e.g. tidal or electromagnetic) may occur. In total, 22 circularly polarised (Stokes $I$ and $V$) spectra have been obtained over a little more than 1 year. Individual observations are separated by several hours to several days or weeks, depending on the run (see the log of the observations in Table \ref{table:summary}). Each polarimetric spectrum has been obtained by combining 4 successive sub-exposures of 780\,s, between which the Fresnel rhombs were rotated by 90\ddeg. The total exposure time of each ESPaDOnS observation was 3120\,s. The data have been reduced at the CFHT using the {\sc Upena} pipeline feeding the {\sc Libre-ESpRIT} package \citep{donati97}. The peak signal-to-noise ratio of the polarised spectra ranges from 420 to 600 depending on the observing conditions.

\subsubsection{Medium- and high-resolution spectroscopy}
Four spectra of HD\,34736 were collected with the Medium Resolution Echelle Spectrograph (MRES) of the 2.4 m Thai National Telescope at Doi Inthanon (Chiang Mai, Thailand) in 2016 and 2021. MRES is a fibre-fed \'echelle spectrograph designed to register spectra from 420 to 900 nm with resolving power $R=16\,000$--$20\,000$ depending on three available modes. This spectrograph is installed in a room with thermal control and is well suited to accurate measurement of radial velocities. One-dimensional spectra were extracted from the CCD frames in a standard way using the Image Reduction and Analysis Facility (IRAF). A Th-Ar lamp was used to calibrate spectra in the wavelength domain. Resulting spectra with $S/N=80$--$170$ at 550 nm were cropped to 440--700 nm and normalized to the continuum.

Sixteen high-resolution spectra were obtained with the High Efficiency and Resolution Mercator Echelle Spectrograph (HERMES) between Nov. 3 2015 and Jan. 28, 2016. The observations were performed at the Roque de los Muchachos Observatory (La Palma, Islas Canarias, Spain) using the 1.2 m Mercator Telescope. HERMES is fed by optical fibres from the telescope. The instrument has a spectral resolution of $R\approx 85\,000$, and covers a spectral range from 377 to 900\,nm \citep{2011A&A...526A..69R}. It is isolated and temperature-controlled, yielding excellent wavelength stability.
For these observations, the high-resolution mode of HERMES was used, and Th-Ar-Ne calibration exposures were made at the beginning, middle, and end of the night. The exposure time was calculated to reach a signal-to-noise ratio ($S/N$) of 25 or higher in the $V$ band. The reduction of the spectra was performed using the fifth version of the HERMES pipeline, which includes barycentric correction.

\subsection{Photometry}\label{photom:obs}
The Kilodegree Extremely Little Telescope (KELT) survey provides time-series photometric data for a large fraction of the sky via two small-aperture (42 mm) wide-field (26$^{\circ}$ $\times$ 26$^{\circ}$) telescopes, with a northern location at Winer Observatory in Arizona in the United States, and a southern location at the South African Astronomical Observatory near Sutherland, South Africa \citep{Pepper2007,Pepper2012}. The pass-band is roughly equivalent to a broadband $R$ filter, and the typical cadence is approximately 30 minutes.

Non-astrophysical trends are corrected and outliers are removed from KELT light curves with the Trend Filtering Algorithm \citep[TFA;][]{Kovacs2005} as implemented in the \textsc{Vartools} package \citep{Hartman2012}. The TFA-processed version of the KELT light curve for HD\,34736 used here contains 2808 observations over a time baseline of $\sim$5 years (from 2010 to 2015).

The field containing HD\,34736 was observed by the Transiting Exoplanet Survey Satellite ({\it TESS}, \citealt{2014SPIE.9143E..20R}) in sectors 05 and 32, correspondingly, in 2018 and 2020. The light curves obtained using the Science Processing Operations Center pipeline (SPOC, only for sector 05) and the MIT Quick-Look Pipeline (QLP, for both sectors) are available for downloading through the interface of The Mikulski Archive for Space Telescopes (MAST)\footnote{\url{https://dx.doi.org/10.17909/T9RP4V}}.

\begin{figure*}
\centering\includegraphics[width=175mm]{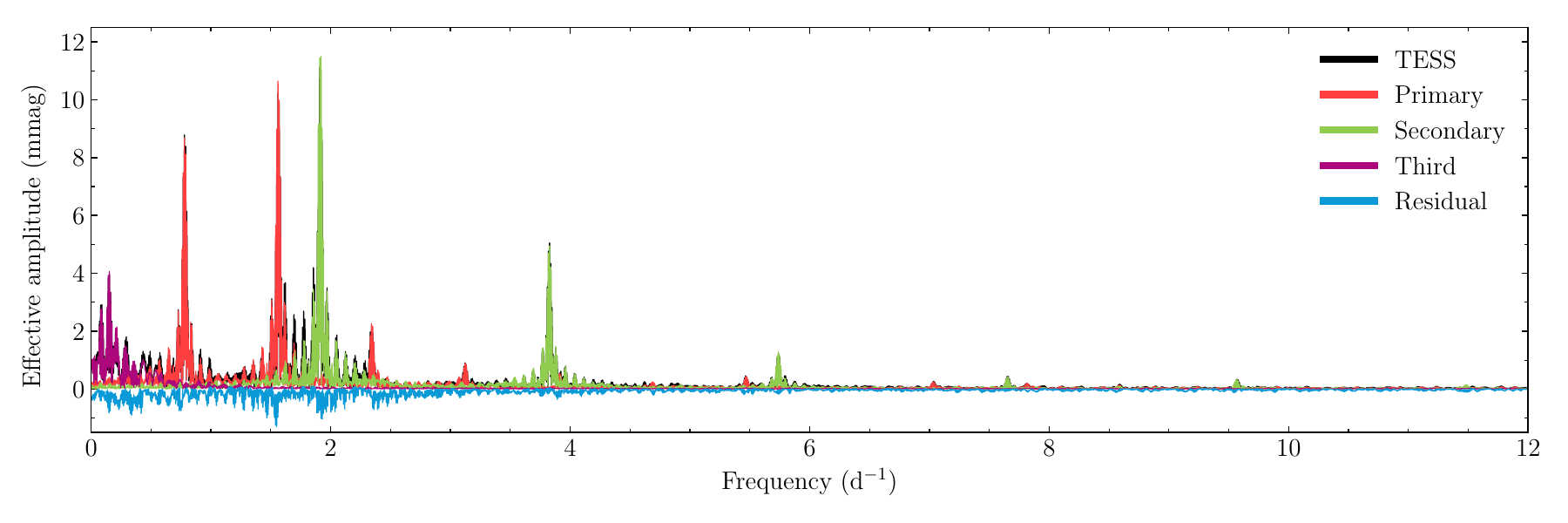}
\caption{Complex amplitude periodogram of the \hvezda\ {\it TESS} variability, in which a total of three sources participate: both components of the SB2 binary \hvezda\ and a highly variable young star located close to the studied system. The light of the primary component of the binary shows a rotational modulation with a period of 1\fd280 (orange line), and the light of the secondary changes with a rotational period of 0\fd523 days (green line). The third component, identified with 
nearby variable UCAC4 414-008437, shows semi-regular changes (purple line) on a scale of several days (details are in Sec.\,\ref{UCAC4}). A solid blue line shows the residuals.}\label{fig:Periodogram}
\end{figure*}

Follow-up observations aimed at the identification of the quasi-periodic signal in the {\it TESS} data (Sec. \ref{UCAC4}) were taken in two standard filters $R_\mathrm{C}$ and $I_\mathrm{C}$ with the DK154 telescope at the La Silla Observatory over 14 nights from December 2022 to January 2023.

\section{Data analysis and results} \label{sec:analysis}
\subsection{Photometric variability}\label{photom}

Frequency analysis of high-quality photometric data collected by TESS Sector 05 (from 2018) confirmed not only the presence of the 1\fd2799 period, compatible with the 1\fd29 period of magnetic field variability \citep{2017ASPC..510..214R}, but also another independent photometric variation with the much shorter period of 0\fd52 \citep{2020EPJWC.24005003S}. The following TESS Sector 32 observations from 2020 fully supported this revelation.

The {\it TESS} frequency spectrum (Fig.~\ref{fig:Periodogram}, black curve) is dominated by two systems of frequencies. The first set (red line) corresponds to a rotationally modulated signal with eleven harmonics. The second set (green line) also carries the signal of rotational modulation with more than seven harmonics.

\begin{figure}
\centering\includegraphics[width=\hsize]{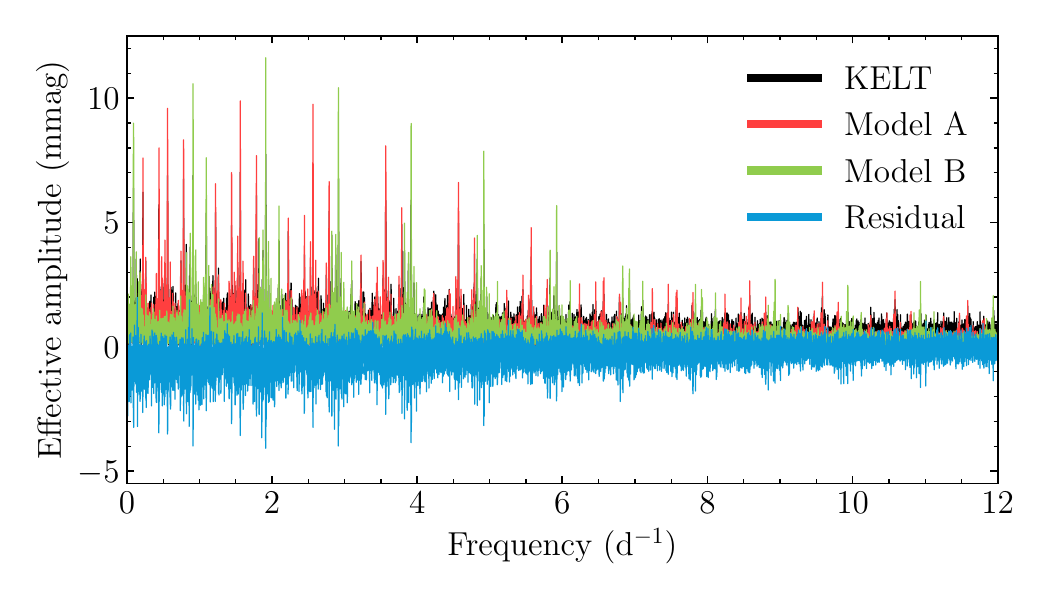}
\caption{The amplitude periodogram of KELT data (black line) displays a dense forest of peaks; nevertheless, the positions of all of them can be explained by two basic frequencies $f_A,\, f_B$ and a lot of harmonics and aliases of the two-component model function (Eq. \ref{eq:2comp}). The model predictions based on the analysis of {\it TESS} data are systematically higher. Residuals are shown by a solid blue line.}
\label{fig:Keltprdg}
\end{figure}

Both periodic signals were found in the five-year KELT photometry. The corresponding frequency spectrum is shown in Fig.~\ref{fig:Keltprdg} as a black curve. We use these data to disentangle photometric variability and determine the ephemeris of both components. It is known that phase light curves of mCP stars obtained in filters with different effective wavelengths generally differ \citep[e.g.][]{2019A&A...625A..34K}. However, we do not consider this because the KELT and {\it TESS} pass bands nearly coincide. The following semi-phenomenological analysis aims to model as accurately as possible the observed photometric variations of \hvezda\ in the KELT and {\it TESS} filters and to isolate the rotational variabilities of both components of the binary star.

\subsubsection{Phenomenological model of light curves of the \hvezda\ binary}\label{LC_double}

The observed, chaotic-looking light curves of \hvezda\ can be satisfactorily interpreted as the sum of two strictly periodic light curves with the instantaneous periods $P_A(t_A,\boldsymbol{\gamma}_A)$ and $P_B(t_B,\boldsymbol{\gamma}_B)$, and corresponding phase functions $\vartheta(t_A,\boldsymbol{\gamma}_A)$ and  $\vartheta(t_B,\boldsymbol{\gamma}_B)$, where $t_A,\,t_B$ are times of observation corrected for {L}ight {T}ravel {T}ime {D}elay (LTTD) of individual binary components $A$ and $B$, as defined in Appendices\,\ref{sec:Phenomenological_models} and \ref{sec:LiTE}.

The thorough analysis of {\it TESS} and KELT data shows that the periods of both components undergo secular, more or less linear changes in time, so we have to use a more complex period model also containing non-zero time derivatives of their periods. The orthogonal ephemeris parameters for individual binary components then have three vector components, especially $\boldsymbol{\gamma}_A=[M_{1A};P_{1A};\dot{P}_A]$ and $\boldsymbol{\gamma}_B=[M_{1B};P_{1B};\dot{P}_B]$ (see Sec.\,\ref{sec:OPF_models}). Then
\begin{align}
    \vartheta_{1A}=\frac{t_A\!-\!M_{1A}}{P_{1A}};\quad \vartheta_A=\vartheta_{1A}\!-\!\frac{\dot{P}_A}{2}(\vartheta_{1A}\!-\!\eta_{2A})(\vartheta_{1A}\!-\!\eta_{3A}); \\
    \vartheta_{1B}=\frac{t_B\!-\!M_{1B}}{P_{1B}};\quad \vartheta_B=\vartheta_{1B}\!\!-\!\frac{\dot{P}_B}{2}(\vartheta_{1B}\!-\!\eta_{2B})(\vartheta_{1B}\!-\!\eta_{3B}),
\end{align}
where $\eta_2$, $\eta_3$ are orthogonalization coefficients expressing data time distributions (Sec.\,\ref{sec:OPF_models} and Table\,\ref{table:parametry}).

The underlying light curves of both components are complex. They can be described by a harmonic polynomial of $m_A=11$ and $m_B=7$ orders, typical of mCP stars with complex surface photometric spot geometries and the presence of semi-transparent structures trapped in co-rotating stellar magnetospheres \citep{2020pase.conf...46M,2022A&A...659A..37K}. The light curve of an individual binary component of such type can be explicitly evaluated using special harmonic polynomials (SHP) $\Xi(\vartheta,\mathbf{b})$ (See Sec.\,\ref{sec:LC_models}).

KELT and {\it TESS} photometry differ in how data are obtained and in the following basic reductions. However, as their effective wavelengths are nearly the same, we can assume the resulting light curve $F(t,\boldsymbol{\alpha})$ of \hvezda\ in the simple form:
\begin{equation}  F(t,\,\boldsymbol{\alpha})=\overline{m}+\Xi(\vartheta_A,\,\mathbf{b}_{A})+\Xi(\vartheta_B,\,\mathbf{b}_{B}).\label{eq:2comp}
\end{equation}

The vector of free parameters $\boldsymbol{\alpha}$ with 44 elements of the model of the observed light curve  $F(t,\boldsymbol{\alpha})$ including their uncertainties, can be determined using standard $\chi^2$ minimization:
\begin{equation}\label{eq:solution}
    \chi^2=\sum_{i=1}^n\,\frac{\hzav{y_i-F(t_i,\boldsymbol{\alpha})}^2}{\sigma_i^2};\quad \frac{\partial \chi^2}{\partial \boldsymbol{\alpha} }=\mathbf{0},
\end{equation}
where $n$ is the total number of the photometric observation used, $t_i$ is the HJD time of the $i-$th individual observation, $y_i$ is its magnitude corrected for instrumental trends, and $\sigma_i$ is the estimate of its internal uncertainty. The vector constraint that the quantity $\chi^2$ is minimal gives 44 non-linear equations of 44 unknowns, which can be solved using standard iterative methods.

\begin{figure}
\centering\includegraphics[width=\hsize]{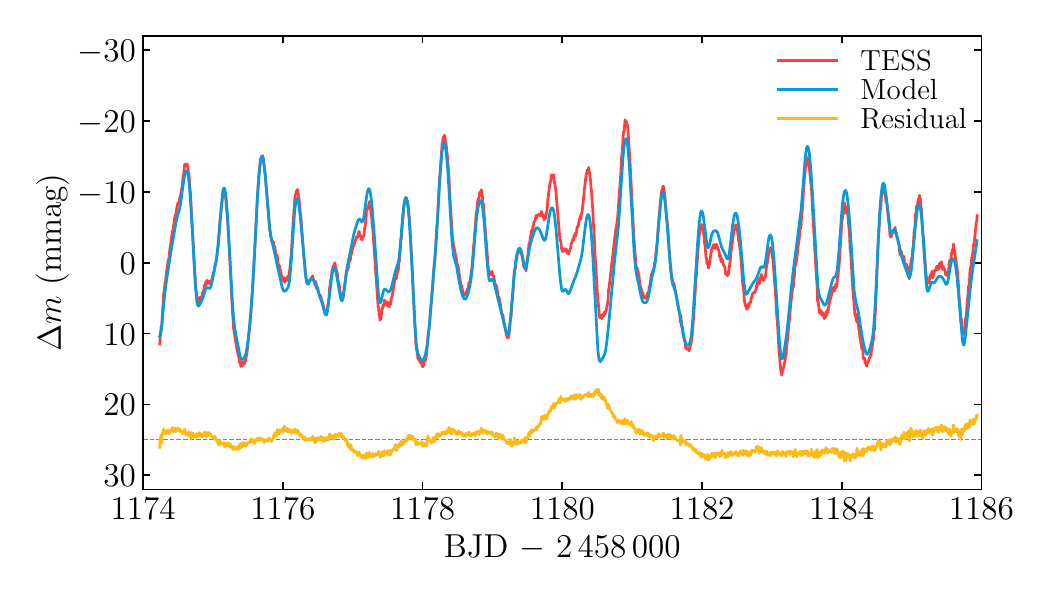}
\caption{The first part of the {\it TESS} light curve of \hvezda\ taken in Sector 32 (red line) modelled as the sum of two strictly periodic variations (blue line) and their residual (orange line) shifted by 25 mmag.}
	\label{fig:Tess_raw}
\end{figure}

\begin{table}
\begin{center}
\caption{The light curve ephemeris of $A$ and $B$ components. The meaning of parameters of the quadratic orthogonal fit are specified in Sec.\,\ref{sec:OPF_models}.}
\label{table:parametry} \normalsize
\begin{tabular}{l | l}
\hline\hline
 $M_{1A}=2\,458\,732.907\,0(5)$ &$M_{1B}=2\,458\,775.300\,0(5)$ \\
 $P_{1A}=1\fd279\,988\,5(1\,1)$  & $P_{1B}=0\fd522\,693\,8(5)$\\
 $\dot{P}_{A}=3.98(17)\times 10^{-8}$ & $\dot{P}_{B}=-4.4(9)\times 10^{-9}$  \\
 $\eta_{2A}=-1191.2$ & $\eta_{2B}=-2172.7$\\
 $\eta_{3A}=169.2$ &$\eta_{3B}=388.6$\\
 $A_{\rm{eff}A}=14.1$ mmag & $A_{\rm{eff}B}=13.0$ mmag\\
 \hline\hline
\end{tabular}
\end{center}
\end{table}

The analysis of residuals of the fit of the observed {\it TESS} light curve by the two-component model function $F(t,\boldsymbol{\alpha})$ shows an unexpectedly high scatter of 2.7 mmag, while the true {\it TESS} photometry accuracy should be at least eight times better. We propose that the cause of this discrepancy is that the light of \hvezda\ is contaminated by a nearby fainter, strongly variable star. The contribution to the variability of \hvezda\ is considerable, and it causes additional semi-regular variations on the time scale of several days, sometimes reaching more than six mmag as shown in Fig.\,\ref{fig:Tess_raw}. The frequencies and amplitudes of the parasitic light variations can also be seen in the amplitude periodogram in Fig.\,\ref{fig:Periodogram} as a purple line. We have identified the source of invading variability as a young red pre-main-sequence star UCAC4 414-008437 (Appendix \ref{UCAC4}).

\subsubsection{Final light curve model solution. Disentangling of the light curve. Dips}\label{sec:dips}

\begin{figure*}
\centering\includegraphics[width=170mm]{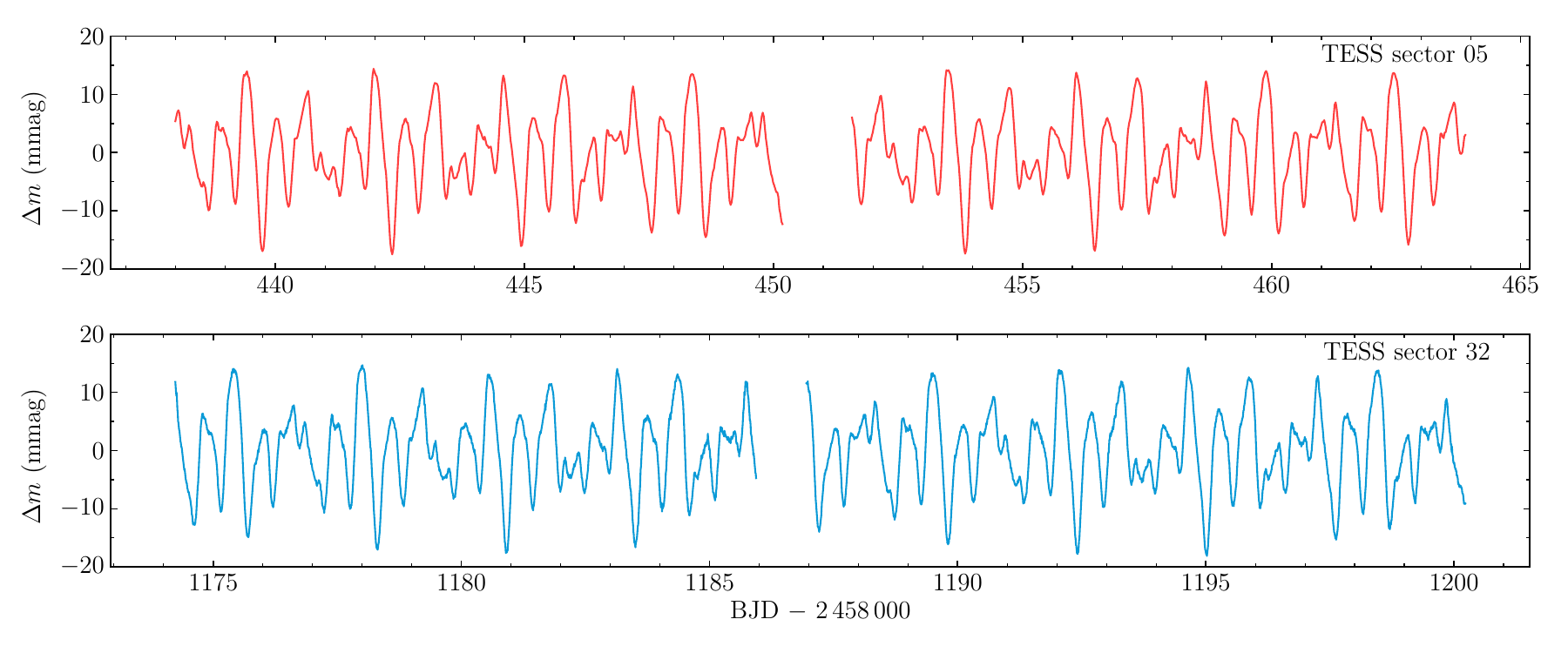}
\caption{{\it TESS} light curves (Sector 05~--- upper part, and Sector 32~--- bottom part) corrected for aperiodic variation of the parasitic light of the third component.}
\label{fig:LCper}
\end{figure*}

We solve the set of equations given by (\ref{eq:solution}) using {\it TESS} magnitudes corrected for the aperiodic variation of the third component to compute a final set of the model parameters. The result is shown in Fig.\,\ref{fig:LCper}. Table\,\ref{table:parametry} gives the final orthogonal ephemeris for both components. Using them, we can, for example, predict the moments of maximum brightness of individual components $\mathit{\Theta}_A(E_A)$, $\mathit{\Theta}_B(E_B)$ in the epochs $E_A$, $E_B$ from the point of view of an observer as follows:
\begin{align}
    \mathit{\Theta}_A=M_{1A}+P_{1A}\,E_A+\frac{P_{1A}\dot{P}_A}{2}(E_A\!-\eta_{2A})(E_A\!-\eta_{3A})-\Delta t\!_A;\\
    \mathit{\Theta}_B=M_{1B}+P_{1B}\,E_B+\frac{P_{1B}\dot{P}_B}{2}(E_B\!-\eta_{2B})(E_B\!-\eta_{3B})-\Delta t\!_B,
\end{align}
where $\Delta t_A$ and  $\Delta t_B$ are corrections for LTTD for individual components orbiting in a binary (Appendix~\ref{sec:LiTE}).

\begin{figure*}
\centering\includegraphics[width=87mm]{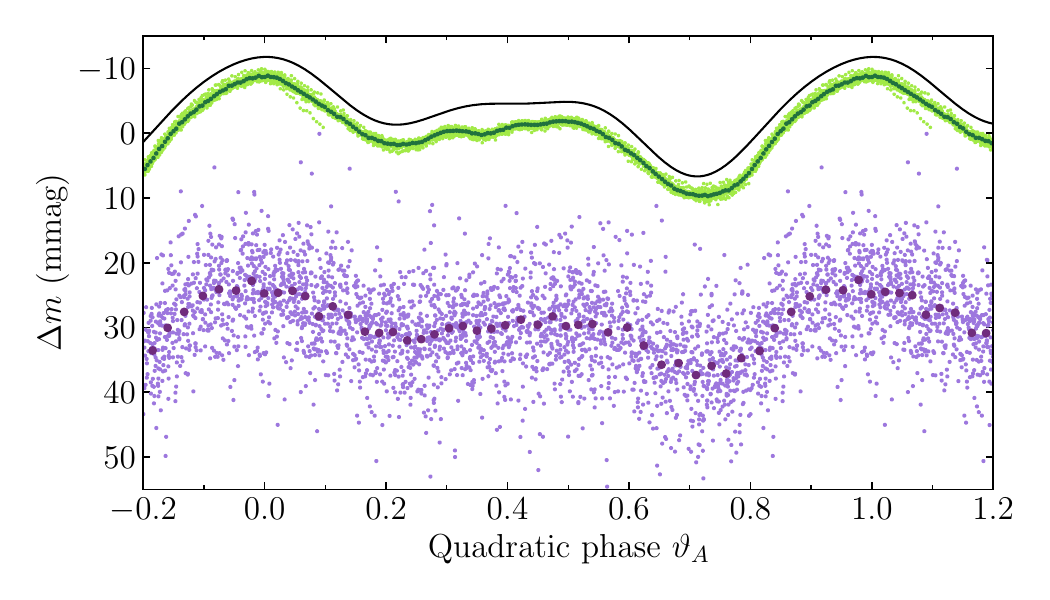}
          \includegraphics[width=87mm]{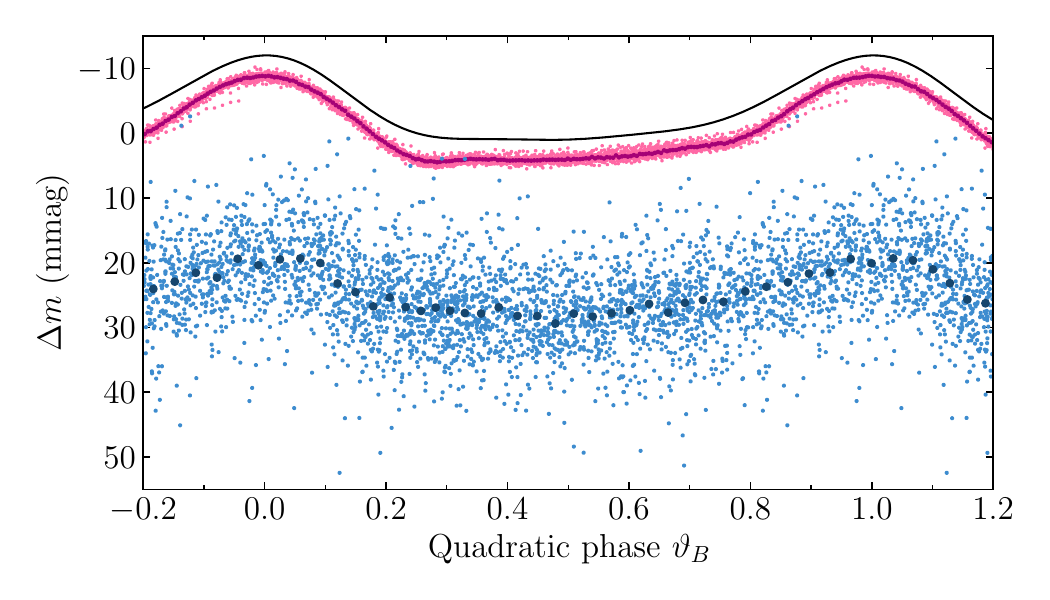}
\caption{Corrected, disentangled, and phased {\it TESS} (upper curve) and KELT (lower curve) photometry of the $A$ (left panel) and $B$ components (right panel). Bins of about a hundred neighbourhood observations (dark dots) represent both components' mean light curves. The quadratic ephemerides of both components are given in Table\,\ref{table:parametry}. Black thin lines shifted by 3\,mmag from corresponding {\it TESS} curves show the variants of fit made as the sum of a harmonic polynomial of the fourth degree, which was used for deriving dips.}
\label{fig:phase_curves}
\end{figure*}

\begin{figure*}
\centering\includegraphics[width=85mm]{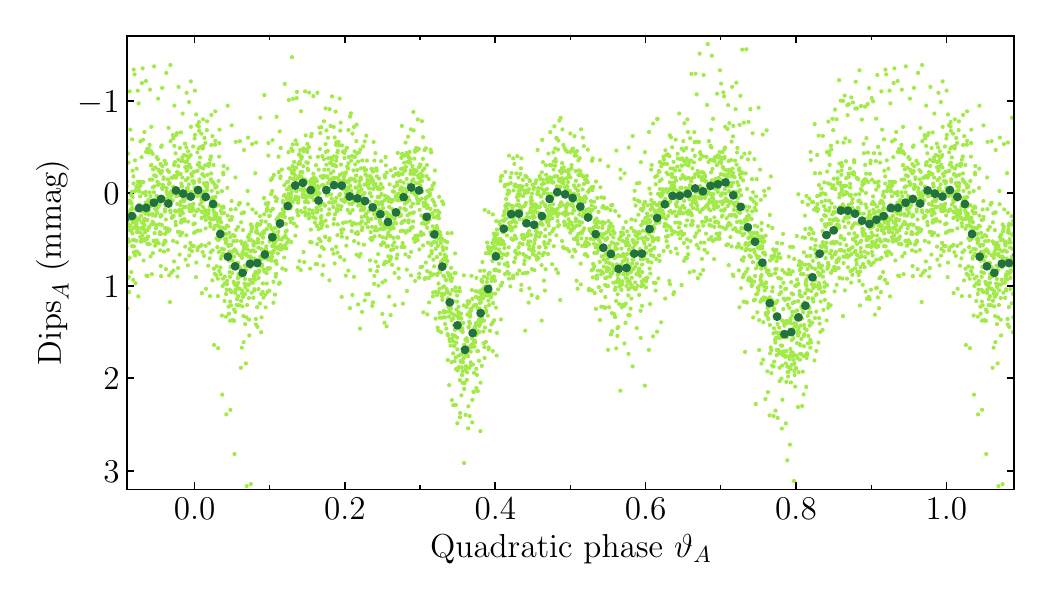}
          \includegraphics[width=85mm]{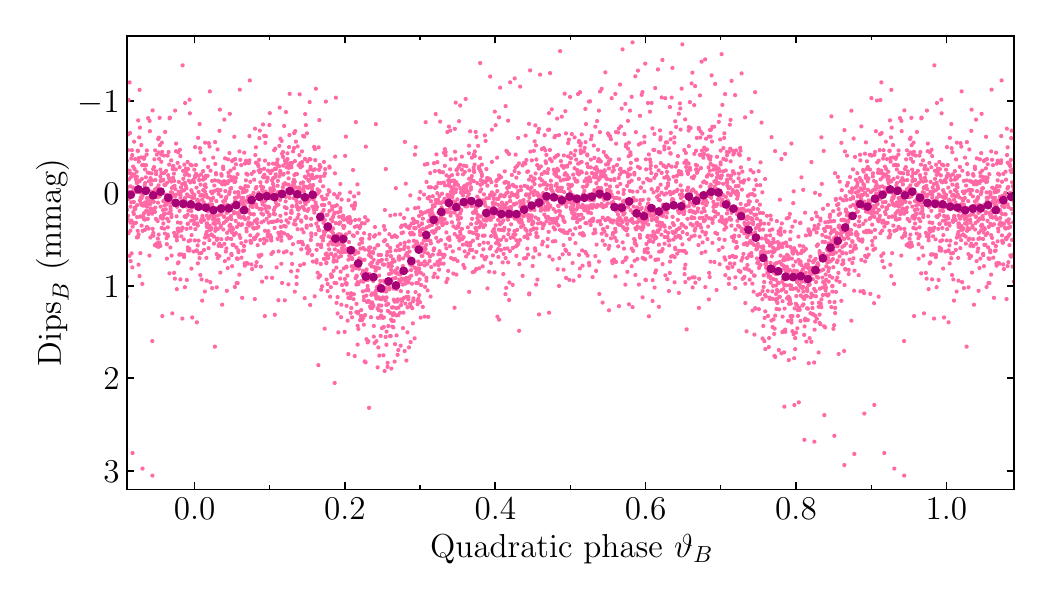}
    \caption{Dips in {\it TESS} light curves of A (left) and B (right) components in mmags. Darker points are bins of reduced photometric observations.}
	\label{fig:warps}
\end{figure*}

Using the final model of the light curve, we can disentangle the light curves of individual components and plot the phased light curves for both components (Fig.\,\ref{fig:phase_curves}). The $A$ phased light curves defined by the observation or bins of neighbouring observations in {\it TESS} and KELT colours are similar; only the amplitude of the latter is a bit smaller. The light curves are double-wave and rather complex, with at least five dips \citep{2020pase.conf...46M} with amplitudes up to 1.5 mmag (Fig.\,\ref{fig:warps}). These details are probably caused by the presence of absorbing semitransparent structures confined in the co-rotating magnetosphere of the star \citep{2022A&A...659A..37K}. To empirically differentiate between surface inhomogeneities and circumstellar environment as two main drivers of the variability of CP stars with magnetospheres, we represent the observed light curves as the sum of fourth-degree harmonic polynomials emulating the contribution from spots and a finite number of relatively symmetrical dips appearing as Gaussian-like profiles and described by the phase of the centre, half-width, and depth. Standard regression analysis techniques can then be used to determine the light curve parameters. The dips depicted in Fig.\,\ref{fig:warps} are obtained in this manner from the {\it TESS} photometry. The fourth-order polynomial fits are shown as the thin black lines in Fig.\,\ref{fig:phase_curves} for both components of \hvezda.

The effective amplitudes of the $A$ and $B$ components' contributions to the {\it TESS} light curve are $A_{\rm{eff}A}=14.1$ mmag and $A_{\rm{eff}B}=13.0$ mmag (Table\,\ref{table:parametry}). It is appropriate to remind at this point that for all phase light curves, especially in Fig.\,\ref{fig:phase_curves} and \ref{fig:warps}, we show only contributions to the total brightness of the system. The amplitude of the intrinsic variation of the sources is naturally different and depends on the luminosity.

\begin{figure*}
\centering\includegraphics[width=85mm]{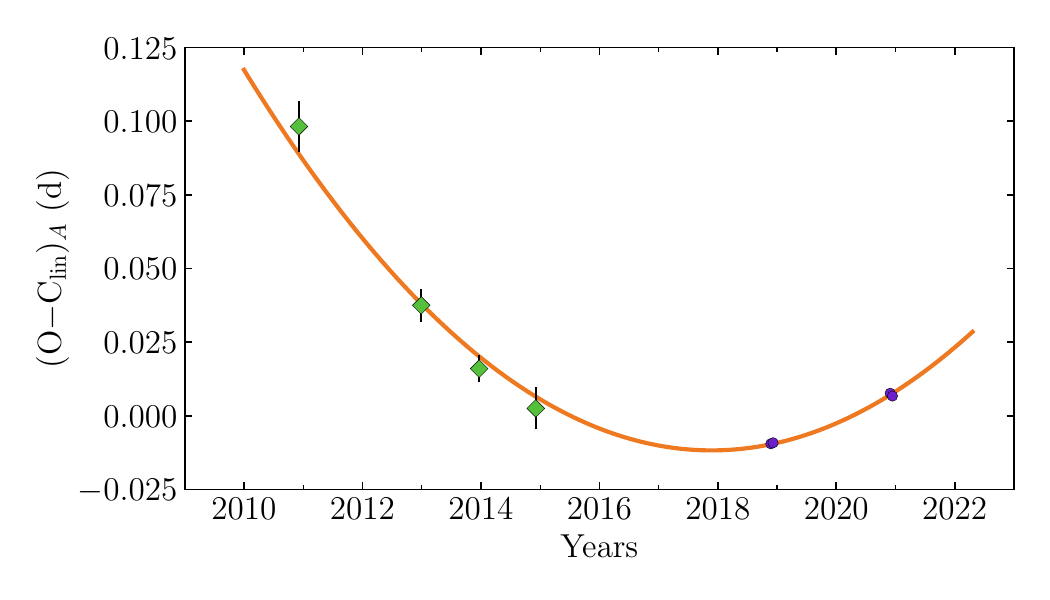}
\centering\includegraphics[width=85mm]{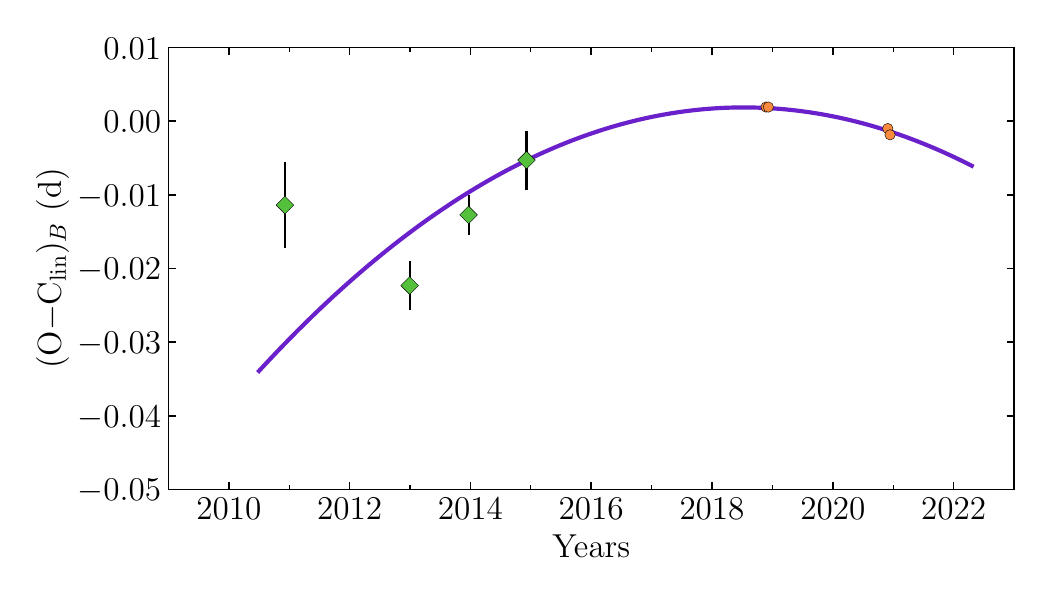}
	\caption{Dependence of phase shift between the observed light curves and the linear ephemeris prediction (\oc$_{\mathrm{lin}})_A$ (left) and (\oc$_{\mathrm{lin}})_B$ (right) in days versus time in years for A, B binary components. The fits of the dependence by parabolae with $\dot{P}\!_A=3.98(17)\times10^{-8}=1.26(6)$\,s\,yr$^{-1}$, $\dot{P}\!_B=-4.4(9)\times10^{-9}=-0.14(3)$\,s\,yr$^{-1}$ are shown as solid curves. TESS observations are denoted as circles, while KELT ones are diamonds.}
	\label{fig:OCAB}
\end{figure*}

\subsubsection{Virtual \oc\ diagrams. Period changes }

By finding non-zero time derivatives of the period of light changes for both system components, it is obvious that their rotation changes with time. To test the adequacy of the linear period change model used above, it is appropriate to visualize the observed period changes using the so-called virtual \oc\ diagrams, introduced and developed by \citet{2006Ap&SS.304..363M,2011ASPC..451..111M,2012IAUS..282..391M}. In the mentioned method, it is assumed that the shape of the light curve is nearly constant, so changes in the instantaneous period will be manifested by a variable phase shift $\Delta \varphi(t)$ of the observed light curve relative to the light curve that the star would have if its period remained constant. The instantaneous phase shift is then connected with the instantaneous $\oc(t)$, as follows:  $\oc(t)=-P\,\Delta \varphi(t)$, where $P$ is a mean period.

To calculate the coordinates of the points plotted in the virtual $\oc_{\mathrm{lin}A}$ and $\oc_{\mathrm{lin}B}$ diagrams in Fig.\,\ref{fig:OCAB} we divided the KELT and {\it TESS} data, sorted by observation time, into $n_k=8$ consecutive groups, characterized by the middle epoch $E_{Ak}$ and $E_{Bk}$ (see Table \ref{table:OCAB}). We also introduced a new model of phase functions $\vartheta_{kA}$,\,$\vartheta_{kB}$ with fixed values of parameters $M_{1A},\,M_{1B},\,P_{1A},\,P_{1B}$, valid for a particular $k$ and $2\,n_k=16$ free parameters: $\oc_{kA},\,\oc_{kB}$:
\begin{align}
&\vartheta_{kA}=\frac{t_{kA}-M_{1A}-\oc_{kA}}{P_{1A}};\quad \vartheta_{kB}=\frac{t_{kB}-M_{1B}-\oc_{kB}}{P_{1B}};\nonumber\\
&   k=1,2,\dots,n_k. \label{eq:varthk}
\end{align}
Parameters $\oc_{kA},\,\oc_{kB}$, including their uncertainties, were calculated using a standard minimalization of the $\chi^2$ (see Eq.\,\ref{eq:solution}). They are given in Table \ref{table:OCAB}, together with virtual times of zero-th phase/light curve maximum moments $O_{Ak}$ and $O_{Bk}$ for epochs $E_{Ak}$ and $E_{Bk}$, respectively, where:
\begin{align}
    O_{Ak}=\oc_{kA}+M_{1A}+P_{1A}E_{Ak};\nonumber \\ O_{Bk}=\oc_{kB}+M_{1B}+P_{1B}E_{Bk}.
\end{align}

The virtual \oc\ diagrams show beyond any doubt that the angular velocities of both components of \hvezda\ are changing, with the $A$ component currently having the highest rate of change among all known mCP stars with variable rotation \citep{2021osvm.confE..17M}. A linear increase in the rotation period derived here is highly credible. In the case of the $B$ component, a monotonous acceleration of the rotation is noticeable, while a linear decrease in the rotation period appears to be a good initial hypothesis. The fact that the changes in periods are opposite essentially excludes any explanation involving the gravitational action of an invisible, distant third component.

\begin{table*}
\begin{center}
\caption{Data from KELT and {\it TESS} 05, 32 photometries, for $A$ and $B$ components, corrected for the variability of the other components, were divided into eight consequent groups of $N$ observations with the averages in `Years' end mean epochs $E_A$, $E_B$ according to the ephemeris of the relevant components. (\oc$_{\mathrm {lin}})_A$ and (\oc$_{\mathrm {lin}})_B$ are mean differences between the observed moment of the particular component light maximum/instantaneous zero-th phase and its prediction according to the linear ephemeris models (see Fig.\,\ref{fig:OCAB}). $O_A$, $O_B$ are the times of the light maxima $A$ and $B$ components for the epoch $E_A$ and $E_B$.}\small
\label{table:OCAB}
\begin{tabular}{c c r r c l r c l}
		\hline\hline
 Year  & Source & $N$ & $E_A$ & (\oc$_{\mathrm{lin}})_A$& $O_A\!-\!2450000$& $E_B$ & (\oc$_{\mathrm{lin}})_B$& $O_B\!-\!2450000$\\
 \hline
2011 &   KELT 1              &   457   & $-2499$ &    0.098(9) &  5\,534.314 &  $-6201$ &  $-0.011$(6) &  5\,534.065    \\
2013 &   KELT 2              &   857   & $-1910$ &    0.038(6) &  6\,288.166 &   $-4758$ &  $-0.022$(4) &  6\,288.301  \\
2014 &   KELT 3              &   955   & $-1631$ &    0.016(5) &  6\,645.262 &   $-4076$ &  $-0.013$(3) &  6\,644.788  \\
2015 &   KELT 4              &   539   & $-1358$ &    0.002(7) &  6\,994.685 &   $-3406$ &  $-0.005$(5) &  6\,995.000  \\
2019 &   {\it TESS} 05,\,I\,\,     &   582   &  $-226$ &   $-0.0095$(12) &  8\,443.6200 &   $-634$ &   0.0019(5) & 8\,443.9141 \\
2019 &   {\it TESS} 05,\,II        &   595   &  $-215$ &   $-0.0092$(12) &  8\,457.7002 &   $-608$ &   0.0019(5) & 8\,457.5042  \\
2021 &   {\it TESS} 32,\,I\,\,     &  1685   &   349 &    0.0076(12) &  9\,179.6306 &    774 &  $-0.0010$(5) & 9\,179.8640  \\
2021 &   {\it TESS} 32,\,II        &  1912   &   360 &    0.0067(12) &  9\,193.7095 &    800 &  $-0.0019$(5) & 9\,193.4532  \\
	\hline\hline
	\end{tabular}
\end{center}
\end{table*}

\subsection{Spectroscopic and spectropolarimetric analysis}

As the \'echelle spectra of \hvezda\ constitute a significant element of the available spectroscopic material, exploiting their wide spectral coverage using a multiline technique to increase the signal-to-noise ratio was natural. Before proceeding to the results, we describe the technique employed in the current study.

\subsubsection{LSD profile analysis of \'echelle spectra}\label{sec:lsdhires}

\begin{figure}
	\centering\includegraphics[width=\hsize]{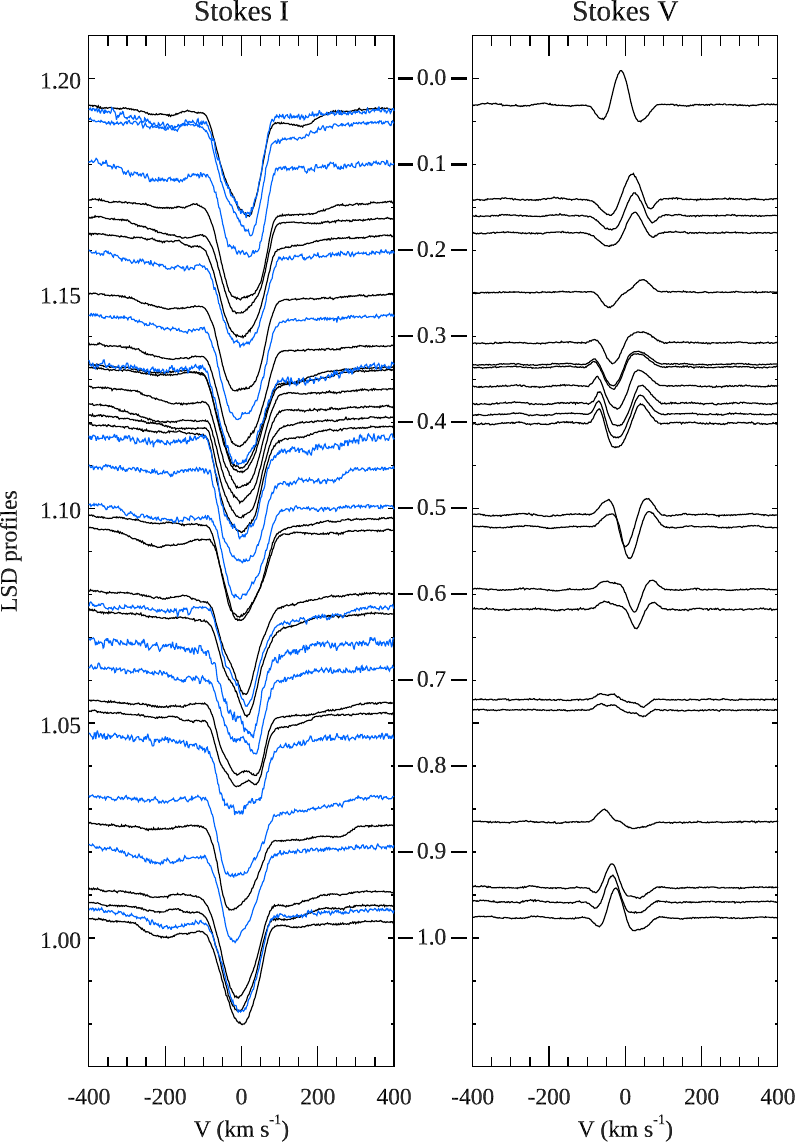}
	\caption{The composite LSD Stokes $I$ (left panel) and Stokes $V$ (right panel) profiles of HD\,34736 derived from high-resolution spectra. The smooth black curves in the left panel correspond to the ESPaDOnS observations, while the blue curves show LSD profiles derived from the HERMES data. The spectra are shifted in velocity to the reference frame of the primary and are offset vertically according to its rotational phase (indicated between the panels).}
	\label{fig:lsdhires}
\end{figure}

We have calculated the least-squares deconvolved (LSD) profiles for the \'echelle spectra of HD\,34736 using the code described by \citet{2010A&A...524A...5K}. The line mask employed for these calculations was extracted from the Vienna Atomic Line Database (VALD, \citealt{1995A&AS..112..525P, 1999A&AS..138..119K, 2015PhyS...90e4005R, 2019ARep...63.1010P}) for the parameters and composition of the magnetic primary presented by \citetalias{2014AstBu..69..191S}. The final mask includes 338 metal lines with a central depth exceeding 10\% of the continuum in the 400--700\,nm wavelength range. The mask is characterized by a mean wavelength of 516.5\,nm and a mean effective Land\'e factor of 1.16. In Fig.~\ref{fig:lsdhires}, we show the resulting Stokes $I$ and $V$ (for ESPaDOnS data) and Stokes $I$ (for HERMES observations) LSD profiles deconvolved from high-resolution spectra. These profiles are arranged according to the rotational phase of the primary calculated following prescriptions given in Sec.~\ref{photom}. The spectra are shifted in velocity to the reference frame of the primary star using the orbital solution discussed in Sec.~\ref{sec:multiple}.

The narrow component of the Stokes $I$ LSD profiles, which corresponds to the contribution of the primary star, shows a moderately coherent variation with rotational phase, compatible with signatures expected for an inhomogeneous surface distribution of chemical elements. The incoherent variation, also evident in the Stokes $I$ panel of Fig.~\ref{fig:lsdhires}, is due to the orbital radial velocity shifts and intrinsic variability of the broad-lined secondary.

The circular polarisation signatures of the primary are detected for all ESPaDOnS observations. These Stokes $V$ LSD profiles exhibit a smooth rotational phase variation, indicating a globally-organised magnetic field topology on the primary star. At the same time, no conclusive evidence of polarisation signatures of the secondary is seen in the Stokes $V$ LSD profile data. Magnetic properties of the components are examined in Sec.\,\ref{sec:mfield}.

\subsubsection{Magnetic field of \hvezda}\label{sec:mfield}

The longitudinal field \bz\ of the narrow lined component was measured from the first moment of the Stokes $V$ LSD profiles deconvolved from ESPaDOnS spectra according to \citet{2000MNRAS.313..851W} and \citet{2010A&A...524A...5K}, or using the techniques described by \citet{2022MNRAS.515..998S} and \citet{2012PASP..124..329M} in the case of low-resolution spectropolarimetry from SAO and DAO, respectively. The three distinct methods yield typical uncertainties ranging from 100 to 800\,G. Table \ref{table:summary} contains the 
full collection of measurements.

\begin{figure}
	\centering\includegraphics[width=80mm]{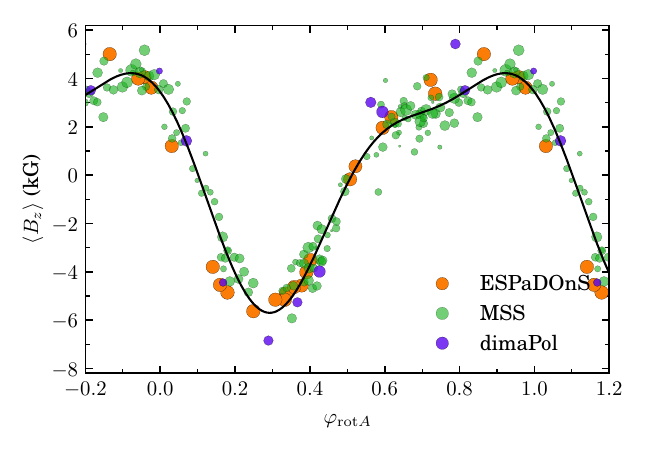}  
	\caption{The observed \bz\ variation versus quadratic rotational phase of the primary component. The areas of the symbols are proportional to the weight of \bz\ measurements. The second-order harmonic polynomial fit is shown with the solid line.}
	\label{fig:magA}
\end{figure}

The observed longitudinal field varies with a period that is compatible with the photometrically-derived period $P_{\rm 1A}$ (Table\,\ref{table:parametry}). Therefore, we interpret this variation as a consequence of the solid-body rotation of a star (chemically peculiar component $A$) with a magnetic field frozen in its outer atmospheric layers. Individual values of \bz\ against the quadratic rotational phase are plotted in Fig.~\ref{fig:magA}.

The phase curve of magnetic field variations is somewhat atypical, indicating a complex configuration of the global magnetic field with non-negligible high-order components and the effect of chemical abundance spots. The effective amplitude of changes in the \bz\ component of the magnetic field is also unusually high: 9.3\,kG.

Assuming a simple oblique dipolar field geometry \citep{1950MNRAS.110..395S, 1967ApJ...150..547P}, and by interpolating the isochrones for $\mathrm{[Fe/H]}=0$ produced by the project MESA Isochrones \& Stellar Tracks\footnote{\url{https://waps.cfa.harvard.edu/MIST/}} (MIST, \citealt{2016ApJS..222....8D, 2016ApJ...823..102C}), we have assessed the strength and obliquity of the magnetic field in the primary component of \hvezda. For this, we use $R=2.05\pm0.06\,R_\odot$ as the appropriate theoretical value for a young star of the age of 4.6\,Myr (average age of the subgroup Ori OB1c, \citealt{1994A&A...289..101B, 2022MNRAS.515..998S}) with an effective temperature of the magnetic primary (\teff$\;=13\,000\pm500$\,K, Sec.\,\ref{sec:phys_par}). Substituting this value for $R$ and $P_\mathrm{rot}=P_{\mathrm{1}A}=1\fd2799885$ into the equation for equatorial rotational velocity
\begin{align}\label{eq:rotation}
\upsilon_\mathrm{e}=\frac{50.6\,R [R_\odot]}{P_\mathrm{rot}[\mathrm{days}]},
\end{align}
we find $\upsilon_\mathrm{e}=81\pm3$~\kms, which together with the spectroscopically measured \vsini$\;=75\pm3$\,\kms\ (Sec.\,\ref{sec:phys_par}) gives us the inclination angle $i=68\pm7^{\circ}$. Then, considering the extrema of the longitudinal magnetic field ($\approx-6/+5$ kG), one can evaluate the polar strength $B_\mathrm{d}$ of the field as $18.9\pm0.8$\,kG and the angle $\beta$ between the magnetic and rotational axes as $83\pm2^\circ$. The approach applied here is not ideal, but it allows us to derive approximate parameters of the stellar magnetic field in the simplest and fastest way.

A more accurate picture of the surface magnetic structure of the primary component of HD\,34736 has been obtained using the Zeeman Doppler imaging (ZDI) magnetic tomography technique \citep{2016LNP...914..177K}. This modelling is based on the mean metal line LSD profiles, illustrated in Fig.~\ref{fig:lsdhires}, derived from the 22 ESPaDOnS circular polarisation observations. Considering the complex composite nature of the spectral variability of HD\,34736, with contributions from the variability due to spots on the primary, secondary, and the orbital motion of the two stars in an eccentric orbit, we chose not to pursue a detailed, simultaneous mapping of individual chemical elements and magnetic field \citep[e.g.][]{2014A&A...565A..83K,2018MNRAS.473.3367O}. Instead, we model the mean Stokes $V$ profiles of the narrow-lined primary, ignoring its surface spots and neglecting blending by the broad-lined secondary but correcting for the orbital radial velocity shifts. This approach is justified considering that variability of the majority of lines in the spectrum of the primary is relatively weak compared to high-amplitude changes seen in well-studied Ap stars with high-contrast chemical spots \citep[e.g.][]{2004A&A...424..935K,2012MNRAS.426.1003S,2016A&A...588A.138R}. Furthermore, its mean Stokes $V$ LSD profiles are smooth. They are characterised by a simple shape, lacking any small-scale features that are typical of polarisation spectra of fast-rotating magnetic stars with highly non-uniform surfaces \citep{2017A&A...605A..13K,2019A&A...621A..47K}. All these factors indicate that chemical inhomogeneities do not significantly affect the shape and variability of the mean metal line Stokes $V$ LSD profile of the primary.

Similar to ZDI studies of cool stars \citep{2016A&A...587A..28H,2016A&A...593A..35R,2019ApJ...873...69K}, we adopt the Unno-Rachkovsky solution of the polarised radiative transfer equation in the Milne-Eddington approximation to describe the local Stokes profiles. The line parameters required by this local line profile model were chosen to match the mean wavelength and Land\'e factor of the LSD line mask, whereas the local equivalent width was adjusted to fit the mean Stokes $I$ spectrum. An inclination angle $i=60\degr$ and projected rotational velocity \vsini$\;=75$~km\,s$^{-1}$ were adopted for the magnetic mapping of the primary. No correction for continuum dilution is required since the decrease of the Stokes $V$ amplitude is compensated by the decrease of the Stokes $I$ line depth.

\begin{figure*}
	\centering\includegraphics[height=0.9\hsize,angle=270]{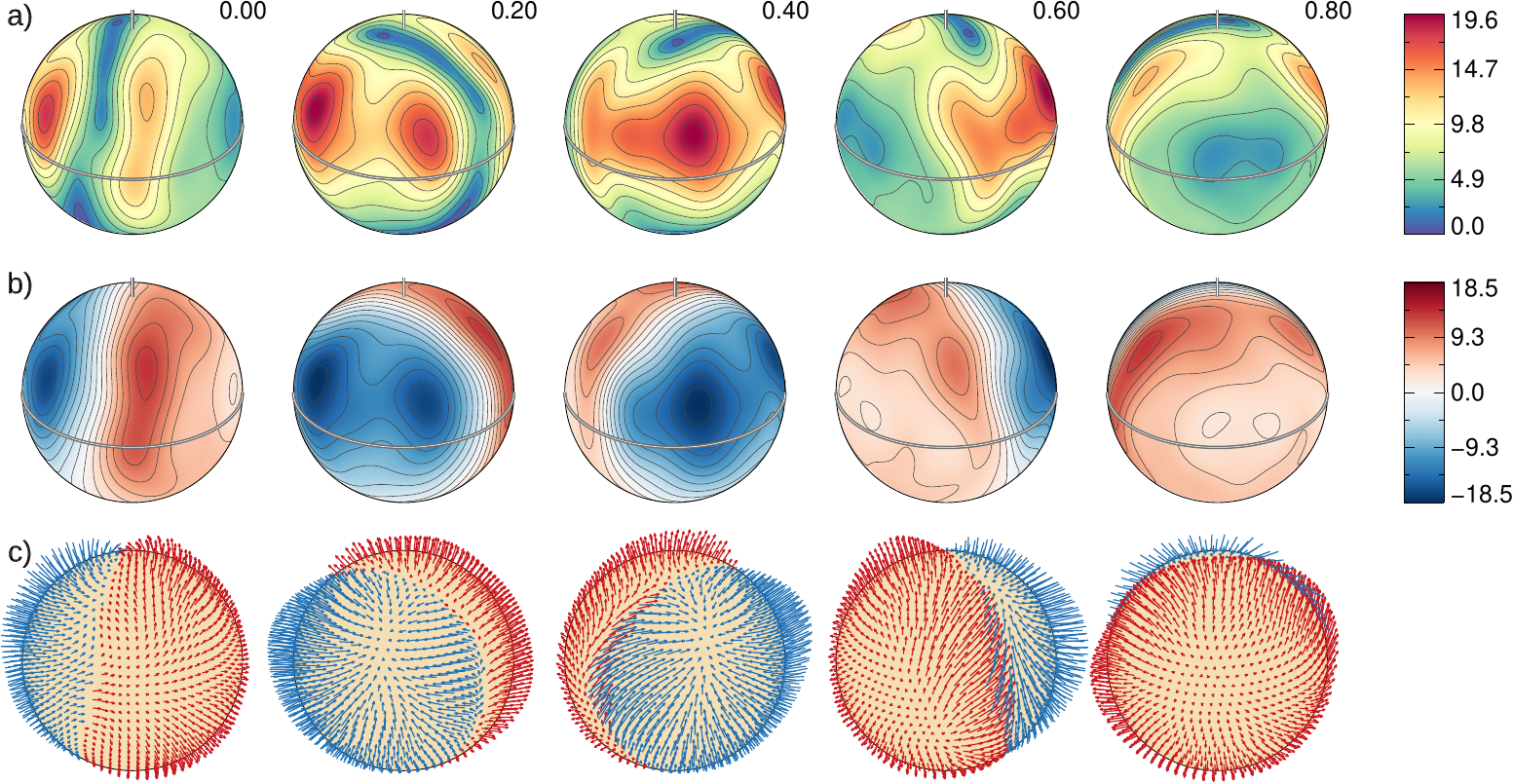}
	\caption{Magnetic field topology of the primary component of HD\,34736 derived using with ZDI. The star is shown at five rotation phases, indicated above each spherical plot column. The spherical plot rows present the maps of a) field modulus, b) radial field, and c) field orientation. The contours over these maps are plotted with a 2~kG step. The vertical bar and thick line indicate the positions of the visible pole and rotational equator, respectively. The colour bars give the field strength in kG. The two colours in the field orientation map correspond to the field vectors directed outwards (red) and inwards (blue).}
	\label{fig:zdi_map}
\end{figure*}

The magnetic field distribution obtained for HD\,34736 with the ZDI code {\sc InversLSD} \citep{2014A&A...565A..83K} is presented in Fig.~\ref{fig:zdi_map}. The primary possesses a strong, distorted dipolar global field geometry, characterised by a large asymmetry between the negative and positive magnetic hemispheres. The strong-field negative magnetic region exhibits a pair of magnetic spots with a local field strength reaching 19.6~kG. The overall mean field strength (averaged over the entire stellar surface) is 7.6~kG. The mean field modulus varies between 6.3 and 11.5~kG, depending on the rotational phase. The phase-averaged value of \bs\ is 8.9~kG.

The ZDI code employed here uses a generalised spherical harmonic expansion to parameterise stellar surface field vector maps \cite[see][]{2014A&A...565A..83K}. This allows us to readily characterise contributions of different spherical harmonic modes to the global field topology of the primary. Regarding the magnetic field energy, the largest contribution comes from the $\ell=1$ (dipole) component, which contributes 63\% of the magnetic energy. All quadrupole ($\ell=2$) and octupole ($\ell=3$) modes are responsible for 22 and 7\% of the energy, respectively. The field of HD\,34736 is predominantly poloidal, with 88\% of the field energy concentrated in the poloidal harmonic modes.

\begin{figure}
	\centering\includegraphics[width=0.9\hsize]{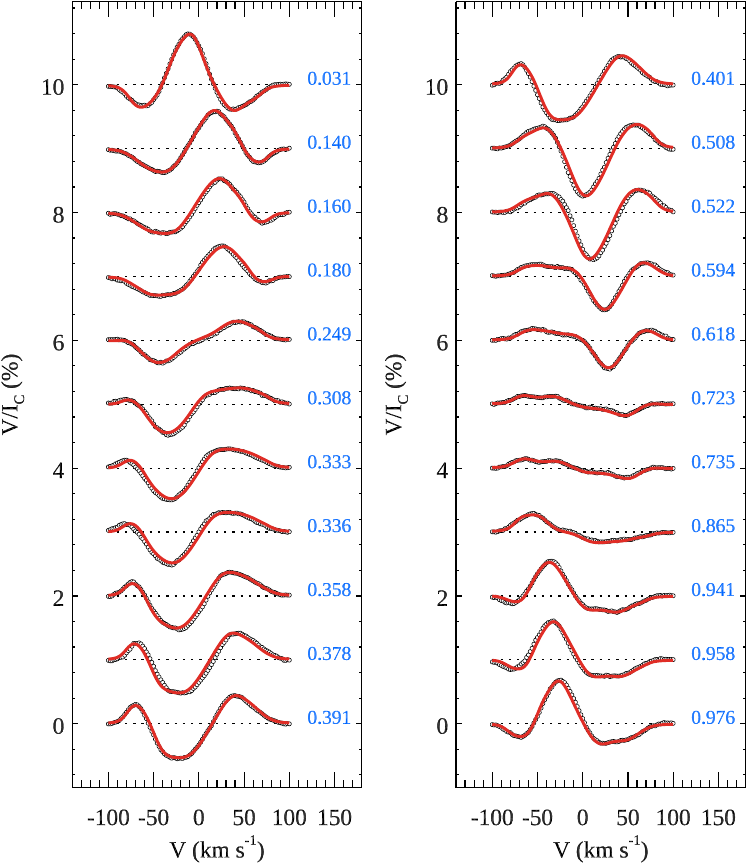}
	\caption{Comparison between the observed Stokes $V$ LSD profiles (open symbols) and the ZDI fit (red solid lines). The spectra are offset vertically and arranged according to the primary's rotational phase, which is indicated to the right of each profile.}
	\label{fig:zdi_prf}
\end{figure}

\begin{figure}
	\centering\includegraphics[width=0.8\hsize]{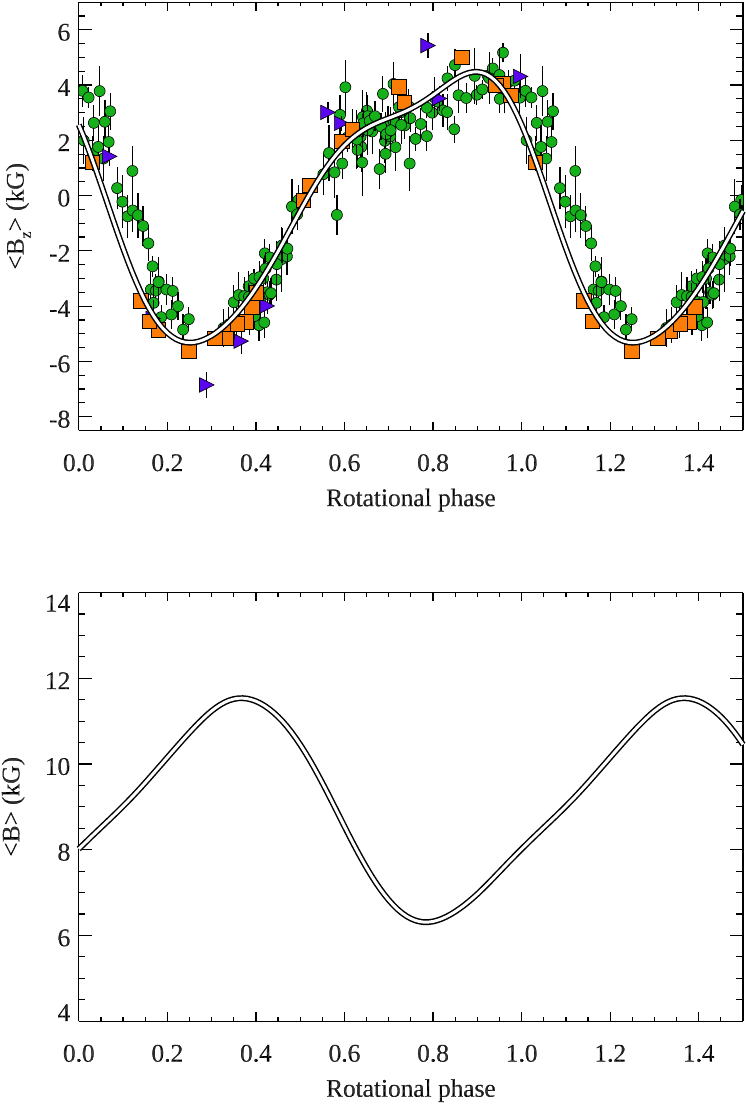}
	\caption{\textit{Upper panel:} Comparison between \bz\ measurements (symbols, colour scheme is to Fig. \ref{fig:magA}) and the longitudinal field variation predicted by the field distribution shown in Fig.~\ref{fig:zdi_map}. \textit{Lower panel:} Predicted variation of the mean field modulus \bs.}
	\label{fig:zdi_bz}
\end{figure}

The final fit achieved by the ZDI code to the observed Stokes $V$ LSD profiles is illustrated in the upper panels of Fig.~\ref{fig:zdi_prf}. The model reproduces the morphology of the observed polarisation profiles well. We also compared the mean longitudinal magnetic field predicted by the ZDI model geometry with \bz\ measurements (Fig.~\ref{fig:zdi_bz}). As expected, the ESPaDOnS \bz\ are very well reproduced. The agreement with other longitudinal field determinations is also reasonably good, considering their scatter. Fig.~\ref{fig:zdi_bz} also shows the predicted \bs\ phase curve.

No conclusive evidence of a magnetic field has been found in the secondary from available spectropolarimetric data. However, we have collected a handful of facts indirectly indicating with a very high probability that the cooler companion star is potentially magnetic.

\begin{figure*}
	\centering\includegraphics[width=0.9\textwidth]{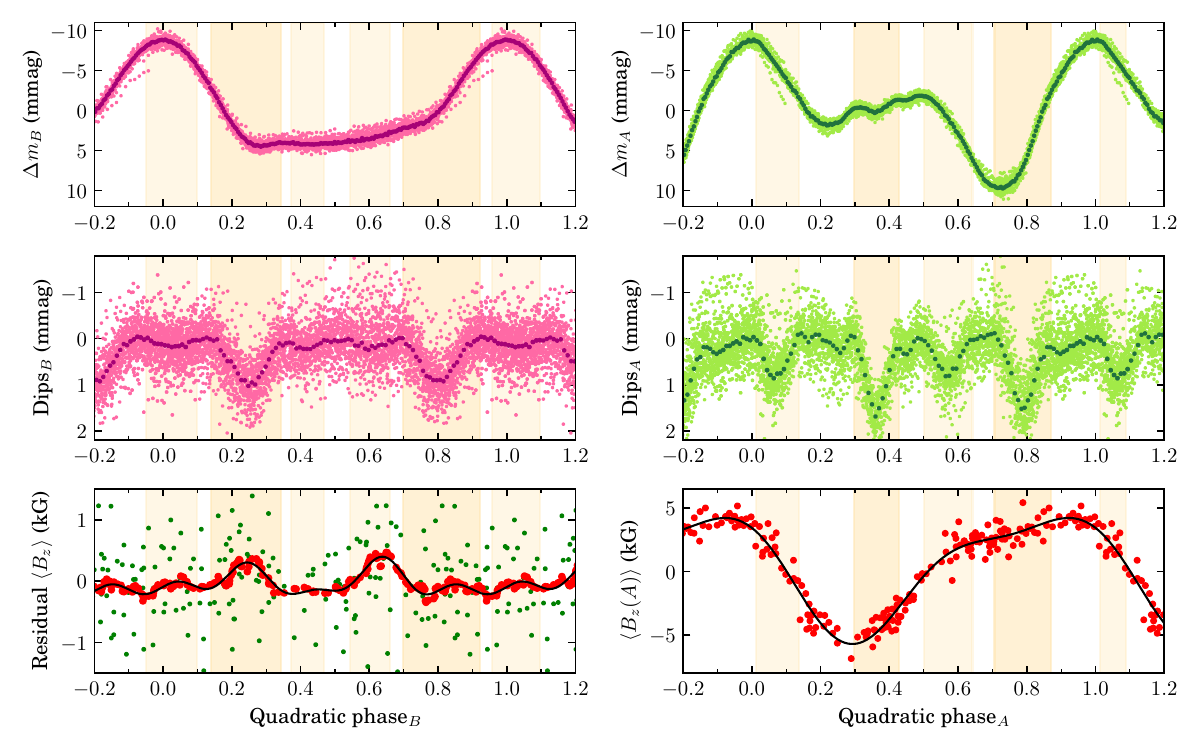}
	\caption{Top panels: {\it TESS} light curves of the components {\it B} (on the left) and {\it A} (on the right) obtained in this study. Middle panels: dips in the light curves of corresponding components. Bottom left panel: The raw (green dots) and smoothed using the running average (red circles) residual longitudinal magnetic field of \hvezda\ as a phase of the rotational period $P_{B}=0.5226938$ days. Bottom right panel: the magnetic curve of the primary component. Thin black lines show the fits made using the low-order harmonic polynomials. Vertical shaded bands indicate the position of dips in the shown light and magnetic curves. More intensive dips are shown in darker colours.}
	\label{fig:BzB}
\end{figure*}

At first, the individual light curves extracted from {\it TESS} photometry show clear periodicities with dips, which are common for chemically peculiar stars harbouring magnetic fields of complex structure. The nature of the dips remains unclear, but as the most plausible explanation, the presence of semitransparent structures confined in the co-rotating magnetosphere of the stars was proposed by \citet{2020pase.conf...46M} and developed by \citet{2022A&A...659A..37K}. In Fig. \ref{fig:BzB}, we combined the {\it TESS} light curves of both components and the dips extracted from them using the techniques described in Sec.\,\ref{sec:dips}. To emphasize the location of the dips in the original light curves, we marked them with shaded bands. The fact that the amplitude and, especially, stability of dips in the light curve of the secondary star (left panels of Fig.\,\ref{fig:BzB}) are comparable to those observed in the hotter component (right panels) with a very strong field supports the hypothesis that a magnetosphere also exists around the cooler component.

The second argument in favour of the magnetic nature of the secondary star is its accelerating rotation. Theoretical modelling of evolution in massive stars predicts a significant impact of the magnetic field, even of the order of a few hundred gauss, on the rotational properties of stars during their evolution on the main sequence \citep[e.g.][]{2011A&A...525L..11M, 2019MNRAS.485.5843K}, whereas for the intermediate-mass stars, such calculations have yet to be performed. Additionally, we cannot ignore evolutionary effects on the rotational rate as a $2.7M_\odot$ star of the age of HD\,34736 still evolves towards the ZAMS in the MIST models.

Eventually, we can examine the residuals of the primary magnetic curve for a possible correlation with the light curve of the secondary star. For this purpose, we have subtracted the smooth fit shown in Fig. \ref{fig:magA} from the measured magnetic field. The residuals have been folded with the rotational period $P_{1B}$ (Table \ref{table:parametry}) and smoothed using the running average. The result of this procedure is shown in the left bottom panel of Fig. \ref{fig:BzB}. The position of two peaks, in this case, coincides well with the beginning and end of the flat section of the photometric curve. Although such coincidence cannot serve as a firm detection of the secondary's magnetic field, we consider it an indirect indicator that the fast-rotating component of \hvezda\ may potentially have a longitudinal field \bz\ order of 500\,G. For comparison, the right bottom panel maps the dips' position on the magnetic curve of the primary. Notably, for this component, the two most intense dips occur close to the extrema of the longitudinal field \bz, while the phases of the remaining two dips cover the moments when the \bz\ curve reverses sign.

In light of these results, we tentatively suggest that the secondary component of \hvezda\ may be a rapidly rotating, weakly magnetic star similar to CU\,Vir \citep{2011A&A...534L...5M, 2014A&A...565A..83K}.

\subsubsection{Physical parameters of the components}\label{sec:phys_par}

The atmospheric parameters of HD\,34736 were first evaluated spectroscopically in \citetalias{2014AstBu..69..191S}. That research led to a two-star solution with the following parameters: \teffA$\;=13\,700$\,K, \teffB$\;=11\,500$\,K, \loggA$\;=\;$\loggB$\;=4.0$, and \vsini(A)$\;=73\pm7$ km\,s$^{-1}$, \vsini(B)$\;\geq90$ km\,s$^{-1}$. With the new observational material covering a broader range of wavelengths and rotational and orbital phases, we decided to revise our previous findings.

First, the projected rotational velocity \vsini\ has been re-evaluated using two different techniques. The spectrum synthesis of two Fe \textsc{ii} lines at 450.8\,nm and 452.2\,nm with low Land\'e factors made using the {\sc SynthMag} code~\citep{2007pms..conf..109K} with the atmospheric parameters adopted from \citetalias{2014AstBu..69..191S} yields \vsini$\;=75\pm3$ km\,s$^{-1}$, in agreement with the previous estimate. Alternatively, fitting the mean Stokes $I$ LSD profiles with the broadening function \citep{2008oasp.book.....G}, we obtain \vsini\ larger by 2--3 km\,s$^{-1}$. Neither of the two approaches can provide an unambiguous estimate for the secondary star. The resulting \vsini\ of the broader-line component ranges from 110 to 180 km\,s$^{-1}$. The wings of the Mg~\textsc{ii} 448.1\,nm line in the composite spectrum at the orbital phase $\varphi_\mathrm{orb}=0.1$ (Sec. \ref{sec:multiple}) argue for the upper limit of the rotational velocity \vsini$\,\approx180$ km\,s$^{-1}$ of the secondary star.

\begin{figure}
	\centering\includegraphics[width=0.9\hsize]{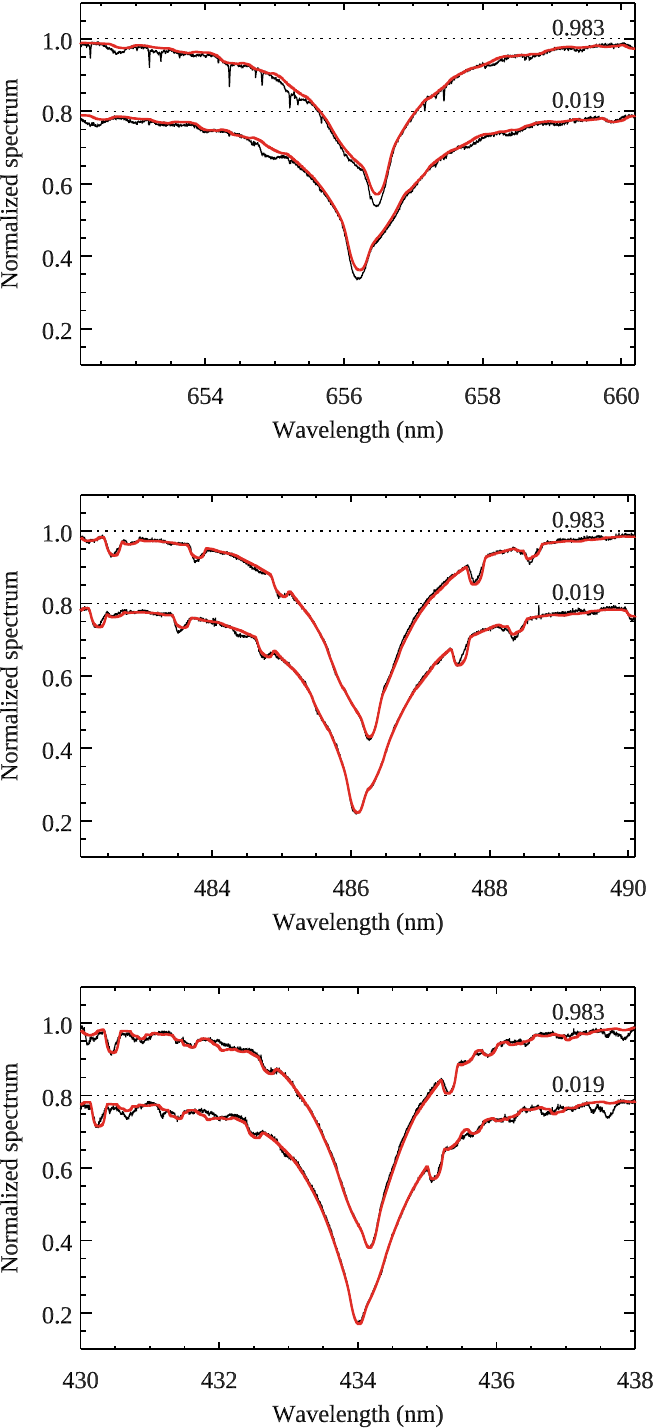}
	\caption{Observed hydrogen Balmer lines (thin black lines) compared with theoretical fit (thick red lines). The observed and model spectra are illustrated for two orbital phases close to periastron, with the second set offset vertically for display purpose.}
\label{hlines}
\end{figure}

Next, we searched for a combination of \teffA, \teffB, and $R_{A}/R_{B}$ which would best fit the observed spectra in three regions containing hydrogen lines H$_\alpha$, H$_\beta$, and H$_\gamma$ at different orbital phases. The analysis was performed on the two ESPaDOnS spectra taken close to the moment of maximum amplitude of the radial velocity \vr. The optimal fit was achieved for \teffA$\;=13\,000\pm500$\,K, \teffB$\;=11\,500\pm1\,000$\,K, and $R_{A}/R_{B}=1.30\pm0.05$  (Fig.~\ref{hlines}). As the hydrogen lines in the spectra of the early-type stars are equally sensitive to both \teff\ and \logg\ and there are no reliable methods of independently constraining both parameters, we adopted \logg$\;=4.0$ as the lower limit at this stage of analysis. Considering the young age of \hvezda, the \logg\ value corresponding to the stellar mass and radius is very likely higher.

The spectroscopic parameters derived from hydrogen lines can be compared with the absolute magnitude computed from the {\it Gaia} Data Release~3 (DR3) parallax, $\pi=2.685\pm0.054$ mas \citep{2022arXiv220800211G}. Assuming no interstellar extinction, the system's total magnitude is $M_{\rm V}=-0.02\pm0.05$. This value can be reproduced with $R_{A}=2.17\pm0.08\,R_\odot$ for the effective temperatures and the radii ratio inferred above. Adopting a moderate reddening of $E(B-V)=0.0248$ based on the Galactic model by \citet{2005AJ....130..659A} increases $R_{A}$ by less than 0.10 $R_\odot$. The resulting radius for the primary is reasonably consistent with evolutionary model predictions for young dwarfs with a \teff\ of $13\,000$ as explained in Sec.~\ref{sec:mfield}. On the other hand, the surface gravity \logg$\;=4.0$ adopted for the hydrogen line fitting is too low for stars at this evolutionary stage. This discrepancy may be explained by the impact of an enhanced metallicity and deficient helium on the hydrogen line profiles, which leads to an underestimation of \logg\ by up to 0.25 dex when not accounted for \citep{1997A&A...320..257L}.

We additionally attempted to cross-check the stellar parameters of the system by fitting theoretical models to the observed spectral energy distribution (SED). To do this, we calculated a grid of model atmospheres using the \textsc{LLmodels} stellar model atmosphere code \citep{2004A&A...428..993S} with average abundances given in Table~\ref{table:abund}. We then optimized model parameters, such as the effective temperatures and stellar radii, to find a model that best fits the observed flux. Observations were taken from the {\it Gaia} DR3 \citep{2022yCat.1355....0G} where, for fitting purposes, we ignored fluxes below the Balmer jump due to calibration inaccuracies. Instead, we used observed broad-band UV fluxes obtained with the S2/68 telescope of the TD1 mission (European Space Research Organization (ESRO) satellite) \citep{1978MNRAS.185..371M}, complemented by data from the 2Micron All-Sky Survey \citep[2MASS,][]{2003tmc..book.....C} for the infrared. Observations were transformed into absolute fluxes using the calibrations given by \citet{2003AJ....126.1090C}.

\begin{figure*}
\centering\includegraphics[width=\textwidth]{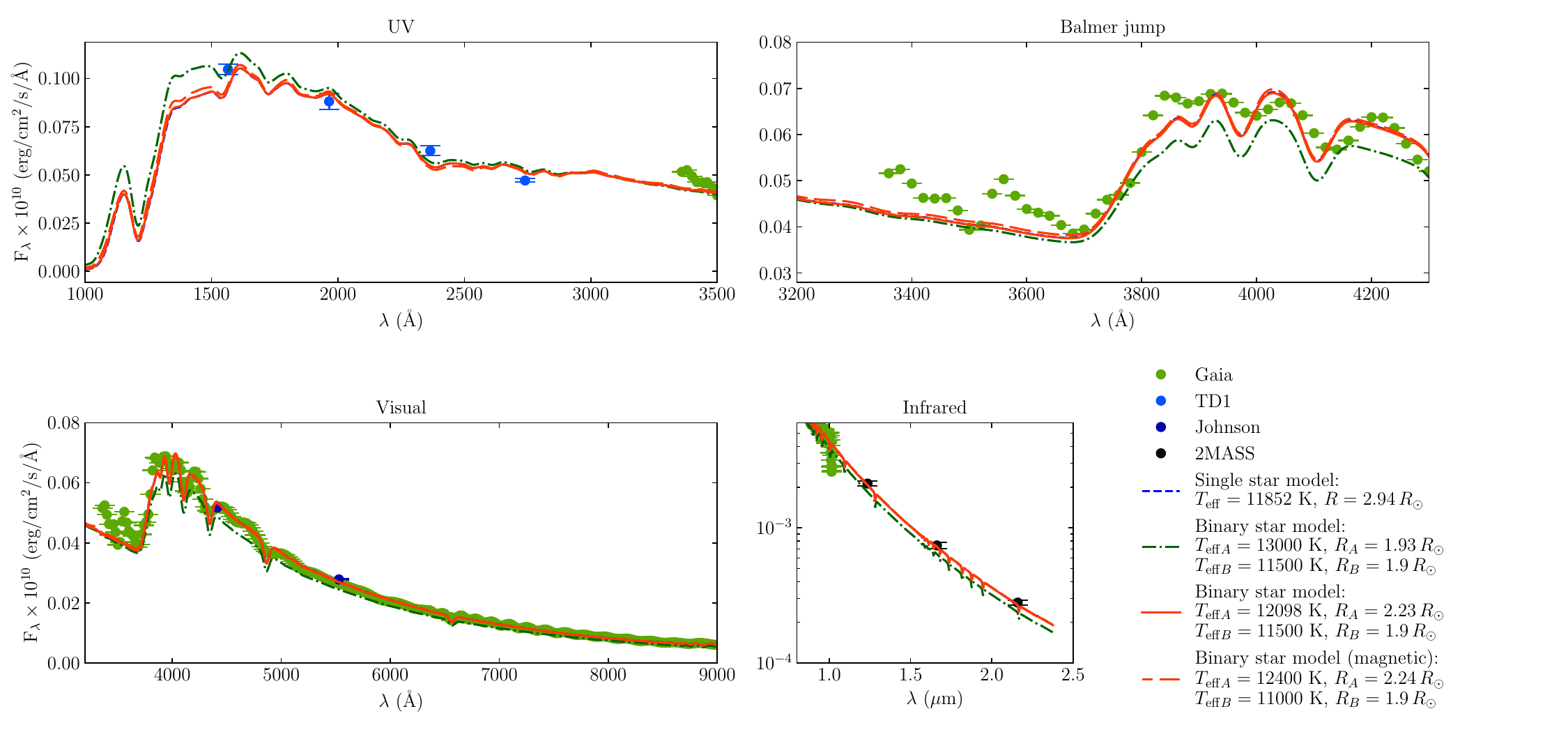}
\caption{Comparison between observed and predicted flux at different wavelength domains. Note the logarithmic y-axis scale for the infrared flux (bottom right panel). See the plot legend for more details.}
\label{fig:sed}
\end{figure*}

The predicted and observed energy distributions are compared in Fig.~\ref{fig:sed}. In our SED fitting, we applied an interstellar extinction correction to $E(B-V)=0.0248$ and $A_{\rm v}=0.0755$, respectively.

First, assuming fixed parameters for the secondary, \teffB$\;=11\,500$\,K and $R_{B} = 1.9$ \Rsun\ (close to the value predicted by the ratio of spectroscopically derived radii) and ignoring the magnetic field, we find the best-fit model for the primary to have \teffA$\;=12\,098\pm100$\,K\footnote{This uncertainty is based solely on the errors of the {\it Gaia} spectrophotometry and does not include the errors for $E(B-V)$. After including the extinction model published by \cite{2005AJ....130..659A}, the combined error of \teffA increases to 365\,K.} and $R_{A}=2.23^{+0.15}_{-0.07}$\,\Rsun\, which includes parallax uncertainty (red solid line in Fig.~\ref{fig:sed}). We could not derive \logg\ from the available observations
and thus kept it similar and fixed to \logg$\;=4.0$ for both components. While the derived radius of the primary agrees well with a previous spectroscopic estimate, the effective temperature that we derive from fitting the SED is significantly lower (by about 900\,K) than the spectroscopically derived value.

Including the magnetic field in the opacity and emerging flux calculation in our model atmospheres results only in a marginal increase of the effective temperature
of the primary to \teffA$\;=12\,152\pm100$\,K (assuming surface average magnetic flux density $\langle B\rangle=7.6$\,kG, Sec. \ref{sec:mfield}), which is still too low compared to the spectroscopic estimate \citep[see, for the details and implementation of the magnetic field in our stellar model atmospheres][]{2008A&A...487..689S,2006A&A...454..933K}.

We could achieve a better match between spectroscopy and SED for the primary
but only assuming the secondary is cooler than 11\,500\,K. For instance, assuming \teffB = 11\,000\,K, $R_{B}=1.9$\,\Rsun\ and calculating magnetic model atmospheres for the primary, we obtain \teffA$\;=12\,400\pm100$\,K, $R_{A}=2.24\pm0.02$\,\Rsun\ with a very similar fit quality as in the previous case of \teffB$\;=11\,500$\,K (red long-dashed line in Fig.~\ref{fig:sed}). 

Finally, assuming a single star model results in a good fit to the observed SED with stellar parameters \teff$\;=11\,852\pm100$\,K, $R=2.94\pm0.01$\,\Rsun\ (blue dashed line in Fig.~\ref{fig:sed}).

The predicted flux calculated assuming spectroscopically derived parameters for the primary (\teffA$\;=13\,000$\,K) and secondary (\teffB$\;=11\,500$\,K)
could not simultaneously match observations in all wavelength ranges (green dash-dot line in Fig.~\ref{fig:sed}), where we again fixed the radius of the secondary
to be $R$(B)$\;=1.9$\,\Rsun, while optimizing for the radius of the primary to match the observed points as closely as possible, which resulted in $R_{A}\;=1.93$\,\Rsun.
We thus conclude that it is impossible to constrain robustly the parameters of both components solely from fitting the SED and a self-consistent approach
similar to that used by, e.g., \citet{2019MNRAS.488.2343R, 2021A&A...655A.106R} and \citet{2013A&A...551A..14S} would be needed, which, however, is out of the scope of the present paper.

\citet{1978AJ.....83..172R} and \citet{RM2009} classified \hvezda\ as a Si-type CP star. The presence of intense lines of singly ionised silicon~(e.g. 412.9--413.0, 504.1, and 634.7--7.1\,nm) in the spectrum of the magnetic component ostensibly supports this classification. However, careful inspection of stellar spectra also reveals variable and strengthened lines of chromium, titanium, and some rare-earth elements, which, together with weak helium lines, implies that the spectrum is more accurately classified as He-wk.

To quantify the peculiarities of the magnetic primary, we analysed its chemical composition at two rotational phases when the magnetic field (Sec. \ref{sec:mfield}, Fig. \ref{fig:magA}) was close to the minimum (HJD 2457331.665, $\varphi_{\mathrm{rot}A}=0.272$, HERMES) and maximum (HJD 2456967.515, $\varphi_{\mathrm{rot}A}=0.773$, MSS).

Spectra of both components were modelled with the {\sc SynthMag} code~\citep{2007pms..conf..109K} in the LTE approximation and using up-to-date atomic data from the VALD database. A homogeneous magnetic field with a radial component $B_\mathrm{r}=17$\,kG was accounted for only in the main component. The microturbulent velocity was set to zero. {\sc Atlas9} atmospheric models for both components were taken from the NEMO database~\citep{2002A&A...392..619H}.

The spectrum of the magnetic component evolves with rotation. The largest variations are found for magnesium, chromium, and silicon. For example, between rotational phases $\varphi_{\mathrm{rot}A}\approx0.272$~(corresponding approximately $\langle B_\mathrm{z}\rangle$ minimum) and $\approx 0.773$~(corresponding to the plateau in the $\langle B_\mathrm{z}\rangle$ curve, Fig. \ref{fig:magA}), the abundance of magnesium varies by 1.1 dex. Iron demonstrates the opposite trend: at $\varphi_{\mathrm{rot}A}=0.773$ its concentration is nearly solar \citep{2021A&A...653A.141A} and increases by 0.5 dex at  $\varphi_{\mathrm{rot}A}=0.272$. The chromium abundance at both phases is approximately the same, but the profiles of the individual Cr lines are variable. We find silicon overabundant by 0.9 dex. Intensity of Si~\textsc{ii} lines at 623.2, 634.7, and 637.1\,nm can be described assuming $B_\mathrm{d} \approx 24$ kG when $\langle B_\mathrm{z}\rangle$ is near minimum. The abundances derived from Si~\textsc{ii} lines differ from those from Si~\textsc{iii} lines in a way that is common for magnetic CP stars \citep{2013A&A...551A..30B}. The chemical composition of the narrow-lined magnetic component is summarized in Table~\ref{table:abund}. We estimate a typical error of about 0.1 dex for the abundances of most elements except praseodymium, which is as high as 0.5 dex. The main sources of error are the spectroscopic variability of both components and uncertainties in atmospheric parameters.

\begin{table}
	\begin{center}
	\caption{Chemical composition of the primary component of HD\,34736 evaluated with respect to the Sun \citep{2021A&A...653A.141A} at the rotational phases 0.773 and 0.272. Only one value is given when the corresponding abundance remains constant. The em-dashes indicate absent data.}
	\label{table:abund}
	\begin{tabular}{l c c}
		\hline\hline
		Element     &   \multicolumn{2}{c}{$\Delta \varepsilon$, dex}                 \\
		            &   $\varphi_{\mathrm{rot}A}=0.272$  &  $\varphi_{\mathrm{rot}A}=0.773$   \\
		\hline
		He     &  \multicolumn{2}{c}{$-1.7$}   \\
		Mg     &  $+0.2$   &  $-1.1$         \\
		Al     &     0      &  ---           \\
		Si     &  \multicolumn{2}{c}{+0.9}   \\
		Ti     &  \multicolumn{2}{c}{+0.66}  \\
		Cr     &  +1.4   & +1.3              \\
		Fe     &  +0.7   & +0.1              \\
		Pr     &   +3.9   & ---              \\
		Nd     &  \multicolumn{2}{c}{+2.8}   \\
		Dy     &  ---     &    +3            \\
	\hline\hline
	\end{tabular}
\end{center}
\end{table}

The broad-lined star is poorly represented in the composite spectra due to its fast rotation, making it impossible to assess its chemical peculiarities. We can only say that with \vsini$\;=180$ km\,s$^{-1}$ adopted for the broad-lined component, this star should have magnesium overabundant by at least 0.5 dex relative to the solar value. The signatures of the intrinsic spectral variability of the companion star are visible only in the Mg~\textsc{ii} 448.1\,nm and selected Si~\textsc{ii} lines at the orbital phases of the maximum Doppler separation.

\subsubsection{Stellar multiplicity}\label{sec:multiple}
\citetalias{2014AstBu..69..191S} depicted \hvezda\ as a double-lined spectroscopic binary (SB2) with an orbital period shorter than one day, which was tentatively proposed for the system based on the limited observations. In the current study, we comprehensively describe this SB2 system, including its orbital solution and possible multiplicity of higher order.

To measure the radial velocity \vr\ of the individual components, we primarily fit the mean LSD Stokes $I$ profiles with a function defining the rotationally broadened profile \citep{2008oasp.book.....G}. For the few instances where two sets of lines were visible, the fitting function was a sum of two profiles. We preferred using a broadening function, as rotation dominates over the other line broadening mechanisms in the spectra of \hvezda. By averaging many lines from different elements, LSD, to some extent, alleviates the impact of a spotted surface on the derived \vr. Also, by using the broadening function for fitting, we give additional weight to the outer parts of the line to minimise the effects of spots, which mostly affect line cores.

At the same time, we approximate the Mg \textsc{ii} 448.1\,nm lines of both components with model spectra synthesised for the components' stellar parameters. Despite the inhomogeneous distribution of magnesium in \hvezda, \vr\ measured from this element shows better accuracy than hydrogen due to the profound blending of the components' hydrogen lines.

Individually measured radial velocities are listed in Table~\ref{table:summary}. For the primary component, we give only \vr\ measured from the LSD profiles \vr(A)$_\mathrm{LSD}$. For the secondary star, where the magnesium line modelling works better in a broader range of orbital phases, we show both types of velocities denoted as \vr(B)$_\mathrm{LSD}$ and \vr(B)$_\mathrm{Mg\textsc{ii}}$. As a conservative upper limit of error, in the case of \vr(B)$_\mathrm{Mg\textsc{ii}}$, we adopted 20 km\,s$^{-1}$. This value includes uncertainties defined by the quality of input data~(e.g., SNR and continuum normalisation) and accuracy of the atmospheric parameters. The radial velocities measured on the same night were averaged.

\begin{figure}
\centering\includegraphics[width=\columnwidth]{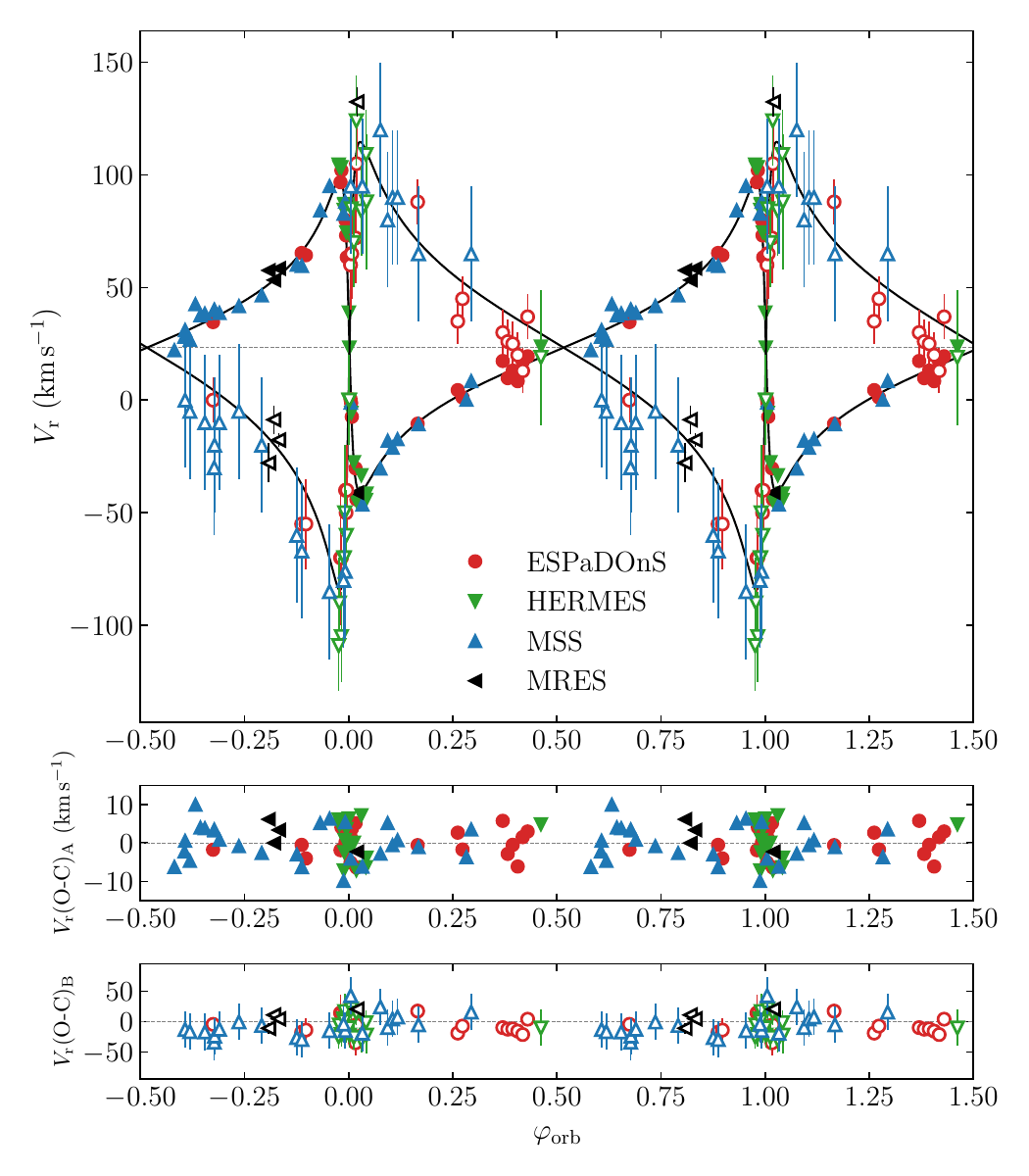}
\caption{Measured radial velocities of the HD\,34736 components plotted against the orbital phase computed for parameters from Table~\ref{table:orbit}. Filled symbols correspond to the magnetic primary, and open symbols mark  $V_\mathrm{r}$ of the secondary.}
\label{orbitfit}
\end{figure}

The final fit of velocities shown in Fig.~\ref{orbitfit} has been made using the programme \textsc{rvfit} \citep{2015PASP..127..567I}. In Table~\ref{table:orbit}, we provide two orbital solutions based on the radial velocities from Table~\ref{table:summary} with each solution based on the different sources of radial velocities of the secondary component.

\begin{table}
	\begin{center}
		\caption{Orbital parameters HD\,34736 derived from the observed radial velocity variation. The \nth{2} and \nth{3} columns show the results of fitting based on velocities measured using the Mg~\textsc{ii} 448.1\,nm line and the LSD profiles, respectively. $T_\mathrm{p}$ is the moment of periastron.}\label{table:orbit}
		\begin{tabular}{lr | r}
			\hline
			\hline
			Parameter & Value (Mg \textsc{ii})   & Value (LSD)  \\
			\hline
			$T_\mathrm{p}$       & 2457415.3460~(0.003)    &  2457415.3481~(0.003) \\
			$K_\mathrm{A}$ (km\,s$^{-1}$)    & 69.74~(0.07)            &  69.74~(0.07)         \\
			$K_\mathrm{B}$ (km\,s$^{-1}$)    & 99.57~(3.15)            &  111.63~(1.12)        \\
			$\gamma$ (km\,s$^{-1}$) & 23.28 ~(0.05)        &  23.32~(0.05)         \\
			$P$ (days) & 83.2193~(0.0030)                  &  83.2183~(0.0035)     \\
			$e$ & 0.8103~(0.0003)                          &  0.8104~(0.0003)      \\
			$\omega$ ($\degr$) & 84.2~(0.1)                &  84.3~(0.1)           \\
			RMS$_\mathrm{A}$ (km\,s$^{-1}$) & 4.86                  &  4.87                 \\
			RMS$_\mathrm{B}$ (km\,s$^{-1}$) & 17.92                 &  19.76                \\
			\hline
			$M_\mathrm{B}/M_\mathrm{A}$ & 0.70~(0.02)                        &  0.62~(0.01)          \\
			$M_\mathrm{A}\sin^3 i$ ($M_{\sun}$) & 4.9~(0.3)         &  6.4~(0.1)            \\
			$M_\mathrm{B}\sin^3 i$ ($M_{\sun}$) & 3.5~(0.1)         &  4.0~(0.1)            \\
			$a_\mathrm{A}\sin i$ ($R_{\sun}$)   & 67.2~(0.1)        &  67.2~(0.1)           \\
			$a_\mathrm{B}\sin i$ ($R_{\sun}$)   & 95.9~(3.0)        &  107.5~(1.0)          \\
			\hline \hline
		\end{tabular}
	\end{center}
\end{table}

According to the best-fit solution, HD\,34736 consists of two hot stars orbiting each other on highly eccentric orbits~($e>0.8$) with a period of 83 days. Interestingly, the rotation of the narrow-lined magnetic primary component is quasi-synchronised with its orbital motion; the ratio $P_\mathrm{orb}/P_{1A}$ is almost equal to 65. The primary has the projected mass $M_\mathrm{A}\sin^3 i=4.9$--$6.4\,M_{\odot}$, where $i$ is the inclination angle of the orbit. The mass $M_\mathrm{B}\sin^3 i$ of the companion is $3.5$--$4.0\,M_{\odot}$. Such values typically characterise early and mid-B main sequence stars and appear systematically larger than those implied by the components' effective temperatures determined in Sec.~\ref{sec:phys_par}. We address this problem in  Sec. \ref{sec:discuss}.

\subsubsection*{Is \hvezda\ an eclipsing binary?}\label{sec:eclipses}

If the angle $i$ were close to 90 degrees, the binary could undergo eclipses. Here, we will try to predict when these eclipses might occur during the orbit and estimate their parameters. To avoid misunderstandings, we will consistently distinguish between so-called \textit{transits} when a smaller component \textit{B} passes over the disc of component \textit{A}, and \textit{occultations} when a more prominent component \textit{A} covers component \textit{B} and can cover it entirely.

 \begin{figure*}
 	\centering\includegraphics[width=70mm]{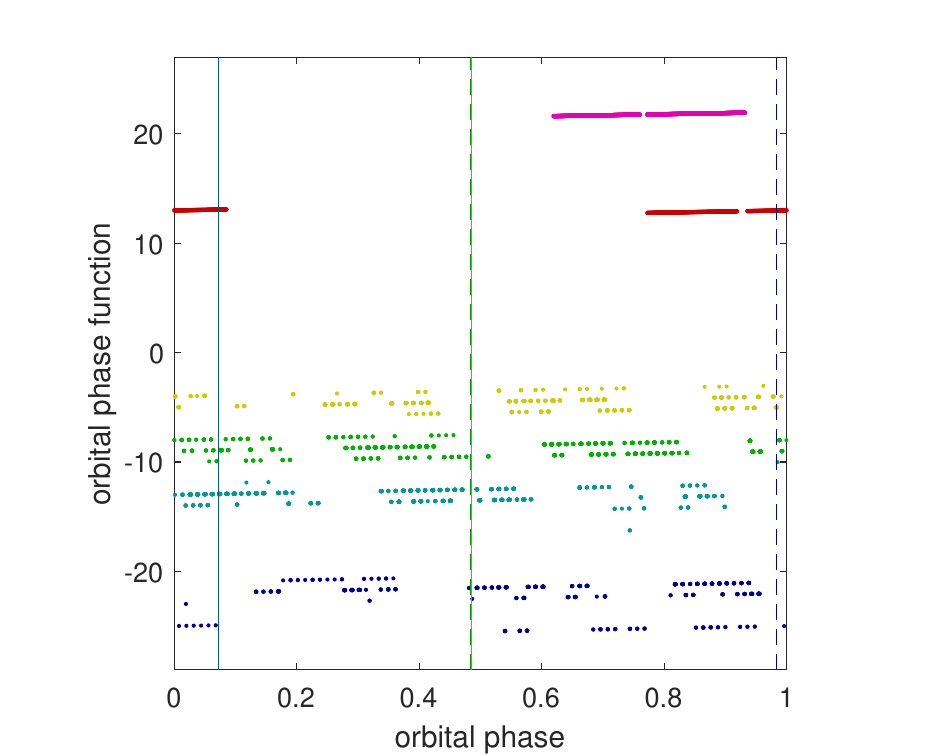}
 	\includegraphics[width=70mm]{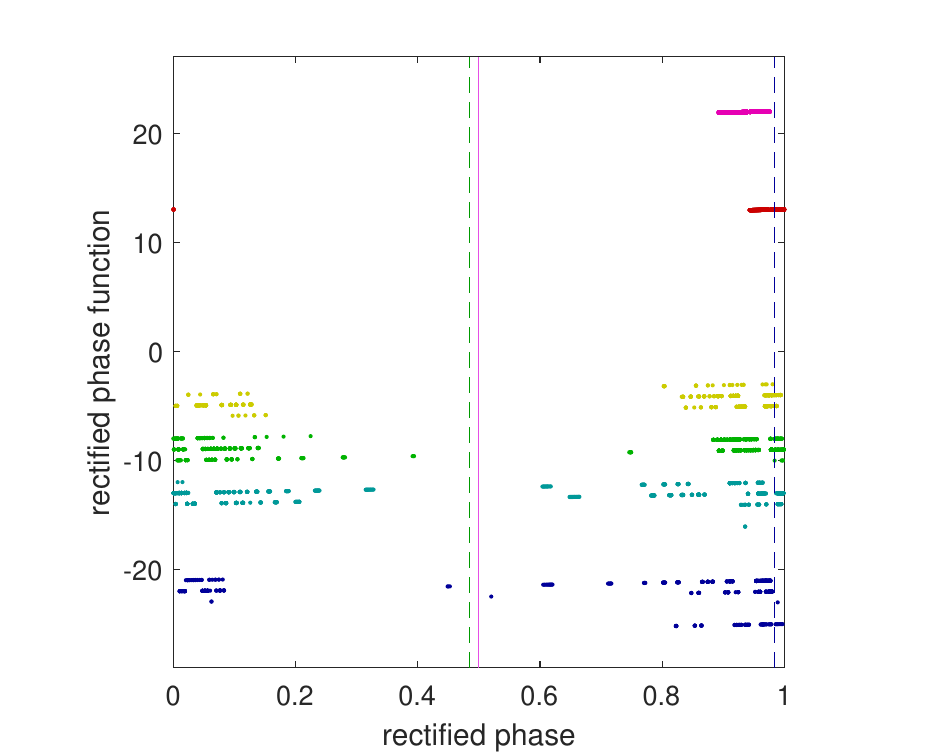}
 	\caption{The distribution of moments of photometric observations with respect to the orbital motion of the components of the spectroscopic binary star \hvezda. The dependencies of (a) the orbital phase function $\vartheta$ on the orbital phase $\varphi$  and (b) the rectified phase function $\vartheta_{\mathrm r}$ on the rectified phase $\varphi_{\mathrm r}$, as these quantities are defined in the equations \ref{eq:orbteta}, \ref{eq:rectteta}, and \ref{eq:orbfi} are used for illustration. Observations KELT 1, 2, 3, and 4 (Table \ref{table:OCAB}) are distinguished by the colour of markers from blue to yellow. {\it TESS} sector 05 and 32 observations are marked in red and magenta. Orbital phases are marked vertically when superior (green) and inferior (pink) conjunction occurs, and the passages of periastron and apastron are indicated in dashed lines. Possible occultations and transits occur at rectified phases 0 and 0.5. From Fig.\,(b), it is obvious that the observation does not cover the possible transit. On the other hand, the observation from {\it TESS} Sector 05 well covers the occultation.}
 	\label{fig:orbphasef}
 \end{figure*}

For the description of the motion of stars in a binary with parameters of orbital period $P_{\rm orb}=83\fd219(3)$, argument of periastron $\omega=1.4696(18)$\,rad, eccentricity $e=0.8103(3)$, and the basic moment of the periastron passage $T_{\mathrm{p}}$, from the viewpoint of a distant observer, it is useful to introduce the orbital phase function  $\vartheta$, the rectified phase function $\vartheta_{\mathrm{r}}$, and their corresponding phases: $\varphi$, and $\varphi_{\mathrm{r}}$, as follows:
 	\begin{align}
 		&M(t)=2\,\pi\,\frac{t-T_{\mathrm{p}}}{P_{\mathrm{orb}}};\label{eq:mean_an}\\
 		&\vartheta(t)=\frac{1}{2\,\pi}\hzav{M(t)+\omega+\frac{\pi}{2}}=\frac{t-M_{\mathrm{orb}}}{P_{\mathrm{orb}}}; \label{eq:orbteta}\\
 		&M_{\mathrm{orb}}=T_{\mathrm{p}}-P_{\mathrm{orb}}\zav{\frac{\omega}{2\,\pi}+\frac{1}{4}}=2\,457\,375.077(23); \nonumber \\
 		&\vartheta_{\mathrm{r}}(t)=\frac{1}{2\,\pi}\hzav{\theta(t)+\omega+\frac{\pi}{2}};\quad \theta=2\,\pi\,\vartheta_{\mathrm{r}}-\omega-\frac{\pi}{2};\label{eq:rectteta}\\
 		&\varphi=\vartheta-\mathrm{floor}(\vartheta);\quad \varphi_{\mathrm{r}}=\vartheta_{\mathrm{r}}-\mathrm{floor}(\vartheta_{\mathrm{r}}); \label{eq:orbfi}\\
 		&r=\frac{a\zav{1-e^2}}{1+e\cos \theta}=\frac{a\zav{1-e^2}}{1+e\sin(2\pi\vartheta_{\mathrm{r}}-\omega)},\\
 		&{V_{\rm r}}_{\mathrm{A,B}}(t)=\gamma \pm K_{\mathrm{A,B}}[\cos(\theta+\omega)+e\,\cos\omega]=\\
 		&\quad\quad\quad =\gamma \pm K_{\mathrm{A,B}}[\sin(2\,\pi\,\vartheta_{\mathrm{r}})+e \cos\omega], \nonumber
 	\end{align}
 	$M(t)$ is the mean anomaly in radians, $\theta$ is the true anomaly, as defined by Eq. \ref{eq:teta}, $r$ is the instantaneous separation of the components, $a$ is the length of the semimajor axis,  $a=(A_{\mathrm{A}}+A_{\mathrm{B}})/\sin i=163(3)/\sin i$\,\Rsun. The ratio between maximum and minimum separation of components is substantial -- $r_{\mathrm {max}}/r_{\mathrm {min}}=(1+e)/(1-e)=9.543(17)$. The distance of components is minimal when the stars pass periastron; this occurs if the both orbital and rectified phases equal $\varphi_{\mathrm{per}}=\varphi_{\mathrm{rper}}=\textstyle{\frac{1}{2\pi}}\omega+\textstyle{\frac{1}{4}}=0.4840(3)$ (see Eqs. (\ref{eq:orbteta}), (\ref{eq:rectteta}), and (\ref{eq:orbfi})). The orbital and rectified phases of the apastron passage are $\varphi_{\mathrm{ap}}=\varphi_{\mathrm{rap}}=\textstyle{\frac{1}{2\pi}}\omega+\textstyle{\frac{3}{4}}=0.9840(3)$.

The moment of minimum brightness during the occultation corresponds to the moment of the \textit{superior conjunction} of the binary star components, i.e., if the rectified phase $\vartheta_{\mathrm{r}}=0$. In contrast, the minimum brightness during the transit occurs when the binary components are in the \textit{inferior conjunction}, and the rectified phase $\vartheta_{\mathrm{r}}$ equals 0.5. Hence the conditions for true anomalies $\theta_{\mathrm{s,i}}$ for superior/inferior conjunctions are:
 	\begin{align}
 		\theta_{\mathrm{s,i}}=2\,\pi\,k -\omega \mp\frac{\pi}{2},\quad \label{eq:th12}
 	\end{align}
 	where $k$ is an integer. In the case of $k=0$, $\theta_{\mathrm s}=-3.0404(18)$\,rad and $\theta_{\mathrm i}=0.1012(18)$\,rad. Combining Eqs.\,(\ref{eq:E}) and (\ref{eq:th12}), we obtain the corresponding values of eccentric anomalies of superior/inferior conjunctions  $E_{\mathrm{s,i}}(\theta_{\mathrm{s,i}})$ consecutively preceding/following the basic passage through periastron.
 	\begin{align}
 		&E_{\mathrm{s,i}}=2\arctan\hzav{\sqrt{\frac{1-e}{1+e}}\,\tan\zav{\frac{\theta_{\mathrm{s,i}}}{2}}};\\
 		&T_{\mathrm{s,i}}=T_{\mathrm{p}}+\frac{P_{\mathrm{orb}}}{2\,\pi}\zav{E_{\mathrm{s,i}}-e\sin E_{\mathrm{s,i}}},\\
 		&\varphi_{\mathrm{s,i}}\!=\frac{T_{\mathrm{s,i}}\!-\!M_{\mathrm{orb}}}{P_{\mathrm{orb}}}\!-\!\mathrm{floor}\zav{\frac{T_{\mathrm{s,i}}\!-\!M_{\mathrm{orb}}}{P_{\mathrm{orb}}}};\quad r_{\mathrm{s,i}}\!=\frac{a\zav{1\!-\!e^2}}{1\mp e\sin\omega}; \label{eq:orbr}
 	\end{align}
 	Eccentric anomalies in superior/inferior conjunctions are $E_{\mathrm {s}}=-2.831(6)$\,rad, $E_{\mathrm {i}}=0.0328(6)$\,rad, times of conjunction close to the basic periastron passage in BJD: $T_{\mathrm {s}}=2\,457\,381.13(13),\ T_{\mathrm {i}}=2\,457\,415.428(3)$, and corresponding phases $\varphi$ according to (\ref{eq:orbteta}) and (\ref{eq:orbfi}) for center of occultations and transits are $\varphi_{\mathrm {s}}=0.0727(16)$, and $\varphi_{\mathrm {i}}=0.4849$.

In the case $i=\pi/2$, $a=A_{\mathrm A}\sin^3i+A_{\mathrm B}\sin^3i=67.2(0.1)+96(3)=163(3)$\,R$_{\odot}$ and according to (\ref{eq:orbr}) the distance of components in occultation is $r_{\mathrm s}=289$\,R$_{\odot}$, while during the transit it is only $r_{\mathrm i}=31.0$\,R$_{\odot}$. To predict the eclipse widths and depths, we used the estimates of stellar parameters: $R_{\mathrm A}=2.1$\,\Rsun,\ $R_{\mathrm B}=1.9$\,\Rsun, \teff$({\mathrm A})=13\,000$\,K, \teff$({\mathrm B})=11\,500$\,K, from Sec.\,\ref{sec:phys_par}. The half-width of the occultation is 1.0 d (!), and its bolometric magnitude depth is 0.44 mag. This part of the light curve was well monitored by {\it TESS} observations in Sector 05; indeed, this occultation would not escape our notice. The transit minimum is even deeper~--- 0.86\,mag, but it would happen literally in a flash~--- its half-width would be only 2.7\,h. Our observations do not sufficiently cover this region of the light curve.

However, we emphasize that the visibility of occultations depends very dramatically on the actual inclination of the orbital plane. An occultation will only occur when $i>88.4^\circ$. In the case of a transit, the situation is more favourable; to observe it, the inclination angle must be greater than $76.5^\circ$. However, there is a high probability that we will miss the transit given its brevity.

\subsection{Radio and X-ray emission of \hvezda}\label{sec:beyondvisual}

\subsubsection{Radio observations}
A radio source in the vicinity of \hvezda\ was detected for the first time in the NRAO VLA Sky Survey (NVSS) and reported by \citet{1998AJ....115.1693C}. An integrated flux density of the object NVSS J051920$-$072048 (field C0520M08) with coordinates $\alpha_{J2000}=05^{h}19^{m}20.98^{s}$, $\delta_{J2000}=-07^\circ20'48.9''$ was measured as $2.3\pm0.4$\,mJy at 1.4\,GHz.

In the Very Large Array Sky Survey \citep[VLASS,][]{lacy2020} archive, we found three observations (epochs 1.1, 2.1, and 3.1) of the sky area encompassing \hvezda\ obtained at $\nu\sim3$\,GHz between 2017 and 2023. Appropriate quick-look images were downloaded from the archive of The Canadian Initiative for Radio Astronomy Data Analysis (CIRADA\footnote{CIRADA Image Cutout Web Service. \url{http://cutouts.cirada.ca/}}) for measurement. The flux densities shown in Tab. \ref{tab:radiofluxes} were obtained by fitting a two-dimensional Gaussian to the point source using the Common Astronomy Software Applications \citep[\textsc{casa},][]{2007ASPC..376..127M}. The error bars include the fitting error, the map rms and 10\% of the flux density (attributed to uncertainty in the absolute flux density scale) added in quadrature. In the table, we included one earlier measurement of the radio flux of \hvezda\ published by \citet{1998AJ....115.1693C}.

\begin{table}
	\begin{center}
		\caption{The flux densities $S_\mathrm{\nu}$ of the radio sources associated with \hvezda\ in NVSS and VLASS observational data. The heliocentric Julian dates are calculated as an average of the first and the last scans' times, as specified in the corresponding archives. $\vartheta_\mathrm{A}$ and $\vartheta_\mathrm{B}$ are quadratic rotational phases for components A and B, respectively. $\varphi_\mathrm{orb}$ is the orbital phase of the spectroscopic binary system.}\label{tab:radiofluxes}
		\begin{tabular}{l l c c c c}
			\hline
			\hline
			Survey & HJD & $\vartheta_\mathrm{A}$ & $\vartheta_\mathrm{B}$  & $\varphi_\mathrm{orb}$ & $S_\mathrm{\nu}$ (mJy)  \\
			\hline
			NVSS      &  2449304.9288  & 0.403 &   0.247  &  0.542  & $2.3\pm0.4$ \\
            VLASS1.1  &  2458085.8949  & 0.527  &  0.049  &  0.058  & $2.2\pm0.3$ \\
            VLASS2.1  &  2459106.1033  & 0.561  &  0.876  &  0.317  & $0.9\pm0.2$ \\
            VLASS3.1  &  2459974.7217  & 0.144  &  0.705  &  0.755  & $1.0\pm0.2$ \\
			\hline \hline
		\end{tabular}
	\end{center}
\end{table}

\begin{figure*}
	\centering\includegraphics[width=\textwidth]{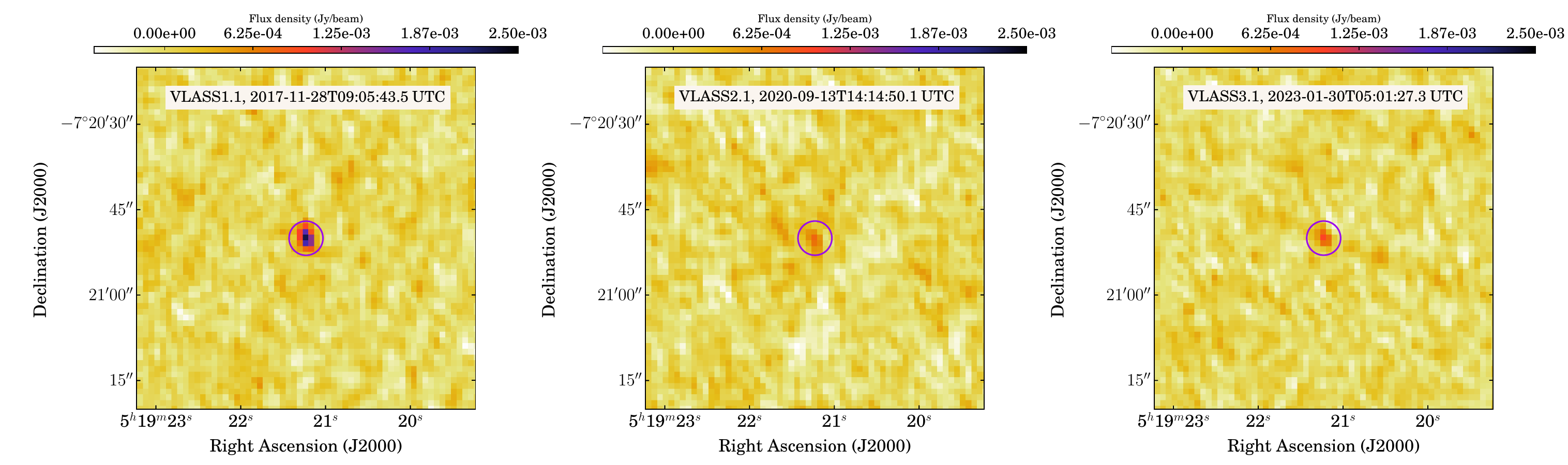}
	\caption{Quick look images of the sky area containing \hvezda\ extracted from the Very Large Array Sky Survey (VLASS) archive. A circle with radius $r=6''$ is centred at the location of the star.}
	\label{fig:vlass}
\end{figure*}

The integrated flux of the radio source VLA J051921.23$-$072049.6 found at the position of the studied star in the VLASS data varies between different observations but remains significant at least within 4$\sigma$ (Fig. \ref{fig:vlass}). The unprecedented pointing accuracy of the VLA leaves no doubt that the detected source is associated with \hvezda. After scaling to the distance to the star, the maximum radio luminosity $L_\mathrm{R}$ of the source is equal to $4.3\times10^{17}$\,erg\,s$^{-1}$\,Hz$^{-1}$.

\subsubsection{X-ray emission}
It was \citet{1992ApJS...81..795G} who first reported the detection of \hvezda\ in X-rays. The authors observed B-type stars with the Imaging Proportional Counter (IPC) of the {\it Einstein} Observatory in the range of energies 0.16--4\,keV, and concluded that X-ray emission with a luminosity $L_\mathrm{X}\geq10^{30}$\,erg\,s$^{-1}$ was common for B0--B3 stars, and became rare or non-existent towards B8--B9. For the studied star, a $\log L_\mathrm{X}$ expressed in erg\,s$^{-1}$ was estimated as 30.59.

In August 1990, the region of the sky with \hvezda\ was observed with the Position Sensitive Proportional Counters (PSPC) of the {\it ROSAT} spacecraft in the 0.1--2.4 keV range, seemingly with zero detection. However, an X-ray source 2SXPS J051921.1$-$072047 in the area of \hvezda\ was detected by the X-ray Telescope (XRT) onboard the Neil Gehrels {\it Swift} Observatory in 2014. During 2.8\,ks of exposure, in the range of 0.3--10\,keV, the mean registered count rate was 0.121\,cts\,s$^{-1}$~\citep{2020ApJS..247...54E} corresponding to a mean luminosity $\log L_\mathrm{X}=29.6$\,[erg\,s$^{-1}$].

Recently, \citet{2024A&A...682A..34M} released the first version of the all-sky X-ray survey eRASS in the western Galactic hemisphere, made with the extended ROentgen Survey with an Imaging Telescope Array ({\it eROSITA}, \citealt{2021A&A...647A...1P}) onboard the ``Spectrum-R\"{o}ntgen-Gamma'' ({\it SRG}) space observatory in the energy range 0.2--8\,keV. According to the data from {\it eROSITA}, an X-ray source 1eRASS J051921.0$-$072049 with coordinates coinciding with the position of \hvezda\ produced 0.268 counts per second in the range 0.2--2.3\,keV (soft X-ray) and was practically inactive beyond 2.3\,keV. The X-ray flux density of 1eRASS J051921.0$-$072049 was estimated as $2.495\times10^{-13}$\,erg\,s$^{-1}$\,cm$^{-2}$. Therefore, the logarithm of the X-ray luminosity ($\log L_\mathrm{X}$) measured in erg\,s$^{-1}$ equals 30.62, consistent with that published by \citet{1992ApJS...81..795G}. Such values imply that \hvezda\ is one of the strongest X-ray emitters among CP stars.

\begin{figure}
	\centering\includegraphics[width=0.9\columnwidth]{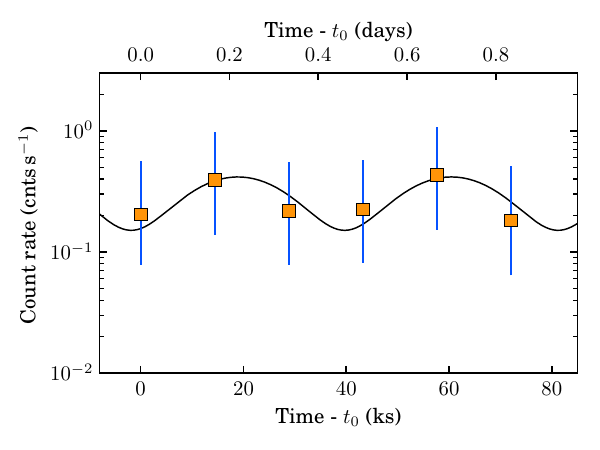}
	\caption{The X-ray light curve of \hvezda\ in the range of $\sim0.2$--10\,keV. The proposed period of variation $P_{\rm X}$ is $0\fd48$ (solid line).}
	\label{fig:xray_lc}
\end{figure}

To study the stability of the X-ray emission of \hvezda\ we have performed a custom analysis of data from the {\it eROSITA} archive with the use of the task \texttt{srctool} of the {\it eROSITA} Science Analysis Software System (eSASS, \citealt{2022A&A...661A...1B}). For X-ray photometry, we set the radius of the object aperture to 30\,arcsec; the background is evaluated within the annulus with radii 90 and 150 arcsec. The X-ray light curve in the range of energies $\sim0.2$--10\,keV is shown in Fig. \ref{fig:xray_lc}. The error bars are evaluated using the Bayesian excess variances implemented in \texttt{bexvar} \citep{2022A&A...661A..18B}.

The length of continuous observations of \hvezda\ with {\it eROSITA} is shorter than the period $P_{{\rm 1}A}=1\fd2799885$ of the hotter magnetic component but is still sufficient to search for variation of the X-ray emission on timescales comparable to the period $P_{{\rm 1}B}=0\fd5226938$ of the cooler secondary component. To our surprise, the best fit of the X-ray data suggests a period $P_\mathrm{X}\approx 0.48$ days, which is even shorter than $P_{{\rm 1}B}$, but the amplitude of this variation is inconclusively small.

\section{Discussion and conclusions} \label{sec:discuss}

Based on the results presented above, one can draw a rather comprehensive picture of the stellar system HD\,34736. In some elements such as the physical parameters of the magnetic component, this picture inherits the conclusions of \citetalias{2014AstBu..69..191S}. But it is only with new dedicated spectroscopic and photometric observations that it becomes possible to uncover the intricate nature of HD\,34736. In this section, we summarise the main outcomes of our research and attempt to place them in the context of modern knowledge about magnetic CP stars.

\subsection{The double-lined binary system of \hvezda}

From the variable radial velocities of lines belonging to two stars observed in its spectrum, HD\,34736 can be described as a stellar system comprising two early-type components. Given the effective temperature and the observed pattern of chemical anomalies, the more massive component can be classified as a CP star of the He-wk type. Notably, subtle spectral variability has probably been detected in Mg and Si lines of the cooler component.

In Sec. \ref{photom}, we have convincingly shown that the complex lightcurve of \hvezda\ obtained by {\it TESS} includes signals from three major contributors. Two of them are the hot visible components with individual lightcurves typical for CP stars.
The rotational period of the hotter primary star $P_{1A}$ is 1.2799885 days, while the secondary component rotates much faster with $P_{1B}=0.5226938$ days.
We have found that these values vary on the timescale of observations. Such a phenomenon is not rare in the world of magnetic CP stars \citep[e.g.][]{2011A&A...534L...5M, 2016CoSka..46...95M, 2019MNRAS.486.5558S, 2022A&A...668A.159M}, but \hvezda\ is to the best of our knowledge the only binary system with both components showing such behaviour. Even more intriguing is that while the main component is slowing down with the largest rate-of-change observed to date, the secondary star appears to be spinning up.

Magnetic fields are a remarkable feature of CP stars. Spectropolarimetry of \hvezda\ shows that the directly observable magnetic field of this star varies with a rotational period $P_\mathrm{1A}$ (Table \ref{table:parametry}) and thus is attributable to the primary component.
Even though the magnetic field of the secondary cannot be observed directly, in Sec. \ref{sec:mfield}, we have collected a number of facts indicating the possible presence of a field in this star. With the longitudinal field \bz\ estimated as 500\,G, in the case of simple dipolar configuration, the mean surface magnetic field can reach $\approx1.5$\,kG. Similar magnetic characteristics are demonstrated by CU Vir \citep{2014A&A...565A..83K}, a CP star that is also known for its fast and variable rotation. Much alike to CU\,Vir, \hvezda\ also attracts attention due to its X-ray and radio emission \citep{2000A&A...362..281T, 2018A&A...619A..33R,2021ApJ...921....9D}. A significant level of emission in the X-ray and radio domains found for the object of our study continues a series of similarities between these two stars. However, this phenomenon may be unrelated to these two stars.

To this moment, we have considered \hvezda\ to be a double-lined binary system. Two variants of the orbital solution presented in Table \ref{table:orbit} give projected masses $M\sin^3 i$ that appear to be much larger than those expected given the \teff\ of the components. In the solution based on the radial velocities derived through the modelling of Mg\textsc{ii} line 448.1\,nm, we have $M\sin^3 i=4.9\,M_\odot$ for component \textit{A}, and 3.5$\,M_\odot$ for component \textit{B}. However, given the average age of 4.6\,Myr of Orion OB1c hosting HD\,34736 \citep{2022MNRAS.515..998S}, from interpolating the MIST isochrones, one expects masses $M_{A}\approx 3.2\,M_\odot$ and $M_{B}\approx2.7\,M_\odot$ if we use \teffA\ and \teffB\ as a reference. These temperatures correspond approximately to spectral types of B7V and B8V\footnote{In the following classification, we use Eric Mamajek's table ``A Modern Mean Dwarf Stellar Color and Effective Temperature Sequence'' for spectroscopic classification. \url{http://www.pas.rochester.edu/~emamajek/EEM_dwarf_UBVIJHK_colors_Teff.txt}}. In contrast, even the lowest admissible masses from the orbital fit turn the primary into a B4V star and the secondary into a B7V star, respectively. On the assumption that the rotational axes of the components are orthogonal to the orbital plane and inclined by approximately $70^\circ$ to the line-of-sight (Sec. \ref{sec:mfield}), both components must be even more massive, implying spectral features of very hot stars which we undoubtedly would have recognised in our data. However, even a quick look at the recorded spectra reveals the features of the late B-type Si and He-wk peculiar stars, e.g. intense hydrogen Balmer and singly ionized silicon lines together with very weak to extinct lines of He\,\textsc{i} and Si\,\textsc{iii}.

Apart from the SED fitting (Sec. \ref{sec:phys_par}), which also implies that the effective temperatures of the components are close to the spectroscopically derived values of 13\,000\,K and 11\,500\,K, an independent test of the fidelity of our results can be achieved using a simple calculation of the apparent brightness of the binary given its distance $d\approx 372$\,pc found from the {\it GAIA} parallax and the adopted interstellar extinction $E(B-V)=0.0248$ mag. The MIST evolutionary models calculated for the age of 4.6\,Myr predict luminosity $\log L_{A}/L_{\odot}\approx 2.3$ and absolute magnitude $M_{\mathrm{V}A} \approx -0.17$ for the primary component with \teffA$\;=13\,000$\,K. After simple calculations, this gives an expected apparent magnitude $V_{A} \approx 7.8^{m}$. Similarly, for the secondary star with \teffB$\;=11\,500$\,K we get $\log L_{B}/L_{\odot}\approx 2.0$, $M_{\mathrm{V}B} \approx 0.24$, and $V_{B} \approx 8.2^{m}$. Thus, the apparent magnitude $V_{AB}$ of such an unresolved system is $7.4^{m}$, which is about $0.4^{m}$ brighter than reported, e.g., by \citet{2022A&A...657A...7K}. Similar calculations allow us to rule out a potential scenario in which the two stars do have larger masses but in which have already evolved to the end of their main sequence stage so that their effective temperatures are close to 13\,000\,K and 11\,500\,K. Ignoring the inherent problems with an explanation of how such an old ($t \gtrsim 100$ Myr) system can appear in the centre of a young association and show kinematic properties indistinguishable from the rest of the association members, this hypothesis about an advanced evolutionary status of \hvezda\ would require higher luminosities. However, any attempt to increase the effective temperature or luminosity of the components will result in even brighter predicted apparent magnitudes.

 At this point, we can conclude that the discrepancy between the stellar dynamical masses and spectral classification is real and can reasonably be explained in terms of higher-order multiplicity. The presence of a third, optically invisible component may also be necessary to explain the activity of \hvezda\ observed in radio and X-ray domains. The levels measured for our target are commonly found in 1) Cool active stars and 2) Some, but not numerous, hot magnetic CP stars. A strong magnetic field is necessary to explain this phenomenon in both cases. We consider both scenarios to be feasible for \hvezda.

\subsubsection{Invisible cool active component?}

The contradiction between the dynamical masses of the components and their spectral appearance found in Sec.~\ref{sec:multiple} could be explained by the presence of a third body in the system. Several scenarios for the architecture of such a hierarchal triple system can be considered. If the two optically-visible components form an SB2 binary with $P_{\rm orb}=83.2$~d and the third star orbits it on a wide orbit with a much longer period, we should observe a long-term trend in the systemic velocity $\gamma$, which is not evident in the \vr\ data. Moreover, in this case, the motion of the SB2 components on the inner orbit is not modified, leaving the dynamical mass problem unsolved. On the other hand, one of the two visible components can itself be a close binary, leading to a total mass exceeding that expected for its spectral type. In this case, we should observe additional modulation of \vr\ on a time scale shorter than $P_{\rm orb}$. This is excluded for the narrow-line primary star but is not out of the question for the secondary given its broad lines and apparent spectral variability, leading to a large scatter of \vr\ measurements and systematic difference between Mg~{\sc ii} and LSD \vr\ results. We explored the scenario where component {\it B} consists of two stars, {\it Ba} corresponding to the hot component visible in the spectrum and a lower mass component {\it Bb} (which we will call component {\it C}), producing no detectable optical spectral contribution. Assuming the orbital inclination is equal to the rotational inclination of the primary determined in Sec. ~\ref{sec:mfield}, $i_{\rm orb}=i=68^{\circ}$ and keeping all orbital parameters except $K_B$ fixed to the values in Table~\ref{table:orbit}, we found that the \vr\ semi-amplitude of the secondary must be reduced to $\approx$\,73~\kms\ to yield a dynamical mass of the primary equal to 3.2~$M_\odot$ as estimated from MIST isochrones. The same estimate suggests 2.7~$M_\odot$ for the secondary, while its dynamical mass with the modified $K_{B}$ is 3.07~$M_\odot$. The difference is attributed to an invisible 0.37~$M_\odot$ component {\it C} orbiting component {\it B}. The \vr\ variation associated with this orbital motion ranges from a few \kms\ to a few tens of \kms, depending on the assumed orbital period. This amplitude is substantially below the width of spectral lines of component {\it B}, essentially leading to additional scatter in the \vr\ measurements of the broad-line component. This scenario leaves the question of why this scatter leads to an apparent overestimation of the \vr\ amplitude of component {\it B} unanswered. It is possible that our measurement procedure is biased to phases with the largest velocity separation of the components {\it A} and {\it B}.

Given the young age of the system and given the detected radio and X-ray emission, we believe that the most probable candidate for the third component is a magnetically active and fast-rotating young stellar object (YSO) like a T\,Tau star.

Combined radio and X-ray emission is a common tracer of activity in a wide variety of objects. \citet{1993ApJ...405L..63G} and \citet{1994A&A...285..621B} find a universal relation, linking luminosity in two mentioned spectral domains: 
\begin{align}\label{eq:GB}
\frac{L_\mathrm{X}}{L_\mathrm{R}} = \kappa\times10^{15.5\pm1} [Hz],
\end{align}
where $\kappa=1$ for late-type active stars and various types of binaries with (sub)giants. For classical Algols, RS CVn and FK Com stars, the authors give $\kappa\approx0.17$. T\,Tau stars and YSOs are significantly overluminous in the radio, and a constant $\kappa$ for them is normally even smaller. For example, in a large-scale radio survey of the star-forming complexes in Ophiuchus, \citet{2013ApJ...775...63D} find that $\kappa=0.03$. Virtually the same value of $\kappa$ has been derived for complexes in Serpens \citep{2015ApJ...805....9O} and Taurus-Auriga \citep{2015ApJ...801...91D}. Young objects in Orion appear overluminous in radio by up to two orders of magnitude and even more according to \citet{2014ApJ...790...49K}.

\citet{2023MNRAS.525.4243S} show that in YSOs of about one solar mass approaching the ZAMS, intense X-ray emission must be the sign of a strong magnetic field of very simple dipolar configuration. Through the modelling of the X-ray and radio emission in flaring T\,Tau stars, \citet{2019MNRAS.483..917W} link the departure from the G\"{u}del-Benz relation to the strength of the surface magnetic field $B_\mathrm{s}$, which causes dramatic increase of the radio luminosity when $B_\mathrm{s}>3$\,kG. For \hvezda, we have $\log L_\mathrm{X}=30.17$, $\log L_\mathrm{R}=17.63$ and, consequently, $\kappa=0.001$. In models by \citet{2019MNRAS.483..917W}, this level corresponds to the activity of a star with $B_\mathrm{s}\approx5$\,kG. Thus, we conclude that the radio and X-ray emission from \hvezda\ can presumably be linked to a single source showing characteristics of young pre-MS objects with a rather strong surface magnetic field.

The X-ray luminosity of T\,Tau stars also depends on their mass. For the objects in the Orion Nebula Cluster, \citet{2005ApJS..160..401P} give a linear dependence between $\log L_\mathrm{X}$ and $\log M$ expressed in solar units as $\log L_\mathrm{X}=30.37 (\pm0.06) + 1.44 (\pm0.10)\log M$. We have used this relation to estimate the mass of the unknown source of X-ray emission and have eventually come to an inconclusive result. The variable X-ray luminosity of \hvezda\ (Sec. \ref{sec:beyondvisual}) brings us to a broad range of masses from about $0.3M_\odot$ when $\log L_\mathrm{X}=29.6$ to almost $1.6M_\odot$ when $\log L_\mathrm{X}=30.62$. Fast rotation, common for YSOs, can also not be neglected since we have a signature of variability in the X-ray data. Different aspects of the rotation problem and its relationship to activity are discussed for the X-ray-active YSOs in several open clusters by \citet{2016A&A...589A.113A} and \citet{2023ApJ...952...63G}.

Additional observations may help to clarify the invisible body's evolutionary status and infer its real physical parameters.

\subsubsection{Or magnetospheres?}
At the same time, one cannot completely rule out the possibility of magnetospheric activity in B-type components.

Magnetic early-type stars with strong surface magnetic field and rapid rotation are extremely likely to produce non-thermal radio emission \citep{leto2021,shultz2022}. It has been recently shown that such emission is driven by magnetic reconnections triggered by centrifugal breakout (CBO) events \citep{owocki2022}. CBOs are small-scale explosions in the magnetosphere during which magnetically confined stellar wind plasma breaks open the field lines temporarily and escapes the star. A necessary condition for CBOs to take place is that the Alfv\'en radius $R_\mathrm{A}$ should be larger than the Kepler radius $R_\mathrm{K}$ \citep[see][for definitions of $R_\mathrm{A}$ and $R_\mathrm{K}$]{ud-doula2008}. The region between the Kepler radius and the Alfv\'en radius is named the centrifugal magnetosphere \citep[CM,][]{petit2013}. For the stellar parameters of the magnetic primary (magnetic field strength is taken as 8.9 kG, Sec. \ref{sec:mfield}), we estimate the two parameters as $R_\mathrm{K} = 3.6\,R_{*}$ and $R_\mathrm{A} = 79\,R_{*}$, establishing that the star's magnetosphere should experience CBOs and can drive non-thermal radio emission. In case of the secondary star, if we assume it to have a surface magnetic field strength of $\approx$ 1.5 kG (Sec. \ref{sec:mfield}), we find $R_\mathrm{K}=1.8\,R_*$, and $R_\mathrm{A}\approx 37\,R_*$, suggesting that the secondary is also capable of producing radio emission if it is indeed magnetic.

The observed violation of the G\"{u}del-Benz relation is actually consistent with the known properties of magnetic hot stars, where the radio and X-ray emission are primarily produced by two distinct channels. The X-ray emission is produced due to the shock resulting from the collision between magnetically channelled stellar winds from the two magnetic hemispheres \citep[e.g.][]{ud-doula2015}, whereas the radio is driven by the CBOs. In addition, hot magnetic stars have been found to be overluminous in radio with respect to the G\"{u}del-Benz relation \citep{leto2017,leto2018,2018A&A...619A..33R}. For CU\,Vir, the ratio between X-ray and spectral radio luminosity was found to be $10^{12}$ Hz \citep{2018A&A...619A..33R}, similar to that observed for the case of HD\,34736.

Finally, the variable radio emission observed between different epochs of observation is also one of the characteristics of non-thermal radio emission observed from magnetic early-type stars. The incoherent radio emission exhibits a rotational modulation that correlates with that observed for the longitudinal magnetic field. In addition to the incoherent emission, some magnetic hot stars also produce coherent radio emission observed as periodic radio pulses \citep[e.g.][]{2000A&A...362..281T, das2022}, adding further variability to the lightcurve. Due to the sparse rotational phase coverage, it is unclear whether or not HD\,34736 also produces coherent radio emission. Future observations around the rotational phases of enhanced flux density will be able to provide conclusive evidence in this direction.

If the secondary star is confirmed to be magnetic, there could be another source of variability, both in radio and X-ray, related to binary magnetospheric interaction. So far, $\varepsilon$ Lupi is the only magnetic hot star binary system that has been investigated for such variability, and it was found to produce enhanced X-ray and radio emission at the periastron phase \citep{das2023,biswas2023}. In particular, the radio lightcurve, which has a better orbital phase coverage, revealed secondary enhancements at orbital phases away from the periastron that turned out to be persistent \citep{biswas2023}. The reason behind those enhancements is not well understood. 

Thus, the combination of binarity, magnetism, variable X-ray, and radio emission make HD\,34736 an important system for follow-up observation in both radio and X-ray wavebands in order to pinpoint the true origin of the emission and their significance for the stellar system itself.

\subsection{Concluding remarks}

The results obtained in our ten-year-long study of \hvezda\ and presented in this paper potentially put this star in a special place among known binary and multiple systems with magnetic components. Not only does the young age and strong magnetic field of the primary make \hvezda\ unique, but it is its unprecedented combination of components on different stages of stellar evolution. Here, we have two MS stars, which just entered the ZAMS or are approaching it, and a potential T\,Tau-like object. Apart from the main component, where the magnetic field is firmly detected using spectropolarimetry, we have indirect evidence of magnetic fields in two other companions. Only three binary systems comprising two components with firmly detected magnetic fields are known to date. Two pairs, namely HD\,156424~\citep{2021MNRAS.504.4850S} and BD +40$^\circ$ 175 \citep{1999AstL...25..809E, 2011AN....332..948S}, belong to wide systems with orbital periods order of years and decades. In this list, the doubly-magnetic $\varepsilon$ Lup \citep{2015MNRAS.454L...1S} is the only system with an orbital period shorter than a year. The formation and evolution of compact magnetic binaries can be used to validate hypotheses explaining the origin of stellar magnetism in the upper main sequence of the Hertzsprung-Russell diagram. Explaining the case of \hvezda\ with more than two magnetic components, it is reasonable to conclude that the magnetic properties of the protostellar environment and the mechanisms of evolution other than stellar mergers (as in the case of some known magnetic hot stars, e.g. the case of HD 148937, \citealt{2024Sci...384..214F}) are responsible for the appearance of at least some multiple magnetic systems.

 \section*{Acknowledgments}

The authors express their gratitude to the anonymous reviewer for their insightful suggestions and comments.
 
 D.S. acknowledges financial support from the project PID2021-126365NB-C21(MCI/AEI/FEDER, UE) and from the Severo Ochoa grant CEX2021-001131-S funded by MCIN/AEI/10.13039/501100011033.

E.A. acknowledges support by the ``Programme National de Physique Stellaire'' (PNPS) of CNRS/INSU co-funded by CEA and CNES.

I.Y. is grateful to the Russian Foundations for Basic Research for financial support (grant no. 19-32-60007).

O.K. acknowledges support from the Swedish Research Council (projects 2019-03548 and 2023-03667), the Swedish National Space Board (projects 185/14, 137/17), and the Royal Swedish Academy of Sciences.

Z.M. \& J.J. are grateful that publication could produced within the framework of institutional support for the development of the research organization of Masaryk University.

G.A.W acknowledges Discovery Grant support from the Natural Sciences and Engineering Research Council (NSERC) of Canada.

The research leading to these results has (partially) received funding from the KU\,Leuven Research Council (grant C16/18/005: PARADISE), from the Research Foundation Flanders (FWO) under grant agreementG089422N, as well as from the BELgian federal Science Policy Office (BELSPO) through PRODEX grant PLATO.

The research was partially supported by the grant 21-12-00147 (Russian Science Foundation).

This work is partially based on observations obtained at the Canada-France-Hawaii Telescope (CFHT), which is operated by the National Research Council of Canada, the Institut National des Sciences de l'Univers (INSU) of the Centre National de la Recherche Scientifique of France, and the University of Hawaii. Observations with the 6-m telescope BTA of the Special Astrophysical Observatory are supported by the Ministry of Science and Higher Education of the Russian Federation. This work has used the VALD, NASA ADS, and SIMBAD databases.

This paper includes data collected by the {\it TESS} mission, publicly available from the Mikulski Archive for Space Telescopes (MAST). Funding for the {\it TESS} mission is provided by the NASA's Science Mission Directorate. This project makes use of data from the KELT survey, including support from The Ohio State University, Vanderbilt University, and Lehigh University. 

This work is based on data from {\it eROSITA}, the soft X-ray instrument aboard SRG, a joint Russian-German science mission supported by the Russian Space Agency (Roskosmos), in the interests of the Russian Academy of Sciences represented by its Space Research Institute (IKI), and the Deutsches Zentrum für Luft- und Raumfahrt (DLR). The {\it SRG} spacecraft was built by Lavochkin Association (NPOL) and its subcontractors, and is operated by NPOL with support from the Max Planck Institute for Extraterrestrial Physics (MPE). The development and construction of the {\it eROSITA} X-ray instrument was led by MPE, with contributions from the Dr. Karl Remeis Observatory Bamberg \& ECAP (FAU Erlangen-Nuernberg), the University of Hamburg Observatory, the Leibniz Institute for Astrophysics Potsdam (AIP), and the Institute for Astronomy and Astrophysics of the University of Tübingen, with the support of DLR and the Max Planck Society. The Argelander Institute for Astronomy of the University of Bonn and the Ludwig Maximilians Universität Munich also participated in the science preparation for {\it eROSITA}. The {\it eROSITA} data shown here were processed using the eSASS software system developed by the German {\it eROSITA} consortium.

\section*{Data Availability}
The authors can provide the extracted 1D spectra and light curves used in this study upon a reasonable request.

\bibliographystyle{mnras}
\bibliography{hd34736}
\label{lastpage}

\newpage
\appendix

\section{Phenomenological model of a rotationally modulated variable} \label{sec:Phenomenological_models}

\subsection{Models of phase function }\label{sec:PF_model}

The following semi-phenomenological analysis aims to model as accurately as possible the observed photometric variations of the \hvezda\ object in the KELT and {\it TESS} filters and to derive the rotation periods of the outer co-rotating layers of both components of the binary star so that it is possible to describe and discuss both the distribution and parameters of the photometric and spectroscopic spots, as well as the geometry of magnetic fields (if any). From long-term observations of well-monitored mCP stars, it follows that the phase curves of photometric, spectroscopic, and spectropolarimetric measurements are unchanged in the time scale of decades or centuries; it is advantageous to introduce and use the concept of a monotonically raising \textit{phase function} $\vartheta(t)$, which is the sum of an epoch $E(t)$ and a common phase $\varphi(t)$, $\vartheta(t)=E(t)+\varphi(t)$ in further studies. The phase function $\vartheta(t)$ and its inversion time-like function $\T(\vartheta)$ are related to an instantaneous period $P(t)$ (or $P(\vartheta)$) through simple differential equations with a boundary condition \citep[for details see in][]{2008A&A...485..585M,2016CoSka..46...95M},:
\begin{align}
&\frac{\rm d\vartheta}{{\mathrm d}t}=\frac{1}{P(t)} ;\quad \vartheta(t=M_0)=0;\quad
\Rightarrow\quad \vartheta(t)= \int_{M_0}^t \frac{\rm d \tau}{P(\tau)};\label{eq:fazovka}\\
&\frac{\mathrm{d} \T(\vartheta)}{\rm d\vartheta}=P(\vartheta);\quad \T(0)=M_0;\quad
\T(\vartheta)= M_0+\int_0^\vartheta P(\zeta)\,\rm{d}\zeta,\label{eq:inverze}
\end{align}
where $\tau$ and $\zeta$ are auxiliary variables. Using $\T(\vartheta)$, we can predict the moment of the zero-th phases $t(\varphi=0,\,E)=\T(E)$ for an chosen epoch $E$. The common phase $\varphi(\vartheta)$ and the epoch $E(\vartheta)$ for a chosen phase function $\vartheta(t)$ are given by the relations: $\varphi(t)=\mathrm{FP}(\vartheta(t))$ and $E=\mathrm{IP}(\vartheta)$, where FP and IP are the operators for the fractional part and the integer part of a number.

It is useful to introduce an auxiliary variable $\vartheta_0(t)$ instead of time
\begin{equation}
    \vartheta_0(t)=\frac{t-M_0}{P_0}, \label{eq:vartheta0_def}
\end{equation}
where $P_0$ is the instantaneous period at the properly chosen origin of epoch counting at the $t=M_0$. Then
\begin{align}
  &  \frac{\mathrm{d}\vartheta(\vartheta_0)}{\mathrm{d}\vartheta_0}=\frac{P_0}{P(\vartheta_0)}; \quad \frac{\mathrm{d}\vartheta_0(\vartheta)}{\mathrm{d}\vartheta}=\frac{P(\vartheta)}{P_0};\label{eq:basictheta}\\
  &\vartheta(\vartheta_0)= \int_{0}^{\vartheta_0}\frac{P_0}{P(\vartheta_0')}\,\mathrm{d}\vartheta_0';\quad \vartheta_0(\vartheta)=\int_{0}^{\vartheta}\frac{P(\vartheta')}{P_0}\,\rm{d}\vartheta'.\label{eq:simpleteta}
\end{align}
\subsection {Linear Maclaurin and orthogonal phase function model}
\label{sec:mclphf}

Period analysis of mCP stars show that the rotational periods of the majority of them are constant \citep{2016CoSka..46...95M} which means that the solutions of the eqs. (\ref{eq:fazovka}) and (\ref{eq:inverze}) for the phase function $\vartheta(t)$ and its inversion $\T(\vartheta)$ are linear and the same as the auxiliary variable $\vartheta_0$ introduced above (\ref{eq:vartheta0_def})
\begin{equation}
P(t)=P_1; \quad \vartheta(t)=\vartheta_0(t)=\frac{t-M_{0}}{P_1};\quad  t(\vartheta)=M_{0}+P_1\,\vartheta,
\end{equation}
with only two parameters of the linear ephemeris: the mean period $P_1$ is the BJD time of one, selected primary maximum of the observed light curve. We can this form of the ephemeris transform into the orthogonal form as:
\begin{align}
&\T(\vartheta)=\zav{M_0+\eta_{1}\,P_1}+P_1(\vartheta-\eta_{1})=M_1+P_1\,(\vartheta-\eta_{1});\label{eq:linaprox} \\
&\eta_{1}=\textrm{round}\zav{\frac{\sum \vartheta_i w_i}{\sum w_i}};\quad M_1=M_0+\eta_1 P_1,\quad \vartheta_1(t)=\frac{t-M_1}{P_1}, \nonumber
\end{align}
where $w_i$ are weights of individual measurements, $\theta_1(t)$ is an orthogonal form of linear phase function. Knowing the uncertainties of orthogonal parameters of linear approximation $\delta M_1, \delta P1$ we can easily estimate uncertainties of quantities $\delta \T(\vartheta)$ and $,\delta \vartheta(t)$.
\begin{equation}
 \delta \T(\vartheta)=\sqrt{(\delta M_1)^2+\hzav{\delta P_1\,(\vartheta-\eta_1)}^2},\quad \delta(\vartheta)=\delta \T(\vartheta)/P_1.
\end{equation}
\subsection {Quadratic Maclaurin and orthogonal phase function model} \label{sec:OPF_models}
Let us now assume that the instantaneous period $P(t)$ at a moment $t$ varies in a linear way such that
\begin{equation}
    P(t)=P_0+\dot{P}\,(t-M_0)=P_0\zav{1+\dot{P}\,\vartheta_0};\ \mathrm{where}\ \ \vartheta_0=\frac{t-M_0}{P_0}.
\end{equation}
Then using equations in (\ref{eq:simpleteta}) we obtain the phase function:
\begin{equation}
     \vartheta(\vartheta_0)=
         \int_{0}^{\vartheta_0}\frac{ \mathrm{d}\vartheta_0' }{ 1+\dot{P}\vartheta_0' }
		=\frac{ \ln(1+\dot{P}\vartheta_0) }{ \dot{P} }
	\simeq \vartheta_0-\textstyle{\frac{1}{2}}\,
		\dot{P}\,\vartheta_0^2,  \label{eq:Maclaurin2}
\end{equation}
which we truncate to the two first term as $\dot{P}$ is generally very small in our context. Now we can isolate the time phase function $\vartheta_0$ by the equation \ref{eq:Maclaurin2} and expending the exponential in a series. Using Eq.\,\ref{eq:basictheta} we can calculate the instant period $P(\vartheta)$ as a function of the phase function $\vartheta$:
\begin{align}
& \vartheta_0(\vartheta)=\frac{e^{\dot{P}\vartheta}-1}{\dot{P}} \simeq \displaystyle\vartheta +  \frac{\dot{P}}{2}\,\vartheta^2, \label{eq:exp}\\
& P(\vartheta)=P_0\,\frac{\mathrm d\vartheta_0}{\mathrm d\vartheta}=P_0\,e^{\dot{P}\vartheta}\simeq P_0\zav{1+\dot{P}\,\vartheta},\\
&\T(\vartheta)=M_0+P_0\,\vartheta_0 \simeq M_{0}+P_0\,\vartheta+\frac{P_0\dot{P}\,\vartheta^2}{2}.  \label{eq:P2}
 \end{align}

A disadvantage of the Maclaurin ephemeris model described by relation is the correlation between ephemeris parameters that hinders the error analysis. Mathematically, the model represented by the simplified equation \ref{eq:P2} is a simple quadratic polynomial with respect $\vartheta$. This allows us to construct an orthogonal version of the phase function model \citep[for details see in][]{2008A&A...485..585M,2016CoSka..46...95M}, allowing for a more robust error estimation, using the standard Gram-Schmidt procedure, as follow:
\begin{align}
&\displaystyle \T(\vartheta)\simeq M_1+P_1(\vartheta-\eta_1)+\frac{P'P_1}{2}(\vartheta^2-\eta_{21}\, \vartheta-\eta_{20})=  \nonumber \\
&\quad \quad = M_1+P_1(\vartheta-\eta_1)+\frac{P'P_1}{2}(\vartheta-\eta_2)(\vartheta-\eta_3),\label{eq:ortefem}
\end{align}
where $M_1$, $P_1$, and $P'$ are parameters of the orthogonal square ephemeris. The orthogonalization coefficients $\eta_1$, $\eta_2$, $\eta_3$, $\eta_{20}$, and $\eta_{21}$ were opted so they fulfill the following orthogonalization  constraints:
\begin{align}
& \overline{(\vartheta-\eta_1)}=\overline{\vartheta^2-\eta_{21}\vartheta-\eta_{20}}=\overline{\vartheta\,(\vartheta-\eta_2)(\vartheta-\eta_3)(\vartheta-\eta_1)}=0; \nonumber\\
& \overline{\vartheta\,(\vartheta^2-\eta_{21}\vartheta-\eta_{20})}=\overline{\vartheta\,(\vartheta-\eta_2)(\vartheta-\eta_3)}=0;\\
& \overline{\vartheta^q}=\frac{\sum \vartheta_i^q\,w_i}{\sum\,w_i},\quad\eta_{21}=\frac{\overline{\vartheta^3}-\overline{\vartheta^2}\,\overline{\vartheta}}{\overline{\vartheta^2}-\overline{\vartheta}^2}; \quad \eta_{20}=\frac{\overline{\vartheta^2}^2-\overline{\vartheta^2}\,\overline{\vartheta}^2}{\overline{\vartheta^2}-\overline{\vartheta}^2};\nonumber\\
&\displaystyle\eta_1=\mathrm{round}(\overline{\vartheta});\quad \eta_{3,2}=\frac{\eta_{21}}{2}\pm\sqrt{\zav{\frac{\eta_{21}}{2}}^2+\eta_{20}},\nonumber\\
&\eta_{20}=-\eta_2\,\eta_3;\quad \eta_{21}=\eta_2+\eta_3. \label{eq:gamy_def}
\end{align}

Using this ephemeris form (\ref{eq:ortefem}) in our model fitting, we obtain the values of the parameters $M_1$, $P_1$, and $P'$, including their uncorrelated uncertainties.

If we set $\vartheta=E$, where $E$ is an integer epoch, in the equation (\ref{eq:ortefem}), we obtain the moment of the zero phase $\varphi$ for this epoch.
\begin{align}
&\displaystyle \T(E)\simeq M_1+P_1(E-\eta_1)+\frac{P'P_1}{2}(E^2-\eta_{21}\, E-\eta_{20})=  \nonumber \\
&\quad \quad = M_1+P_1(E-\eta_1)+\frac{P'P_1}{2}(E-\eta_2)(E-\eta_3).\label{eq:ortEfem}
\end{align}
The instantaneous periods for the phase function $\vartheta$, or the epoch $E$, $P(\vartheta),\ P(E)$, equal to:
\begin{equation}
    P(\vartheta)=\frac{\mathrm{d}t}{\mathrm{d}\vartheta}=P_1+P'P_1\zav{\vartheta-\frac{\eta_{21}}{2}};\quad
 P(E)=P(\vartheta=E).\label{eq:ortP}
\end{equation}
Introducing time-like quantity $\vartheta_1(t)$ we can compute the phase function $\vartheta(t)$ and the instantaneous period $P(t)$ for any moment $t$
\begin{align}
& \displaystyle\vartheta_1=\frac{t-M_1}{P_1}+\eta_1;\quad\vartheta(\vartheta_1)=\vartheta_1-\frac{P'}{2}(\vartheta_1^2-\eta_{21}\, \vartheta_1-\eta_{20}),\label{eq:orttheta}\\
&P(t)=P_1\frac{\mathrm d \vartheta_1}{\mathrm d \vartheta}=\frac{P_1}{1\!-\!P'(\vartheta_1\!-\!\frac{\eta_{21}}{2})}\simeq P_1\hzav{1+P'\zav{\vartheta_1\!-\!\frac{\eta_{21}}{2}}};\label{eq:ortPt}\\
&\dot{P}(t)=\frac{1}{P_1}\frac{\mathrm{d}P}{\mathrm{d}\vartheta_1}=P'=\dot{P}.\label{eq:dotka}
\end{align}
If we know uncertainties of parameters $M_1,\,P_1$, and $P'$ ($\delta M_1,\,\delta P_1$, and $\delta P'$) we can simply compute the uncertainty of prediction of the time of the phase function $\vartheta$, $t(\vartheta)$ (eq \ref{eq:ortefem}), the predicted phase function $\delta \vartheta(t)$ (eq \ref{eq:orttheta}) and the uncertainty of the instantaneous period estimate $\delta P(t)$ (eq \ref{eq:ortPt})
\begin{align}
    & \delta \T(\vartheta)=\sqrt{(\delta M_1)^2+[\delta P_1(\vartheta\!-\!\eta_1)]^2+\hzav{\frac{P_1 \delta P'(\vartheta\!-\!\eta_2)(\vartheta\!-\!\eta_3)}{2}}^2}, \nonumber\\
    &\delta P(t)=\sqrt{(\delta P_1)^2+\hzav{P_1\delta P'\zav{\vartheta-\frac{\eta_{21}}{2}}}^2};\quad \delta \vartheta(t)\simeq \frac{\delta \T(\vartheta)}{P_1}.\label{eq:delta_T}
\end{align}
Using the following relations we can easily return to Maclaurin ephemeris:
\begin{equation}
    M_0=M_1\!-\!\eta_1P_1\!-\!\frac{P' P_1\eta_{20}}{2},\ \ P_0=P_1\!-\!\frac{P_1P'\eta_{21}}{2},\ \ \dot{P}=P'.
\end{equation}

If we put the origin of epochs near to $M_1$, so that $\eta_1=0$, a lot of relations become simpler:
\begin{align}
&\displaystyle \T(\vartheta)\simeq M_1+P_1(\vartheta)+\frac{P'P_1}{2}(\vartheta^2-\eta_{21}\, \vartheta-\eta_{20})=  \nonumber \\
&\quad \quad = M_1+P_1(\vartheta)+\frac{P'P_1}{2}(\vartheta-\eta_2)(\vartheta-\eta_3),\quad\mathrm{where} \label{eq:ortefems}\\
& \eta_{21}=\frac{\overline{\vartheta^3}}{\overline{\vartheta^2}}; \quad \eta_{20}=\overline{\vartheta^2};\quad \eta_{3,2}=\frac{\overline{\vartheta^3}\pm\sqrt{\,\overline{\vartheta^3}^2+4\,\overline{\vartheta^2}^3}}{2\,\overline{\vartheta^2}},\\
&P(\vartheta)=\frac{\mathrm{d}t}{\mathrm{d}\vartheta}=P_1\hzav{1+P'\zav{\vartheta-\frac{\eta_{21}}{2}}};\\
&M_0=M_1-\frac{P'P_1\eta_{20}}{2},\quad P_0=P_1\!-\!\frac{P_1P'\eta_{21}}{2},\\
& \displaystyle\vartheta_1=\frac{t-M_1}{P_1};\quad\vartheta(\vartheta_1)=\vartheta_1-\frac{P'}{2}(\vartheta_1^2-\eta_{21}\, \vartheta_1\!-\!\eta_{20}),\label{eq:ortthetas}\\
&P(t)=P_1\!\frac{\mathrm d \vartheta_1}{\mathrm d \vartheta}\simeq P_1\hzav{1+P'\zav{\vartheta_1-\frac{\eta_{21}}{2}}};\ \ \dot{P}(t)= P'.\label{eq:dotkas}
\end{align}
\begin{table}
	\begin{center}
	\caption{Coefficients $\beta_{ij}$ of the asymmetrical part of the light curve model harmonic polynomial till $m=11$-th order.}
	\label{table:betty} \scriptsize
	\begin{tabular}{c| c c c c c c c}
\hline
 \hline
 $i$ &  $\beta_{i1}$&$\beta_{i2}$&$\beta_{i3}$&$\beta_{i4}$ &$\beta_{i5}$  &$\beta_{i6}$ & $\beta_{i7}$\\
 \hline
2  &  0.8944 &  -0.4472 &        0 &        0 &        0 &        0 &        0  \\
3  &  0.3586 &   0.7171 &  -0.5976 &        0 &        0 &        0 &        0  \\
4  &  0.1952 &   0.3904 &   0.5855 &  -0.6831 &        0 &        0 &        0  \\
5  &  0.1231 &   0.2462 &   0.3693 &   0.4924 &  -0.7385 &        0 &        0  \\
6  &  0.0848 &   0.1696 &   0.2544 &   0.3392 &   0.4241 &  -0.7774 &        0  \\
7  &  0.0620 &   0.1240 &   0.1861 &   0.2481 &   0.3101 &   0.3721 &  -0.8062  \\
8  &  0.0473 &   0.0947 &   0.1420 &   0.1894 &   0.2367 &   0.2840 &   0.3314  \\
9  &  0.0373 &   0.0747 &   0.1120 &   0.1493 &   0.1866 &   0.2240 &   0.2613  \\
10 &  0.0302 &   0.0604 &   0.0906 &   0.1208 &   0.1509 &   0.1811 &   0.2113  \\
11 &  0.0249 &   0.0498 &   0.0748 &   0.0997 &   0.1246 &   0.1495 &   0.1745  \\
\hline
 $i$ & $\beta_{i8}$& $\beta_{i9}$& $\beta_{i10}$& $\beta_{i11}$&&&\\
\hline
$2\div 7$ & 0 & 0 & 0 & 0 & & & \\
8  & -0.8284 &       0  &       0  &       0 & & & \\
9  &  0.2986 &  -0.8460 &       0  &       0 & & &\\
10 &  0.2415 &   0.2717 &  -0.8604 &       0 & & & \\
11 &  0.1994 &   0.2243 &   0.2492 &  -0.8723& & & \\
 \hline
 \hline
\end{tabular}
\end{center}
\end{table}
\subsection{Modelling light curves of chemically peculiar stars} \label{sec:LC_models}

The observed light curves of chemically peculiar stars can be easily described as strictly periodic harmonic polynomials of the order $m=2 \div 18$ with a period of 0\fd5 to several hundred days, corresponding to the rotation periods of studied CP stars. The underlying light curves sometimes needed to be expressed by a harmonic polynomial of about tenth order, typical of mCP stars with the complicated appearance of surface photometric spots and semi-transparent structures trapped in co-rotating stellar magnetospheres.

\subsubsection{Monochromatic light curves} \label{sec:MonoLC_models}

For an explicit description of a monochromatic light curve with the effective wavelength $\lambda$, it is advantageous to use special harmonic polynomials (SHP). SHP of the $m$-th order, $\boldsymbol{\Xi}(\vartheta,m)=[\mathit{\Xi}_1,\mathit{\Xi}_2,\dots,\mathit{\Xi}_{2m-1}]$, is a row vector with the length 2\,$m$-1, which represent a base of mutually orthonormal harmonic functions with zero time derivative at the phase $\varphi=0$, while $\mathbf{b(\lambda})=[b_1,b_2,\ldots,b_{2m-1}]'$ parameters. $m$ of them are simple symmetric functions with an extreme at phase 0, and $(m-1)$ are antisymmetric functions with zero derivatives at phase 0. Such polynomials have one of their extremes in the phase $\varphi=0$:
\begin{align}
    &\mathit{\Xi}_1(\vartheta,m)=\cos(2\,\pi\,\vartheta);\quad\mathit{\Xi}_{2i-2}(\vartheta,m)=\cos(2\,\pi\,i\,\vartheta);\nonumber\\
    &\mathit{\Xi}_{2i-1}(\vartheta,m)=\sum_{j=1}^i\beta_{ij}\sin(2\,\pi\,j\,\vartheta);\quad i=2, 3,\ldots,m;
\end{align}
when the coefficients $\beta_{ij}$ (given in Table \ref{table:betty}) fulfils the following constraints
\begin{equation}
    \sum_{j=1}^i\,j\,\beta_{ij}(\lambda)=0;\quad \sum_{j=1}^i\,\beta_{ij}\,\beta_{kj}= \delta_{ik};\quad i\geq k. \label{constr2}
\end{equation}
Parameters $\beta_{ij}$ that fulfill the orthonomalization constraints (Eq: \ref{constr2}) are in Table\,\ref{table:betty}. The model of the monochromatic light curve $F(\vartheta,m)$ then may be expressed in the form
\begin{equation}
    F(\vartheta,m)=m_k+\boldsymbol{\Xi}(\vartheta,m)\cdot\mathbf{b}(\lambda)=m_{k}+\sum_{i=1}^{2m-1}\,b_i(\lambda)\,\mathit{\Xi_i}(\vartheta,m),
\end{equation}
where $m_k$ are the mean magnitudes of observational subsets. In the case of \hvezda\ we have divided observations into eight segments -- see Table \ref{table:OCAB}.

A robust measure of monochromatic variability of a periodic light curve is the so-called \textit{effective amplitude} $A_{\mathrm{eff}}(\lambda)$, which can be easily expressed thanks to the orthonormality of the basis of special harmonic polynomials:
\begin{equation}
    A_{\mathrm{eff}}(\lambda)=2\,\mathrm{norm}(\mathbf{b}(\lambda))=2\,\sqrt{\,\sum_{i=1}^{2m-1}b_i^2(\lambda)}. \label{eq:effampl}
\end{equation}

\section{Light-travel time delay} \label{sec:LiTE}

The orbital motion of the stars in the binary affects the photometric behavior of the system. The light-travel time (Roemer) delay applies in the variability of the individual components; with a suitable inclination of the orbit, mutual eclipses of binary members can also occur. Since we know the parameters of the spectroscopic path with extraordinary precision, we can (see Table\,\ref{table:orbit}) reliably calculate and predict the mentioned effects.

Following the table, it will assume that the orbital period (the time between two consecutive passages through the same anomaly) is
$P_{\mathrm{orb}}=83\fd219(3)$; the numerical eccentricity is $e=0.810\,3(3)$, the fundamental moment of the periastron passage is $T_{\mathrm{p}}=2\,457\,415.346(3)$; the projections of the semiaxes of the individual components in light days are: $A_\mathrm{A}=a_{\mathrm{A}}\sin\,i =0.001\,83$\,ld; $A_{\mathrm{B}}=a_{\mathrm{B}}\sin i=0.002\,56$\,ld ($i$ being the unknown inclination angle of the orbit), and the argument of the periastron (the angle from the orbital ascending node to its periastron, measured in the direction of motion), in radians: $\omega=1.470(2)$\,rad. Using these parameters we can compute for any time $t$ the following quantities: a true anomaly $\theta$, $E$ an eccentric anomaly, and $M$ a mean anomaly.

\begin{align}
&E(M)=M+e\sin E; \quad \mathrm{where}\ M(t)=2\,\pi\,\frac{t-T_{\mathrm p}}{P_{\mathrm{orb}}};\quad \\
&\theta(E)=2\arctan\hzav{\sqrt{\frac{1+e}{1-e}}\,\tan\zav{\frac{E}{2}}}+2\,\pi\ \mathrm{round}\zav{\frac{E}{2\,\pi}}; \label{eq:teta}
\\&E(\theta)=2\arctan\hzav{\sqrt{\frac{1-e}{1+e}}\,\tan\zav{\frac{\theta}{2}}}+2\,\pi\ \mathrm{round}\zav{\frac{\theta}{2\,\pi}}. \label{eq:E}
\end{align}

\begin{figure}
	\centering\includegraphics[width=\hsize]{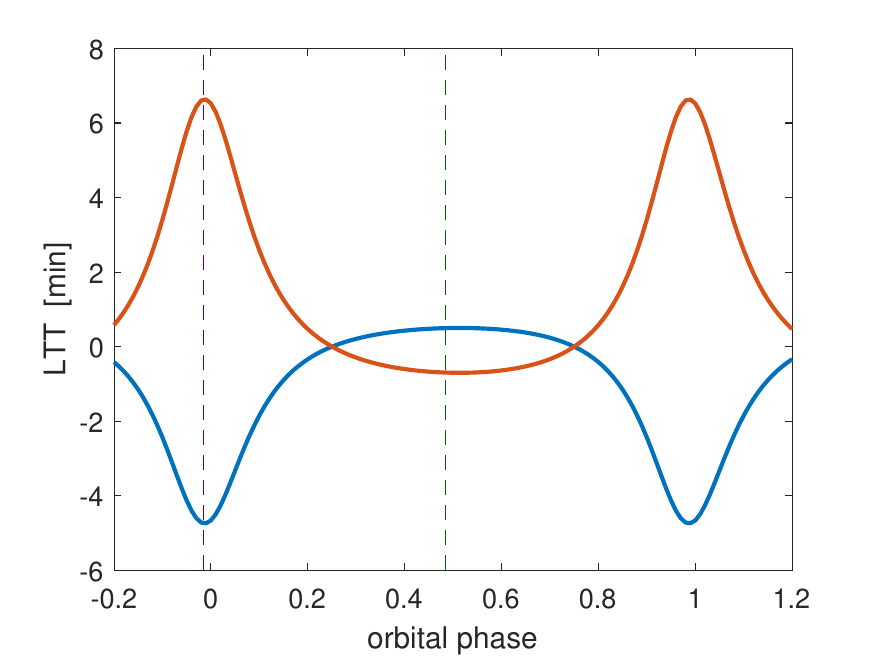}
	\caption{The dependence of light-travel time delay on the orbital phase as introduced by (\ref{eq:orbfi}) for $A$ (blue line) and $B$ (red line) components. The green dashed line signs the phase of the periastron passage, while the magenta line signs the apastron passage.}
	\label{fig:LTT}
\end{figure}

Suppose we want to clean the timing of the events on the individual components of the double stars from their orbital motion. In that case, we can do it by offsetting the time corrections of the light-travel time delay \citep{2016MNRAS.455.4136B}\ $\Delta_\mathrm{A}(t)$ and $\Delta_\mathrm{B}(t)$, where $t_\mathrm{A}$ and $t_\mathrm{B}$ are the times related to the gravity center of the system.
\begin{align}
&\Delta_\mathrm{A}(t)=A_\mathrm{A}\,\frac{(1-e^2)\,\sin\hzav{\theta(t)+\omega}}{1+e\cos \theta};\quad t_\mathrm{A}=t-\Delta_\mathrm{A}(t);\label{Delty}\\
&\Delta_\mathrm{B}(t)=-A_\mathrm{B}\,\frac{(1-e^2)\,\sin\hzav{\theta(t)+\omega}}{1+e\cos \theta};\quad t_\mathrm{B}=t-\Delta_\mathrm{B}(t).\nonumber
\end{align}
The above relations for specific values of its eccentricity and argument of periastron show the binary's components spend most of their time near the apastron, with component A being on average seven light minutes closer to us than the less massive and smaller component B (see Fig. \ref{fig:LTT}).

If, on the other hand, we want to know the prediction of the time of some significant moment from the observer's point of view (e.g. time of maximum $\mathit{\Theta(E)}$) for the epoch $E$, the moment of the prediction relative to the center of gravity of the system must be corrected by the corresponding LTT delay:
\begin{equation}
t_{\mathrm{max}A}\! = \! \mathit{\Theta(E_A)}\!+\!\Delta_A(\mathit{\Theta(E_A)}),
\quad t_{\mathrm{max}B}\! = \! \mathit{\Theta(E_B)}\!+\!\Delta_B(\mathit{\Theta(E_B)}).\label{back}
\end{equation}

\section{Elimination of UCAC4 414-008437 light variation} \label{UCAC4}

\begin{figure}
\centering \includegraphics[width=85mm]{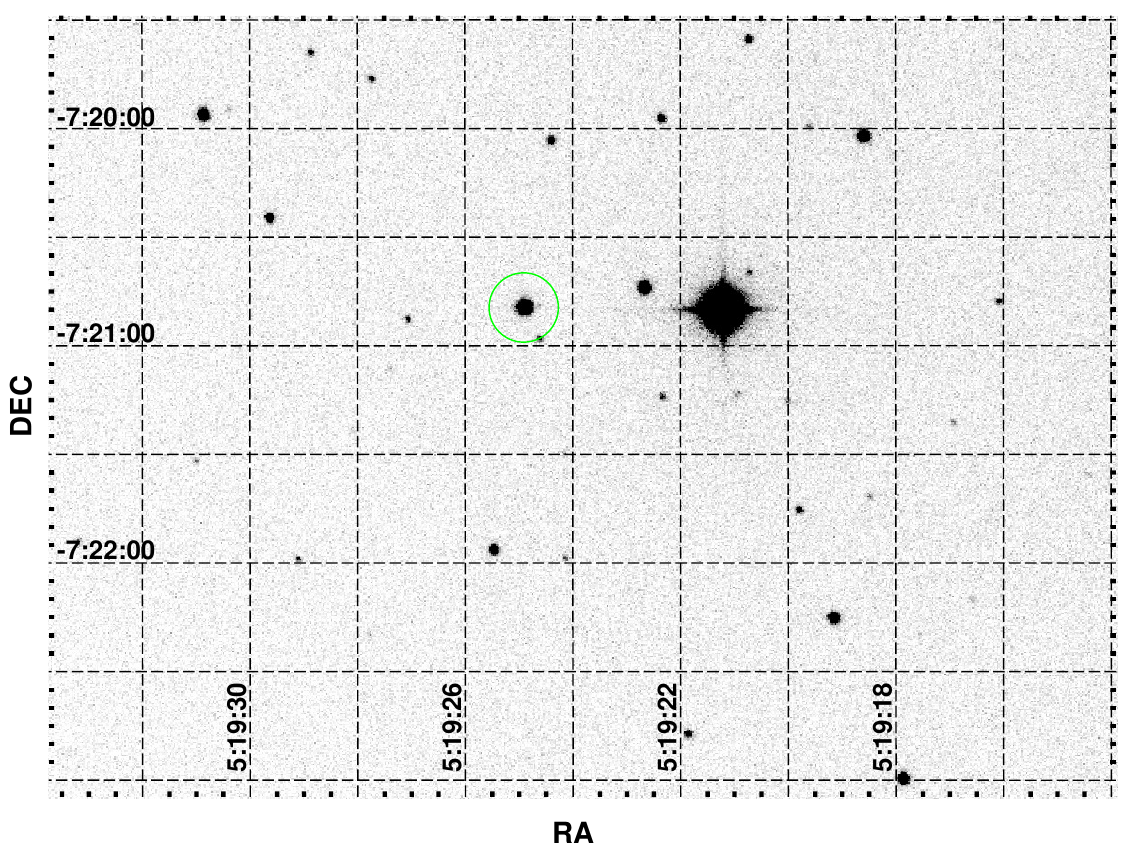}
\caption{Field of view of the DK154 telescope with the brightest star \hvezda\ and the star UCAC4 414-008437 (green circle) located at an apparent distance $54.83''$ corresponding to about 2.5 pixels of the detector of the {\it TESS} satellite.}
\label{fig:DK154}
\end{figure}

\begin{figure}
\centering\includegraphics[width=\columnwidth]{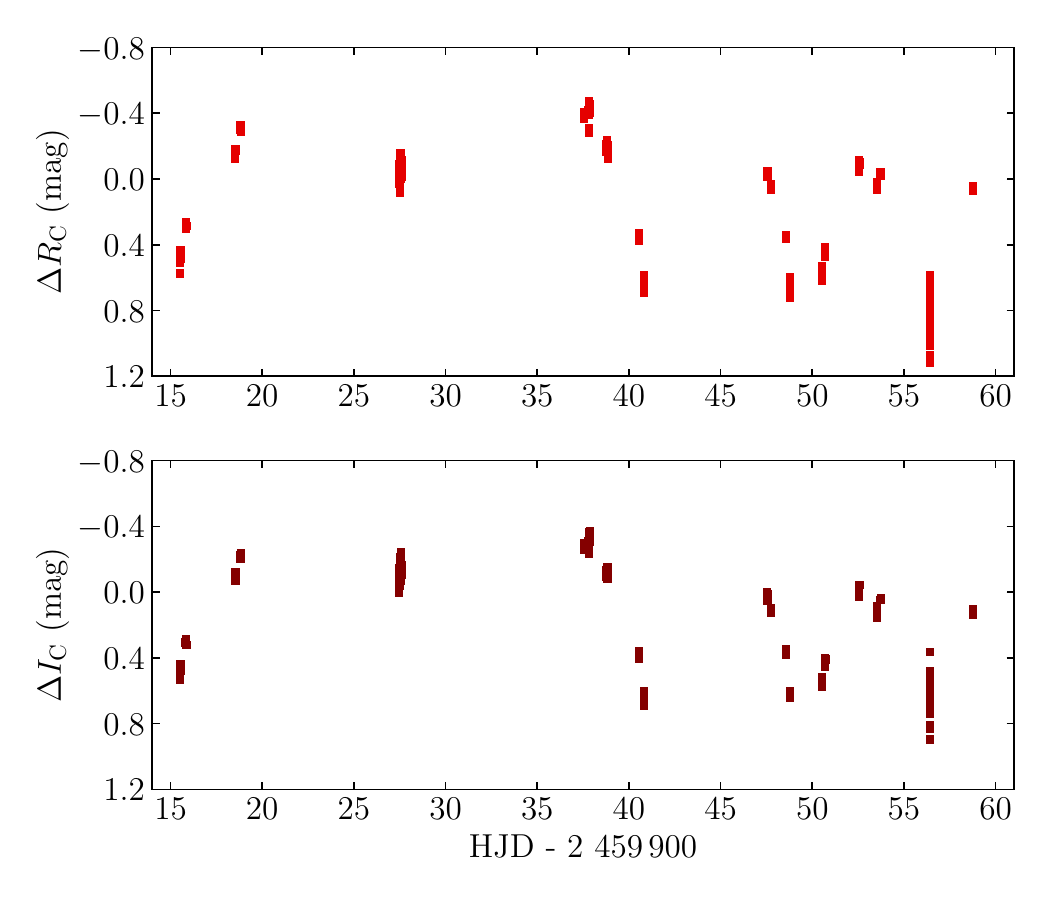}
\caption{Light variation of UCAC4 414-008437 in filters $R_\mathrm{C}$ and $I_\mathrm{C}$ registered with the DK154 telescope at La Silla Observatory.}
\label{fig:3compCDK154}
\end{figure}

\begin{figure}
\centering\includegraphics[width=\columnwidth]{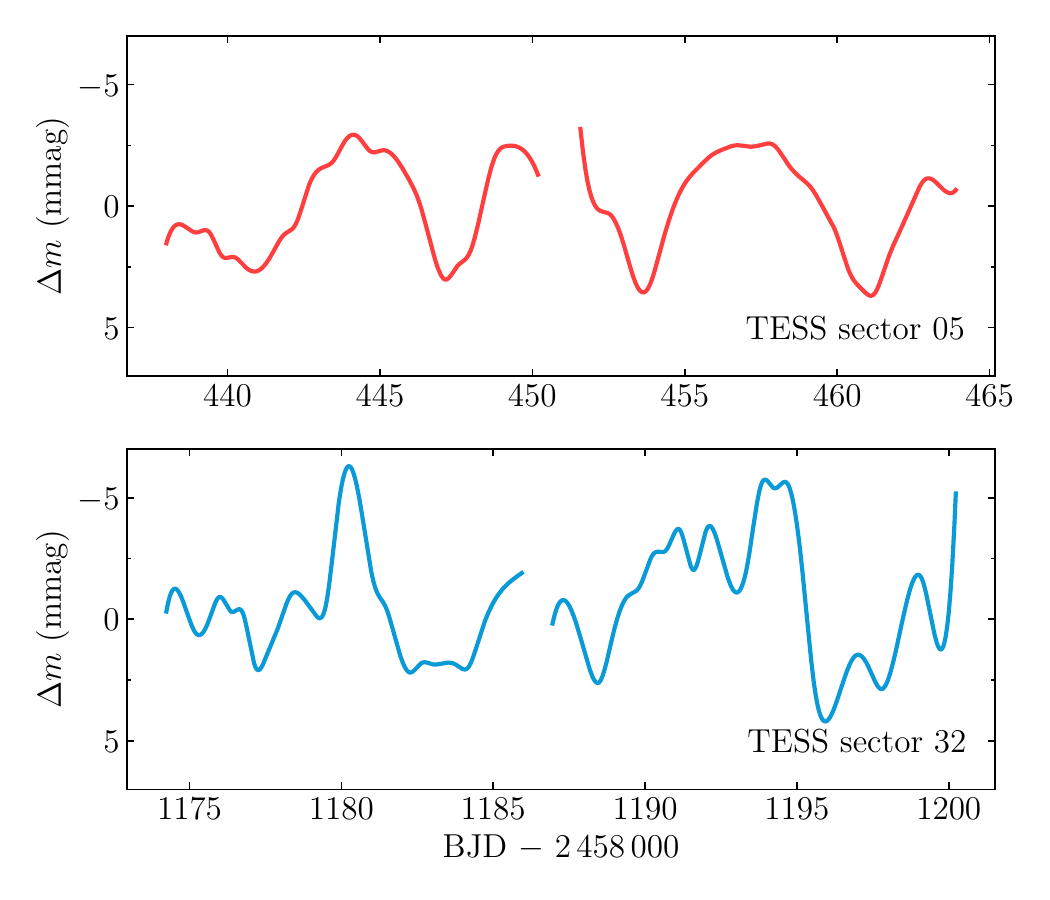}
\caption{The third component TESS light curve contribution in mmags for Sectors 05 and 32.} \label{fig:3component}
\end{figure}

Our follow-up photometry of a young red pre-main-sequence star UCAC4 414-008437 separated by 2.5 {\it TESS} pixels from \hvezda\ (Fig. \ref{fig:DK154}) shows that its light curve remains relatively smooth on the timescale of weeks (Fig. \ref{fig:3compCDK154}). This fact makes it possible to identify this star as a source of an additional signal in the {\it TESS} data (Sec. \ref{LC_double}) and correct these light curves for the contribution of specific UCAC4 414-008437 variations, assuming that all aperiodic variations (Fig.\,\ref{fig:3component}) longer than 0.1 days are caused solely by this object.

The relevance and accuracy of this correction, which decreases the scatter of the fit to 0.32 mmag, was independently confirmed by a custom treatment of the original {\it TESS} data using a smaller numerical aperture lowering several times the contribution of the parasitic light from the third component. We found that the corrected light curves obtained by both methods agree very well, but we opted for the first one as it is a bit more accurate.

\newpage
\onecolumn
\section*{Online material}
\setcounter{table}{0}
\renewcommand{\thetable}{\arabic{table}}
\begin{center}
	\begin{longtable}{l c c c c c c c c c}
 \captionsetup{width=.95\linewidth}
 \caption[Sumary]{Summary table with individual measurements and corresponding errors. Heliocentric Julian Date (HJD) is given for the middle of exposures. The second column contains the name of the used instrument. The longitudinal magnetic field \bz\ of the primary component and the method of its evaluation are in the third and fourth columns. The radial velocity of components A and B measured from the LSD profiles or from the modelling of Mg {\sc ii} 448.1 nm line are given as \vr(A) and \vr(B) with corresponding subscripts. By three asterisks in the sixth column, we marked the spectra averaged within the nights grouped by blank lines. The last three columns contain quadratic rotational phases $\vartheta_{A}$ and $\vartheta_{B}$, and the orbital phase $\varphi_\mathrm{orb}$. The full table is available online.} \label{table:summary} \\

\hline\hline
HJD       &  Instrument & $\langle B_\mathrm{z}\rangle$ & Method &  \vr(A)$_\mathrm{LSD}$   & \vr(B)$_\mathrm{Mg\textsc{ii}}$  & \vr(B)$_\mathrm{LSD}$ & $\vartheta_\mathrm{A}$ & $\vartheta_\mathrm{B}$  & $\varphi_\mathrm{orb}$ \\
2450000+  &             &  (G) &        &   (km\,s$^{-1}$)  &  ($\pm20$\,km\,s$^{-1}$) &  (km\,s$^{-1}$) & & & \\
\hline
\endfirsthead

\multicolumn{10}{c}%
{{\bfseries \tablename\ \thetable{} -- continued from previous page}} \\
\hline
HJD       &  Instrument & $\langle B_\mathrm{z}\rangle$ & Method &  \vr(A)$_\mathrm{LSD}$   & \vr(B)$_\mathrm{Mg\textsc{ii}}$  & \vr(B)$_\mathrm{LSD}$ & $\vartheta_\mathrm{A}$ & $\vartheta_\mathrm{B}$  & $\varphi_\mathrm{orb}$ \\
2450000+  &             &  (G) &        &   (km\,s$^{-1}$)  &  ($\pm20$\,km\,s$^{-1}$) &  (km\,s$^{-1}$) & & & \\
\hline
\endhead

\hline \multicolumn{10}{c}{{Continued on next page}} \\ \hline\hline
\endfoot

\hline\hline
\endlastfoot

6589.4929   &   MSS      &   $-3500\phantom{\pm}440$    &  COG  &    $-30.39\phantom{\pm}0.95$   &  $-$    & $-$ & 0.426 &  0.206 &  0.560  \\
& & & & & & & & \\
6639.4980   &   MSS      &   $-160\phantom{\pm}530$     &  COG  &    $40.14\phantom{\pm}0.50$    &  $-30$  & $-$ & 0.496 &  0.867 & 0.161   \\
& & & & & & & & \\
6644.4389   &   MSS      &   $-4580\phantom{\pm}560$    &  COG  &    $41.79\phantom{\pm}0.85$    &  $-5$   & $-$ & 0.356 &  0.320 & 0.220   \\
& & & & & & & & \\
6732.1696   &   MSS      &   $3650\phantom{\pm}360$     &  COG  &    $46.35\phantom{\pm}0.65$    &  $***$  & $-$ & 0.899 &  0.162 & 0.275 \\
6732.1855   &   MSS      &   $3841\phantom{\pm}340$     &  COG  &    $47.98\phantom{\pm}0.61$    &  $***$  & $-$ & 0.911 &  0.192 & 0.275 \\
6732.2008   &   MSS      &   $4338\phantom{\pm}270$     &  COG  &    $47.58\phantom{\pm}0.60$    &  $***$  & $-$ & 0.923 &  0.222 & 0.275 \\
6732.2154   &   MSS      &   $4600\phantom{\pm}360$     &  COG  &    $46.85\phantom{\pm}0.56$    &  $-20$  & $-$ & 0.935 &  0.250 & 0.275 \\
6732.2313   &   MSS      &   $4250\phantom{\pm}300$     &  COG  &    $46.53\phantom{\pm}0.53$    &  $***$  & $-$ & 0.947 &  0.280 & 0.275 \\
6732.2452   &   MSS      &   $5170\phantom{\pm}340$     &  COG  &    $45.88\phantom{\pm}0.54$    &  $***$  & $-$ & 0.958 &  0.307 & 0.275 \\
6732.2605   &   MSS      &   $4100\phantom{\pm}430$     &  COG  &    $44.67\phantom{\pm}0.49$    &  $***$  & $-$ & 0.970 &  0.336 & 0.276 \\
& & & & & & & & \\
6739.1905   &   MSS      &   $-3280\phantom{\pm}550$    &  COG  &    $56.19\phantom{\pm}0.43$    &  $***$  & $-$ & 0.384 &  0.596 & 0.359 \\
6739.2058   &   MSS      &   $-3000\phantom{\pm}340$    &  COG  &    $59.16\phantom{\pm}0.46$    &  $***$  & $-$ & 0.396 &  0.625 & 0.359 \\
6739.2204   &   MSS      &   $-4690\phantom{\pm}570$    &  COG  &    $60.42\phantom{\pm}0.46$    &  $-60$  & $-$ & 0.407 &  0.653 & 0.359 \\
6739.2356   &   MSS      &   $-4590\phantom{\pm}550$    &  COG  &    $62.10\phantom{\pm}0.46$    &  $***$  & $-$ & 0.419 &  0.682 & 0.359 \\
6739.2509   &   MSS      &   $-3590\phantom{\pm}500$    &  COG  &    $63.00\phantom{\pm}0.46$    &  $***$  & $-$ & 0.431 &  0.712 & 0.360 \\
& & & & & & & & \\
6740.1876   &   MSS      &   $-3400\phantom{\pm}630$    &  COG  &    $57.67\phantom{\pm}0.41$    &  $***$  & $-$ & 0.163 &  0.504 & 0.371 \\
6740.2022   &   MSS      &   $-3440\phantom{\pm}590$    &  COG  &    $59.67\phantom{\pm}0.44$    &  $***$  & $-$ & 0.174 &  0.532 & 0.371 \\
6740.2175   &   MSS      &   $-4400\phantom{\pm}450$    &  COG  &    $58.78\phantom{\pm}0.47$    &  $-67$  & $-$ & 0.186 &  0.561 & 0.371 \\
6740.2328   &   MSS      &   $-3400\phantom{\pm}570$    &  COG  &    $58.98\phantom{\pm}0.47$    &  $***$  & $-$ & 0.198 &  0.591 & 0.371 \\
6740.2473   &   MSS      &   $-4300\phantom{\pm}520$    &  COG  &    $61.16\phantom{\pm}0.48$    &  $***$  & $-$ & 0.209 &  0.618 & 0.372 \\
& & & & & & & & \\
6939.5133   &   MSS      &   $4330\phantom{\pm}1000$    &  COG  &    $-8.72\phantom{\pm}1.25$    &  $-$    & $-$ & 0.894 &  0.839 & 0.766 \\
6939.5551   &   MSS      &   $4250\phantom{\pm}930$     &  COG  &    $-1.52\phantom{\pm}1.13$    &  $-$    & $-$ & 0.927 &  0.919 & 0.767 \\
6939.5849   &   MSS      &   $4360\phantom{\pm}990$     &  COG  &    $4.89\phantom{\pm}1.09$     &  $-$    & $-$ & 0.950 &  0.976 & 0.767 \\
6939.6002   &   MSS      &   $3660\phantom{\pm}690$     &  COG  &    $6.04\phantom{\pm}1.02$     &  $-$    & $-$ & 0.962 &  0.006 & 0.767 \\
& & & & & & & & \\
6940.4503   &   MSS      &   $2170\phantom{\pm}480$     &  COG  &    $9.83\phantom{\pm}0.54$     &  $***$  & $-$ & 0.626 &  0.632 & 0.777 \\
6940.4655   &   MSS      &   $1760\phantom{\pm}900$     &  COG  &    $9.93\phantom{\pm}0.54$     &  $***$  & $-$ & 0.638 &  0.661 & 0.778 \\
6940.4815   &   MSS      &   $3070\phantom{\pm}670$     &  COG  &    $9.64\phantom{\pm}0.52$     &  $***$  & $-$ & 0.651 &  0.691 & 0.778 \\
6940.4967   &   MSS      &   $2320\phantom{\pm}820$     &  COG  &    $9.59\phantom{\pm}0.55$     &  $***$  & $-$ & 0.663 &  0.721 & 0.778 \\
6940.5183   &   MSS      &   $960\phantom{\pm}740$      &  COG  &    $8.73\phantom{\pm}0.56$     &  $65$   & $-$ & 0.679 &  0.762 & 0.778 \\
6940.5336   &   MSS      &   $1980\phantom{\pm}810$     &  COG  &    $8.31\phantom{\pm}0.59$     &  $***$  & $-$ & 0.691 &  0.791 & 0.778 \\
6940.5489   &   MSS      &   $2400\phantom{\pm}800$     &  COG  &    $6.88\phantom{\pm}0.56$     &  $***$  & $-$ & 0.703 &  0.820 & 0.779 \\
6940.5641   &   MSS      &   $1750\phantom{\pm}850$     &  COG  &    $6.95\phantom{\pm}0.62$     &  $***$  & $-$ & 0.715 &  0.849 & 0.779 \\
6940.5808   &   MSS      &   $3020\phantom{\pm}1150$    &  COG  &    $6.49\phantom{\pm}0.60$     &  $***$  & $-$ & 0.728 &  0.881 & 0.779 \\
6940.6051   &   MSS      &   $1160\phantom{\pm}980$     &  COG  &    $10.69\phantom{\pm}1.54$    &  $***$  & $-$ & 0.747 &  0.928 & 0.779 \\
& & & & & & & & \\
6966.5068   &   MSS      &   $4160\phantom{\pm}370$     &  COG  &    $26.46\phantom{\pm}0.51$    &  $***$  & $-$ & 0.984 &  0.480 & 0.090 \\
6966.5221   &   MSS      &   $3540\phantom{\pm}540$     &  COG  &    $28.22\phantom{\pm}0.48$    &  $***$  & $-$ & 0.996 &  0.510 & 0.091 \\
6966.5381   &   MSS      &   $3790\phantom{\pm}590$     &  COG  &    $29.33\phantom{\pm}0.46$    &  $***$  & $-$ & 0.009 &  0.540 & 0.091 \\
6966.5554   &   MSS      &   $3550\phantom{\pm}390$     &  COG  &    $30.65\phantom{\pm}0.47$    &  $0$    & $-$ & 0.022 &  0.573 & 0.091 \\
6966.5707   &   MSS      &   $2630\phantom{\pm}650$     &  COG  &    $32.15\phantom{\pm}0.45$    &  $***$  & $-$ & 0.034 &  0.603 & 0.091 \\
6966.5874   &   MSS      &   $3780\phantom{\pm}900$     &  COG  &    $33.04\phantom{\pm}0.44$    &  $***$  & $-$ & 0.047 &  0.635 & 0.091 \\
6966.6026   &   MSS      &   $2670\phantom{\pm}770$     &  COG  &    $34.29\phantom{\pm}0.42$    &  $***$  & $-$ & 0.059 &  0.664 & 0.092 \\
6966.6179   &   MSS      &   $3050\phantom{\pm}660$     &  COG  &    $34.84\phantom{\pm}0.45$    &  $***$  & $-$ & 0.071 &  0.693 & 0.092 \\
& & & & & & & & \\
6967.4062   &   MSS      &   $3680\phantom{\pm}650$     &  COG  &    $35.65\phantom{\pm}0.57$    &  $***$  & $-$ & 0.687 &  0.201 & 0.101 \\
6967.4207   &   MSS      &   $2650\phantom{\pm}610$     &  COG  &    $35.01\phantom{\pm}0.56$    &  $***$  & $-$ & 0.698 &  0.229 & 0.101 \\
6967.4367   &   MSS      &   $4040\phantom{\pm}790$     &  COG  &    $33.20\phantom{\pm}0.61$    &  $***$  & $-$ & 0.711 &  0.259 & 0.102 \\
6967.4527   &   MSS      &   $3200\phantom{\pm}830$     &  COG  &    $32.76\phantom{\pm}0.61$    &  $***$  & $-$ & 0.723 &  0.290 & 0.102 \\
6967.4694   &   MSS      &   $2540\phantom{\pm}480$     &  COG  &    $31.34\phantom{\pm}0.58$    &  $***$  & $-$ & 0.736 &  0.322 & 0.102 \\
6967.4853   &   MSS      &   $2800\phantom{\pm}510$     &  COG  &    $29.34\phantom{\pm}0.55$    &  $***$  & $-$ & 0.749 &  0.352 & 0.102 \\
6967.5006   &   MSS      &   $2050\phantom{\pm}430$     &  COG  &    $27.32\phantom{\pm}0.56$    &  $***$  & $-$ & 0.761 &  0.382 & 0.102 \\
6967.5159   &   MSS      &   $2590\phantom{\pm}580$     &  COG  &    $25.90\phantom{\pm}0.56$    &  $-5$   & $-$ & 0.773 &  0.411 & 0.103 \\
6967.5332   &   MSS      &   $2150\phantom{\pm}540$     &  COG  &    $23.42\phantom{\pm}0.58$    &  $***$  & $-$ & 0.786 &  0.444 & 0.103 \\
6967.5485   &   MSS      &   $3000\phantom{\pm}640$     &  COG  &    $21.49\phantom{\pm}0.61$    &  $***$  & $-$ & 0.798 &  0.473 & 0.103 \\
6967.5638   &   MSS      &   $3400\phantom{\pm}480$     &  COG  &    $21.51\phantom{\pm}0.60$    &  $***$  & $-$ & 0.810 &  0.503 & 0.103 \\
6967.5798   &   MSS      &   $3080\phantom{\pm}600$     &  COG  &    $22.26\phantom{\pm}0.63$    &  $***$  & $-$ & 0.822 &  0.533 & 0.103 \\
6967.5917   &   MSS      &   $3020\phantom{\pm}640$     &  COG  &    $21.13\phantom{\pm}0.59$    &  $***$  & $-$ & 0.832 &  0.556 & 0.103 \\
6967.6124   &   MSS      &   $2400\phantom{\pm}490$     &  COG  &    $21.28\phantom{\pm}0.59$    &  $***$  & $-$ & 0.848 &  0.596 & 0.104 \\
6967.6248   &   MSS      &   $3630\phantom{\pm}660$     &  COG  &    $21.18\phantom{\pm}0.61$    &  $***$  & $-$ & 0.858 &  0.619 & 0.104 \\
& & & & & & & & \\
6968.6215   &   MSS      &   $2120\phantom{\pm}770$     &  COG  &    $42.60\phantom{\pm}1.61$    &  $-$    & $-$ & 0.636 &  0.526 & 0.116 \\
& & & & & & & & \\
6969.6202   &   MSS      &   $-3000\phantom{\pm}1100$   &  COG  &    $37.71\phantom{\pm}1.32$    &  $-$    & $-$ & 0.417 &  0.437 & 0.128 \\
& & & & & & & & \\
6970.3813   &   MSS      &   $2000\phantom{\pm}850$     &  COG  &    $36.35\phantom{\pm}1.04$    &  $***$  & $-$ & 0.011 &  0.893 & 0.137 \\
6970.4077   &   MSS      &   $1520\phantom{\pm}640$     &  COG  &    $34.41\phantom{\pm}0.40$    &  $***$  & $-$ & 0.032 &  0.944 & 0.137 \\
6970.4230   &   MSS      &   $1760\phantom{\pm}800$     &  COG  &    $35.75\phantom{\pm}0.41$    &  $***$  & $-$ & 0.044 &  0.973 & 0.137 \\
6970.4382   &   MSS      &   $1340\phantom{\pm}820$     &  COG  &    $37.48\phantom{\pm}0.45$    &  $***$  & $-$ & 0.056 &  0.002 & 0.138 \\
6970.4535   &   MSS      &   $1940\phantom{\pm}600$     &  COG  &    $40.00\phantom{\pm}0.40$    &  $***$  & $-$ & 0.068 &  0.031 & 0.138 \\
6970.4778   &   MSS      &   $270\phantom{\pm}770$      &  COG  &    $40.58\phantom{\pm}0.41$    &  $***$  & $-$ & 0.087 &  0.078 & 0.138 \\
6970.4931   &   MSS      &   $-220\phantom{\pm}970$     &  COG  &    $41.55\phantom{\pm}0.40$    &  $***$  & $-$ & 0.098 &  0.107 & 0.138 \\
6970.5084   &   MSS      &   $-750\phantom{\pm}770$     &  COG  &    $41.48\phantom{\pm}0.41$    &  $-10$  & $-$ & 0.110 &  0.136 & 0.139 \\
6970.5230   &   MSS      &   $-540\phantom{\pm}790$     &  COG  &    $40.22\phantom{\pm}0.41$    &  $***$  & $-$ & 0.122 &  0.164 & 0.139 \\
6970.5383   &   MSS      &   $-710\phantom{\pm}820$     &  COG  &    $39.46\phantom{\pm}0.39$    &  $***$  & $-$ & 0.134 &  0.194 & 0.139 \\
6970.5528   &   MSS      &   $-1100\phantom{\pm}740$    &  COG  &    $38.35\phantom{\pm}0.39$    &  $***$  & $-$ & 0.145 &  0.221 & 0.139 \\
6970.5681   &   MSS      &   $-1730\phantom{\pm}650$    &  COG  &    $37.64\phantom{\pm}0.37$    &  $***$  & $-$ & 0.157 &  0.251 & 0.139 \\
6970.5834   &   MSS      &   $-3880\phantom{\pm}800$    &  COG  &    $36.90\phantom{\pm}0.38$    &  $***$  & $-$ & 0.169 &  0.280 & 0.139 \\
6970.5980   &   MSS      &   $-3110\phantom{\pm}880$    &  COG  &    $36.24\phantom{\pm}0.36$    &  $***$  & $-$ & 0.180 &  0.308 & 0.140 \\
& & & & & & & & \\
6972.3932   &   MSS      &   $-700\phantom{\pm}725$     &  COG  &    $39.22\phantom{\pm}0.46$    &  $***$  & $-$ & 0.583 &  0.742 & 0.161 \\
6972.4085   &   MSS      &   $1160\phantom{\pm}550$     &  COG  &    $40.16\phantom{\pm}0.47$    &  $***$  & $-$ & 0.595 &  0.772 & 0.161 \\
6972.4231   &   MSS      &   $2100\phantom{\pm}530$     &  COG  &    $39.37\phantom{\pm}0.49$    &  $***$  & $-$ & 0.606 &  0.800 & 0.162 \\
6972.4383   &   MSS      &   $2420\phantom{\pm}700$     &  COG  &    $40.15\phantom{\pm}0.49$    &  $***$  & $-$ & 0.618 &  0.829 & 0.162 \\
6972.4529   &   MSS      &   $1650\phantom{\pm}660$     &  COG  &    $40.32\phantom{\pm}0.51$    &  $-20$  & $-$ & 0.630 &  0.857 & 0.162 \\
6972.4682   &   MSS      &   $2840\phantom{\pm}980$     &  COG  &    $40.53\phantom{\pm}0.51$    &  $***$  & $-$ & 0.642 &  0.886 & 0.162 \\
6972.4828   &   MSS      &   $2360\phantom{\pm}880$     &  COG  &    $39.44\phantom{\pm}0.51$    &  $***$  & $-$ & 0.653 &  0.914 & 0.162 \\
6972.5476   &   MSS      &   $2140\phantom{\pm}580$     &  COG  &    $41.18\phantom{\pm}0.54$    &  $***$  & $-$ & 0.704 &  0.038 & 0.163 \\
& & & & & & & & \\
6973.3453   &   MSS      &   $-4790\phantom{\pm}670$    &  COG  &    $37.84\phantom{\pm}0.39$    &  $***$  & $-$ & 0.327 &  0.564 & 0.173 \\
6973.3599   &   MSS      &   $-4670\phantom{\pm}650$    &  COG  &    $38.87\phantom{\pm}0.41$    &  $***$  & $-$ & 0.338 &  0.592 & 0.173 \\
6973.3752   &   MSS      &   $-3860\phantom{\pm}630$    &  COG  &    $38.67\phantom{\pm}0.37$    &  $***$  & $-$ & 0.350 &  0.621 & 0.173 \\
6973.3898   &   MSS      &   $-3600\phantom{\pm}830$    &  COG  &    $39.07\phantom{\pm}0.44$    &  $***$  & $-$ & 0.362 &  0.649 & 0.173 \\
6973.4050   &   MSS      &   $-3640\phantom{\pm}690$    &  COG  &    $38.65\phantom{\pm}0.44$    &  $-20$  & $-$ & 0.373 &  0.678 & 0.173 \\
6973.4196   &   MSS      &   $-4430\phantom{\pm}610$    &  COG  &    $39.40\phantom{\pm}0.45$    &  $***$  & $-$ & 0.385 &  0.706 & 0.174 \\
6973.4349   &   MSS      &   $-4360\phantom{\pm}460$    &  COG  &    $38.48\phantom{\pm}0.43$    &  $***$  & $-$ & 0.397 &  0.735 & 0.174 \\
6973.4495   &   MSS      &   $-2950\phantom{\pm}590$    &  COG  &    $38.78\phantom{\pm}0.47$    &  $***$  & $-$ & 0.408 &  0.763 & 0.174 \\
6973.4648   &   MSS      &   $-2090\phantom{\pm}520$    &  COG  &    $38.25\phantom{\pm}0.44$    &  $***$  & $-$ & 0.420 &  0.793 & 0.174 \\
6973.4794   &   MSS      &   $-2240\phantom{\pm}490$    &  COG  &    $39.33\phantom{\pm}0.46$    &  $***$  & $-$ & 0.432 &  0.821 & 0.174 \\
6973.4981   &   MSS      &   $-3040\phantom{\pm}780$    &  COG  &    $40.12\phantom{\pm}1.26$    &  $***$  & $-$ & 0.446 &  0.856 & 0.174 \\
6973.5134   &   MSS      &   $-2300\phantom{\pm}1180$   &  COG  &    $40.08\phantom{\pm}1.26$    &  $***$  & $-$ & 0.458 &  0.886 & 0.175 \\
6973.5287   &   MSS      &   $-2200\phantom{\pm}660$    &  COG  &    $40.10\phantom{\pm}1.25$    &  $***$  & $-$ & 0.470 &  0.915 & 0.175 \\
6973.5432   &   MSS      &   $-400\phantom{\pm}1000$    &  COG  &    $40.94\phantom{\pm}1.24$    &  $***$  & $-$ & 0.481 &  0.943 & 0.175 \\
6973.5585   &   MSS      &   $-680\phantom{\pm}550$     &  COG  &    $40.39\phantom{\pm}1.19$    &  $***$  & $-$ & 0.493 &  0.972 & 0.175 \\
& & & & & & & & \\
6993.5633   &   MSS      &   $890\phantom{\pm}920$      &  COG  &    $84.28\phantom{\pm}0.92$    &  $-$    & $-$ & 0.121 &  0.250 & 0.416 \\
& & & & & & & & \\
6995.3952   &   MSS      &   $770\phantom{\pm}760$      &  COG  &    $94.03\phantom{\pm}1.09$    &  $***$    & $***$                  & 0.552 & 0.755 & 0.438 \\
6995.4119   &   MSS      &   $1540\phantom{\pm}1000$    &  COG  &    $94.69\phantom{\pm}0.91$    &  $***$  & $***$                  & 0.565 & 0.787 & 0.438 \\
6995.4279   &   MSS      &   $840\phantom{\pm}920$      &  COG  &    $95.29\phantom{\pm}1.10$    &  $-20$  & $-88.6\phantom{pm}7.7$ & 0.578 & 0.818 & 0.438 \\
6995.4438   &   MSS      &   $2920\phantom{\pm}720$     &  COG  &    $95.25\phantom{\pm}1.14$    &  $***$  & $***$                  & 0.590 & 0.848 & 0.438 \\
6995.4591   &   MSS      &   $3920\phantom{\pm}970$     &  COG  &    $95.98\phantom{\pm}1.02$    &  $***$  & $***$                  & 0.602 & 0.878 & 0.438 \\
6995.4924   &   MSS      &   $2030\phantom{\pm}1080$    &  COG  &    $102.41\phantom{\pm}1.59$   &  $***$  & $***$                  & 0.628 & 0.941 & 0.439 \\
6995.5077   &   MSS      &   $1200\phantom{\pm}1200$    &  COG  &    $102.14\phantom{\pm}1.56$   &  $***$  & $***$                  & 0.640 & 0.971 & 0.439 \\
6995.5237   &   MSS      &   $2870\phantom{\pm}760$     &  COG  &    $103.12\phantom{\pm}1.53$   &  $***$  & $***$                  & 0.652 & 0.001 & 0.439 \\
& & & & & & & & \\
7085.1727   &   MSS      &   $2510\phantom{\pm}520$     &  COG  &    $-41.91\phantom{\pm}1.44$   &  $***$  & $***$                  & 0.693 & 0.514 & 0.516 \\
7085.1949   &   MSS      &   $2700\phantom{\pm}330$     &  COG  &    $-44.80\phantom{\pm}1.42$   &  $***$  & $***$                  & 0.711 & 0.556 & 0.517 \\
7085.2172   &   MSS      &   $2555\phantom{\pm}395$     &  COG  &    $-48.32\phantom{\pm}1.43$   &  $95$   & $111.9\phantom{pm}6.3$ & 0.728 & 0.599 & 0.517 \\
7085.2394   &   MSS      &   $3230\phantom{\pm}635$     &  COG  &    $-50.32\phantom{\pm}1.36$   &  $***$  & $***$                  & 0.745 & 0.641 & 0.517 \\
& & & & & & & & \\
7090.2264   &   MSS      &   $2610\phantom{\pm}490$     &  COG  &    $-12.64\phantom{\pm}0.57$   &  $***$  &  $-$ & 0.642 & 0.180 & 0.577 \\
7090.2452   &   MSS      &   $2700\phantom{\pm}300$     &  COG  &    $-16.01\phantom{\pm}0.58$   &  $***$  &  $-$ & 0.657 & 0.216 & 0.577 \\
7090.2605   &   MSS      &   $2870\phantom{\pm}540$     &  COG  &    $-18.74\phantom{\pm}0.59$   &  $80$   &  $-$ & 0.669 & 0.245 & 0.578 \\
7090.2778   &   MSS      &   $2500\phantom{\pm}500$     &  COG  &    $-21.27\phantom{\pm}0.59$   &  $***$  &  $-$ & 0.683 & 0.278 & 0.578 \\
7090.2931   &   MSS      &   $2220\phantom{\pm}370$     &  COG  &    $-21.36\phantom{\pm}0.60$   &  $***$  &  $-$ & 0.695 & 0.307 & 0.578 \\
& & & & & & & & \\
7091.1757   &   MSS      &   $-3660\phantom{\pm}500$    &  COG  &    $-17.28\phantom{\pm}0.38$   &  $***$  &  $-$ & 0.384 & 0.996 & 0.589 \\
7091.1930   &   MSS      &   $-3850\phantom{\pm}430$    &  COG  &    $-20.46\phantom{\pm}0.38$   &  $***$  &  $-$ & 0.398 & 0.029 & 0.589 \\
7091.2090   &   MSS      &   $-3850\phantom{\pm}620$    &  COG  &    $-21.64\phantom{\pm}0.38$   &  $***$  &  $-$ & 0.410 & 0.059 & 0.589 \\
7091.2243   &   MSS      &   $-2640\phantom{\pm}600$    &  COG  &    $-22.07\phantom{\pm}0.40$   &  $***$  &  $-$ & 0.422 & 0.089 & 0.589 \\
7091.2396   &   MSS      &   $-3530\phantom{\pm}540$    &  COG  &    $-21.77\phantom{\pm}0.42$   &  $90$   &  $-$ & 0.434 & 0.118 & 0.589 \\
7091.2562   &   MSS      &   $-2480\phantom{\pm}840$    &  COG  &    $-21.84\phantom{\pm}0.43$   &  $***$  &  $-$ & 0.447 & 0.150 & 0.589 \\
7091.2715   &   MSS      &   $-1800\phantom{\pm}630$    &  COG  &    $-21.88\phantom{\pm}0.43$   &  $***$  &  $-$ & 0.459 & 0.179 & 0.590 \\
7091.2868   &   MSS      &   $-1920\phantom{\pm}630$    &  COG  &    $-22.09\phantom{\pm}0.45$   &  $***$  &  $-$ & 0.471 & 0.208 & 0.590 \\
& & & & & & & & \\
7092.1770   &   MSS      &   $-2560\phantom{\pm}390$    &  COG  &    $-18.72\phantom{\pm}0.40$   &  $***$  &  $-$ & 0.167 & 0.911 & 0.601 \\
7092.1943   &   MSS      &   $-3130\phantom{\pm}610$    &  COG  &    $-17.16\phantom{\pm}0.40$   &  $***$  &  $-$ & 0.180 & 0.944 & 0.601 \\
7092.2353   &   MSS      &   $-3450\phantom{\pm}510$    &  COG  &    $-16.30\phantom{\pm}0.38$   &  $90$   &  $-$ & 0.212 & 0.022 & 0.601 \\
7092.2506   &   MSS      &   $-4000\phantom{\pm}500$    &  COG  &    $-17.02\phantom{\pm}0.39$   &  $***$  &  $-$ & 0.224 & 0.052 & 0.601 \\
7092.2652   &   MSS      &   $-4850\phantom{\pm}550$    &  COG  &    $-17.47\phantom{\pm}0.37$   &  $***$  &  $-$ & 0.236 & 0.080 & 0.602 \\
7092.2818   &   MSS      &   $-4470\phantom{\pm}430$    &  COG  &    $-17.38\phantom{\pm}0.37$   &  $***$  &  $-$ & 0.248 & 0.111 & 0.602 \\
& & & & & & & & \\
7331.4324   &   MSS      &   $-$                        &  $-$  &    $88.12\phantom{\pm}0.70$    &  $-76$  &  $-$ & 0.090 & 0.646 & 0.476 \\
& & & & & & & & \\
7332.5339   &   MSS      &   $3500\phantom{\pm}520$     &  COG  &    $-1.07\phantom{\pm}0.43$    &  $95$   &  $-$ & 0.951 & 0.754 & 0.489 \\
& & & & & & & & \\
7414.2401   &   MSS      &   $3180\phantom{\pm}380$     &  COG  &    $95.17\phantom{\pm}1.82$    &  $***$  &  $***$                  & 0.786 & 0.070 & 0.471 \\
7414.2623   &   MSS      &   $3550\phantom{\pm}740$     &  COG  &    $92.24\phantom{\pm}1.86$    &  $***$  &  $***$                  & 0.803 & 0.112 & 0.471 \\
7414.2998   &   MSS      &   $4240\phantom{\pm}450$     &  COG  &    $83.89\phantom{\pm}1.04$    &  $***$  &  $***$                  & 0.833 & 0.184 & 0.471 \\
7414.3214   &   MSS      &   $4720\phantom{\pm}580$     &  COG  &    $82.74\phantom{\pm}1.18$    &  $-80$  & $-107.8\phantom{pm}8.6$ & 0.849 & 0.225 & 0.472 \\
7414.3540   &   MSS      &   $3530\phantom{\pm}550$     &  COG  &    $81.76\phantom{\pm}1.11$    &  $***$  &  $***$                  & 0.875 & 0.287 & 0.472 \\
& & & & & & & & \\
8178.2653   &   MSS      &   $1510\phantom{\pm}680$     &  COG  &    $-10.43\phantom{\pm}0.35$   &  $65$   & $-$ & 0.693 & 0.765 & 0.651 \\
8178.2799   &   MSS      &   $2360\phantom{\pm}670$     &  COG  &    $-10.92\phantom{\pm}0.36$   &  $***$  & $-$ & 0.704 & 0.793 & 0.652 \\
& & & & & & & & \\
9211.3463   &   MSS      &   $3360\phantom{\pm}570$     &  COG  &    $22.19\phantom{\pm}0.95$    &  $-$    & $-$ & 0.780 & 0.223 & 0.065 \\
& & & & & & & & \\
9213.3581   &   MSS      &   $-5930\phantom{\pm}480$    &  COG  &    $28.18\phantom{\pm}0.73$    &  $-$    & $-$ & 0.352 & 0.072 & 0.090 \\
& & & & & & & & \\
6972.1470   &   ESPaDOnS &   $-4020\phantom{\pm}96$     &  LSD  &    $34.66\phantom{\pm}0.21$    &  $0$    & $-$                      & 0.391 &  0.271 & 0.158 \\
7013.0363   &   ESPaDOnS &   $-4905\phantom{\pm}98$     &  LSD  &    $-10.46\phantom{\pm}0.17$   &  $88$   & $-$                      & 0.336 &  0.502 & 0.650 \\
7021.0463   &   ESPaDOnS &   $1958\phantom{\pm}99$      &  LSD  &    $4.44\phantom{\pm}0.21$     &  $35$   & $-$                      & 0.594 &  0.824 & 0.746 \\
7021.9910   &   ESPaDOnS &   $-5158\phantom{\pm}77$     &  LSD  &    $1.20\phantom{\pm}0.16$     &  $45$   & $-$                      & 0.333 &  0.631 & 0.757 \\
7030.0351   &   ESPaDOnS &   $2403\phantom{\pm}108$     &  LSD  &    $17.39\phantom{\pm}0.21$    &  $30$   & $-$                      & 0.618 &  0.020 & 0.854 \\
7031.0379   &   ESPaDOnS &   $-3517\phantom{\pm}104$    &  LSD  &    $9.77\phantom{\pm}0.22$     &  $26$   & $-$                      & 0.401 &  0.938 & 0.866 \\
7032.0343   &   ESPaDOnS &   $-4858\phantom{\pm}81$     &  LSD  &    $13.11\phantom{\pm}0.17$    &  $25$   & $-$                      & 0.180 &  0.844 & 0.878 \\
7033.0303   &   ESPaDOnS &   $4054\phantom{\pm}109$     &  LSD  &    $8.43\phantom{\pm}0.29$     &  $20$   & $-$                      & 0.958 &  0.750 & 0.890 \\
7034.0247   &   ESPaDOnS &   $3371\phantom{\pm}81$      &  LSD  &    $17.04\phantom{\pm}0.20$    &  $13$   & $-$                      & 0.735 &  0.652 & 0.902 \\
7035.0138   &   ESPaDOnS &   $-174\phantom{\pm}115$     &  LSD  &    $19.46\phantom{\pm}0.25$    &  $37$   & $-$                      & 0.508 &  0.544 & 0.914 \\
7405.9464   &   ESPaDOnS &   $-5158\phantom{\pm}96$     &  LSD  &    $65.32\phantom{\pm}0.17$    &  $-55$  & $-59.4\phantom{pm}4.4$   & 0.308 &  0.198 & 0.371 \\
7406.8023   &   ESPaDOnS &   $3622\phantom{\pm}120$     &  LSD  &    $64.35\phantom{\pm}0.24$    &  $-$    & $-$                      & 0.976 &  0.836 & 0.381 \\
7413.7178   &   ESPaDOnS &   $-4572\phantom{\pm}123$    &  LSD  &    $96.76\phantom{\pm}0.19$    &  $-70$  & $-60.7\phantom{pm}4.3$   & 0.378 &  0.070 & 0.464 \\
7413.9017   &   ESPaDOnS &   $365\phantom{\pm}131$      &  LSD  &    $102.02\phantom{\pm}0.20$   &  $-70$  & $-89.3\phantom{pm}2.9$   & 0.522 &  0.422 & 0.467 \\
7414.7185   &   ESPaDOnS &   $-4543\phantom{\pm}99$     &  LSD  &    $79.92\phantom{\pm}0.15$    &  $-40$  & $-48.0\phantom{pm}3.6$   & 0.160 &  0.985 & 0.476 \\
7414.8326   &   ESPaDOnS &   $-5632\phantom{\pm}88$     &  LSD  &    $73.13\phantom{\pm}0.17$    &  $-50$  & $-64.5\phantom{pm}3.9$   & 0.249 &  0.203 & 0.478 \\
7414.9724   &   ESPaDOnS &   $-4632\phantom{\pm}109$    &  LSD  &    $63.44\phantom{\pm}0.17$    &  $-40$  & $-48.5\phantom{pm}4.4$   & 0.358 &  0.471 & 0.479 \\
7415.7189   &   ESPaDOnS &   $3996\phantom{\pm}110$     &  LSD  &    $0.42\phantom{\pm}0.28$     &  $60$   & $-$                      & 0.941 &  0.899 & 0.488 \\
7415.8335   &   ESPaDOnS &   $1202\phantom{\pm}116$     &  LSD  &    $-1.22\phantom{\pm}0.24$    &  $65$   & $-$                      & 0.031 &  0.118 & 0.490 \\
7415.9739   &   ESPaDOnS &   $-3798\phantom{\pm}105$    &  LSD  &    $-7.39\phantom{\pm}0.20$    &  $65$   & $-$                      & 0.140 &  0.387 & 0.491 \\
7416.7190   &   ESPaDOnS &   $3945\phantom{\pm}97$      &  LSD  &    $-30.27\phantom{\pm}0.23$   &  $72$   & $-$                      & 0.723 &  0.812 & 0.500 \\
7416.9011   &   ESPaDOnS &   $5010\phantom{\pm}105$     &  LSD  &    $-44.10\phantom{\pm}0.22$   &  $105$  & $123.2\phantom{pm}3.7$   & 0.865 &  0.160 & 0.503 \\
6972.9243   &   dimaPol  &   $4310\phantom{\pm}796$     &  Pol  &    $-$  &  $-$  &  $-$   &  0.998 & 0.759 & 0.168 \\
6973.9366   &   dimaPol  &   $5429\phantom{\pm}438$     &  Pol  &    $-$  &  $-$  &  $-$   &  0.789 & 0.695 & 0.180 \\
6974.9273   &   dimaPol  &   $3012\phantom{\pm}380$     &  Pol  &    $-$  &  $-$  &  $-$   &  0.563 & 0.591 & 0.192 \\
6975.9563   &   dimaPol  &   $-5266\phantom{\pm}473$    &  Pol  &    $-$  &  $-$  &  $-$   &  0.367 & 0.560 & 0.204 \\
6991.8906   &   dimaPol  &   $3512\phantom{\pm}448$     &  Pol  &    $-$  &  $-$  &  $-$   &  0.815 & 0.049 & 0.395 \\
6992.8882   &   dimaPol  &   $2617\phantom{\pm}269$     &  Pol  &    $-$  &  $-$  &  $-$   &  0.594 & 0.958 & 0.407 \\
6994.9034   &   dimaPol  &   $-4446\phantom{\pm}651$    &  Pol  &    $-$  &  $-$  &  $-$   &  0.168 & 0.814 & 0.432 \\
7084.6555   &   dimaPol  &   $-6850\phantom{\pm}475$    &  Pol  &    $-$  &  $-$  &  $-$   &  0.289 & 0.525 & 0.510 \\
7085.6545   &   dimaPol  &   $1419\phantom{\pm}351$     &  Pol  &    $-$  &  $-$  &  $-$   &  0.070 & 0.435 & 0.522 \\
7088.6693   &   dimaPol  &   $-3993\phantom{\pm}279$    &  Pol  &    $-$  &  $-$  &  $-$   &  0.426 & 0.202 & 0.558 \\
7330.6680   &   HERMES   &   $-$                        &  $-$  &    $103.44\phantom{\pm}0.51$   &  $-105$  & $-78.7\phantom{pm}4.9$  & 0.493 & 0.183 &  0.466 \\
7331.6653   &   HERMES   &   $-$                        &  $-$  &    $74.51\phantom{\pm}0.43$    &  $-60$   & $-$                     & 0.272 & 0.092 &  0.478 \\
7332.6568   &   HERMES   &   $-$                        &  $-$  &    $-5.65\phantom{\pm}0.38$    &  $85$    & $-$                     & 0.047 & 0.989 &  0.490 \\
7333.6613   &   HERMES   &   $-$                        &  $-$  &    $-44.59\phantom{\pm}0.55$   &  $124$   & $-$                     & 0.832 & 0.910 &  0.502 \\
7334.6584   &   HERMES   &   $-$                        &  $-$  &    $-33.35\phantom{\pm}0.27$   &  $84$    & $-$                     & 0.611 & 0.817 &  0.514 \\
7335.5781   &   HERMES   &   $-$                        &  $-$  &    $-44.35\phantom{\pm}0.46$   &  $-$     & $-$                     & 0.330 & 0.576 &  0.525 \\
7335.6840   &   HERMES   &   $-$                        &  $-$  &    $-41.51\phantom{\pm}0.65$   &  $-$     & $-$                     & 0.412 & 0.779 &  0.527 \\
7370.5438   &   HERMES   &   $-$                        &  $-$  &    $23.82\phantom{\pm}0.40$    &  $19$    & $-$                     & 0.650 & 0.463 &  0.946 \\
7413.3551   &   HERMES   &   $-$                        &  $-$  &    $104.55\phantom{\pm}0.45$   &  $-109$  & $-70.8\phantom{pm}4.0$  & 0.095 & 0.376 &  0.460 \\
7413.4886   &   HERMES   &   $-$                        &  $-$  &    $103.41\phantom{\pm}0.46$   &  $-90$   & $-$                     & 0.199 & 0.631 &  0.462 \\
7414.3730   &   HERMES   &   $-$                        &  $-$  &    $84.48\phantom{\pm}0.43$    &  $-70$   & $-108.3\phantom{pm}4.1$ & 0.890 & 0.324 &  0.472 \\
7414.4669   &   HERMES   &   $-$                        &  $-$  &    $86.76\phantom{\pm}0.46$    &  $-70$   & $-92.7\phantom{pm}4.3$  & 0.963 & 0.504 &  0.473 \\
7414.5563   &   HERMES   &   $-$                        &  $-$  &    $87.07\phantom{\pm}0.48$    &  $-50$   & $-90.1\phantom{pm}5.3$  & 0.033 & 0.675 &  0.474 \\
7415.3874   &   HERMES   &   $-$                        &  $-$  &    $38.90\phantom{\pm}0.30$    &  $0$     & $-$                     & 0.682 & 0.265 &  0.484 \\
7415.4866   &   HERMES   &   $-$                        &  $-$  &    $23.31\phantom{\pm}0.35$    &  $0$     & $-$                     & 0.760 & 0.455 &  0.486 \\
7416.3684   &   HERMES   &   $-$                        &  $-$  &    $-27.62\phantom{\pm}0.73$   &  $70$    & $138.7\phantom{pm}3.4$  & 0.449 & 0.142 &  0.496 \\
7417.0260   &   MRES     &   $-$                        &  $-$  &    $-41.08\phantom{\pm}0.19$   &  $100$   & $132.4\phantom{pm}6.7$  & 0.963 & 0.399 &  0.504 \\
9563.0630   &   MRES     &   $-$                        &  $-$  &    $57.6\phantom{\pm}0.8$      &  $-$     & $-27.9\phantom{pm}8.7$  & 0.549 & 0.125 &  0.292 \\
9564.1195   &   MRES     &   $-$                        &  $-$  &    $53.3\phantom{\pm}0.9$      &  $-$     & $-8.7\phantom{pm}6.2$   & 0.375 & 0.146 &  0.305 \\
9565.0816   &   MRES     &   $-$                        &  $-$  &    $58.5\phantom{\pm}0.5$      &  $-$     & $-17.7\phantom{pm}3.0$  & 0.126 & 0.987 &  0.316 \\
\end{longtable}
\end{center}
\twocolumn

\end{document}